\documentclass[11pt]{article}
\pdfoutput=1
\usepackage{jcapmod}

\usepackage{booktabs}
\usepackage[english]{babel}
\usepackage{amsmath,amssymb,amsbsy,amstext, amsthm, simplewick}
\usepackage{hyperref}
\usepackage{graphicx}
\usepackage{amsfonts}
\usepackage{amssymb}
\usepackage[small]{caption}
\usepackage{siunitx}
\usepackage{upgreek}
  \usepackage{framed}
  \usepackage{wrapfig}
  \usepackage{multirow}
  \usepackage{subfig}

\usepackage[usenames]{xcolor}

\usepackage{colortbl}
\definecolor{lightgreen}{cmyk}{0.2, 0, 0.2, 0.2}
\definecolor{lightgray}{cmyk}{0.1,0.2,0,0.1}
\definecolor{lightgray2}{cmyk}{0.1,0.1,0,0.1}

\makeatletter
\newlength{\apb@width}
\newcommand{\autoparbox}[2][c]{\settowidth{\apb@width}{#2}\parbox[#1]{\apb@width}{#2}}
\newcommand{\includegraphicsbox}[2][]{\autoparbox{\includegraphics[#1]{#2}}}
\makeatother

\numberwithin{equation}{section}

\def\beq{\begin{equation}}
\def\eeq{\end{equation}}

\def\bea{\begin{eqnarray}}
\def\eea{\end{eqnarray}}

\def\Beq{\begin{equation}\begin{aligned}}
\def\Eeq{\end{aligned}\end{equation}}

\def\d{{\rm d}}

\def\beq{\begin{equation}}
\def\eeq{\end{equation}}
\def\bea{\begin{eqnarray}}
\def\eea{\end{eqnarray}}

\def\d{{\rm d}}

\def\Mp{M_{\rm pl}}
\def\d{{\rm d}}

\def\k{{\bf k}}
\def\n{{\bf n}}
\def\q{{\bf q}}

\def\x{{\bf x}}

\def\z{{\bf z}}
\def\cs{c_\pi}
\def\ct{c_\gamma}

\DeclareRobustCommand{\SkipTocEntry}[4]{}

\begin{document}

\begin{titlepage}

\setcounter{page}{1} \baselineskip=15.5pt \thispagestyle{empty}

\bigskip\

\vspace{1cm}
\begin{center}

{\fontsize{20}{24}\selectfont  \sffamily \bfseries  Non-Gaussianity as a Particle Detector}
\end{center}

\vspace{0.2cm}
\begin{center}
{\fontsize{13}{30}\selectfont  Hayden Lee,$^{\bigstar}$ Daniel Baumann,$^{\bigstar,\hskip 1pt\spadesuit}$ and Guilherme L. Pimentel$^{\bigstar,\hskip 1pt\spadesuit}$} 
\end{center}

\begin{center}

\vskip 8pt
\textsl{$^{\bigstar}$ Department of Applied Mathematics and Theoretical Physics,\\ 
Cambridge University, Cambridge, CB3 0WA, UK}

\vskip 8pt
\textsl{$^{\spadesuit}$ Institute of Physics, Universiteit van Amsterdam,\\ Science Park, Amsterdam, 1090 GL, The Netherlands}  

\vskip 7pt

\end{center}

\vspace{1.2cm}
\hrule \vspace{0.3cm}
\noindent {\sffamily \bfseries Abstract} \\[0.1cm]
We study the imprints of massive particles with spin on cosmological correlators. Using the framework of the effective field theory of inflation, we classify the couplings of these particles to the Goldstone boson of broken time translations and the graviton. We show that it is possible to generate observable non-Gaussianity within the regime of validity of the effective theory, as long as the masses of the particles are close to the Hubble scale and their interactions break the approximate conformal symmetry of the inflationary background.  We derive explicit shape functions for the scalar and tensor bispectra that can serve as templates for future observational searches.  
\vskip 10pt
\hrule
\vskip 10pt

\vspace{0.6cm}
 \end{titlepage}

\tableofcontents

\newpage
\section{Introduction}

Establishing the field content during inflation is a fundamental challenge of primordial cosmology. Minimal inflationary models have two massless fields: the Goldstone boson of broken time translations,\footnote{Strictly speaking, $\pi$ is only massless in the decoupling limit $\Mp \to \infty$. However, for adiabatic fluctuations, $\pi$ is directly related to the comoving curvature perturbation, $\zeta=-H\pi +{\cal O}(\pi^2)$, which is the true massless degree of freedom even away from the decoupling limit.} $\pi$, and the graviton, $\gamma_{ij}$. While at present there is no evidence for additional degrees of freedom~\cite{Ade:2015ava}, the imprints of extra particles can be subtle, so it remains important to fully characterize their effects and compare them to observations.  Moreover, massive particles are important probes of the ultraviolet completion of inflation. For example, in string theory, massive particles in the low-energy effective theory encode physics at the string and Kaluza-Klein scales~\cite{Baumann:2014nda}. If these scales aren't too far from the inflationary Hubble scale, then their influence may be observable (although the experimental challenge could be enormous).

\vskip 4pt
 \begin{figure}[h!]
\centering
\includegraphics[scale=0.45]{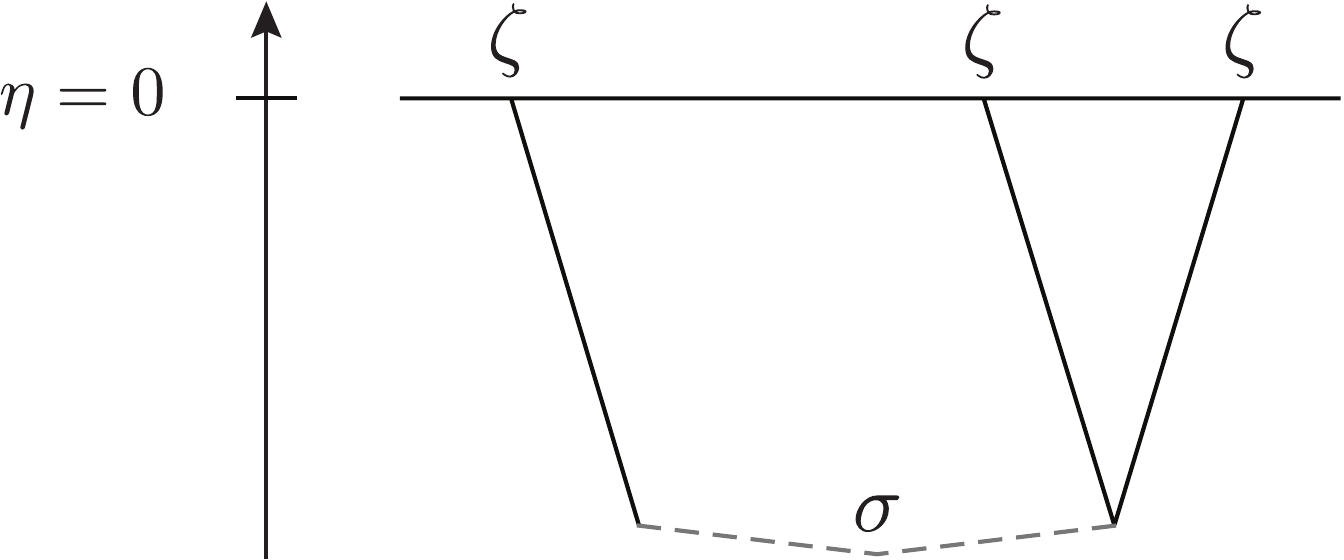}\qquad
\includegraphics[scale=0.45]{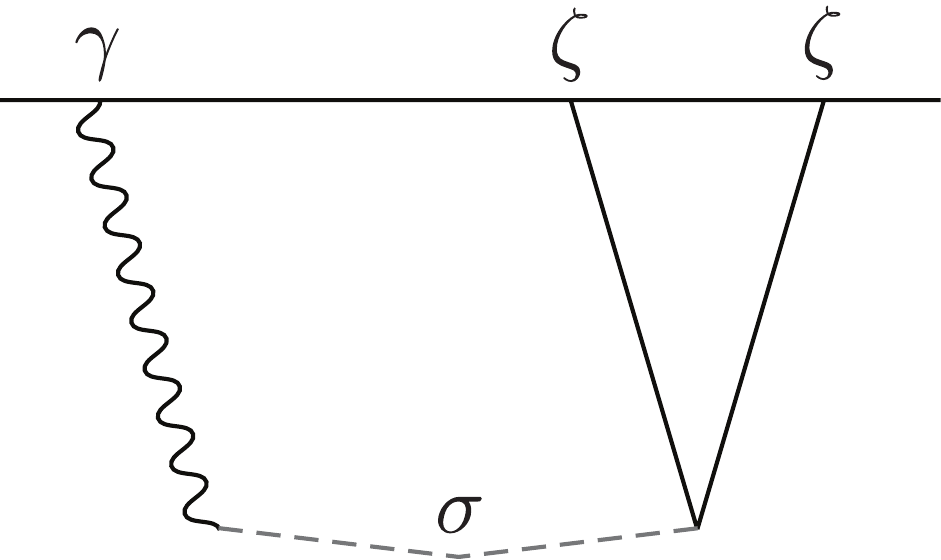}
\caption{Diagrams contributing to $\langle\zeta\zeta\zeta\rangle$ and $\langle\gamma\zeta\zeta\rangle$. The solid, dashed, and wavy lines represent the curvature perturbation $\zeta$, a massive spin-$s$ field $\sigma_{i_1\cdots i_s}$, and the graviton $\gamma_{ij}$, respectively. \label{fig:diagrams0} }
\end{figure}

Since massive particles decay outside of the horizon during inflation, they cannot be observed directly in late-time correlation functions. Instead, the presence of massive particles has to be inferred from their indirect effects on the correlation functions of $\zeta = -H\pi$ and $\gamma_{ij}$ (see Fig.~\ref{fig:diagrams0}). 
Some of these effects can be mimicked by adding a local vertex in the low-energy effective Lagrangian, which is the result of integrating out the heavy fields.  
On the other hand, massive particles may spontaneously be created in an expanding spacetime~\cite{parker1968particle, parker1969quantized,parker1971quantized}, an effect which cannot be represented by adding a local vertex to the effective Lagrangian~\cite{Arkani-Hamed:2015bza}.  The role of these non-local effects as a means of detecting massive particles during inflation was recently highlighted by Arkani-Hamed and Maldacena (AHM)~\cite{Arkani-Hamed:2015bza}:
the spontaneous particle creation allows us to probe massive fields during inflation, even though we are only observing the late-time expectation values of light fields. The rate of particle production in de Sitter space is exponentially suppressed as a function of mass, $e^{-m/T_{\rm dS}}$, with $T_{\rm dS} \equiv H/2\pi$, so their imprints will only be detectable if their masses are not too far above the Hubble rate $H$.\footnote{If the extra fields have strongly time-dependent masses, whose Fourier transforms have support at a frequency~$\hat \omega$, then non-adiabatic particle production occurs at a rate proportional to $e^{- m/\hat \omega}$~\cite{Flauger:2016idt}. The scale $\hat \omega$ may be as large as $\dot \phi^{1/2} = 58 \hskip 1pt H$ without spoiling the slow-roll dynamics. In models with these types of time-dependent couplings, the detectable range of particle masses is somewhat enlarged.} Since the inflationary scale may be as high as $10^{14}\hskip 1pt{\rm GeV}$, this nevertheless provides an opportunity to probe massive particles far beyond the reach of conventional particle colliders.

\vskip 4pt
Nonlinearities in the decay of the massive particles lead to a non-Gaussianity in the late-time correlation functions of $\zeta$ and $\gamma_{ij}$. The form of this non-Gaussianity will depend on the masses and the spins of the extra particles. 
The effects of additional scalar fields during inflation have been explored in many previous works, e.g.~in the context of quasi-single-field inflation~\cite{Chen:2009zp, Baumann:2011nk, Noumi:2012vr}. A characteristic signature of these fields are non-analytic scalings in the soft momentum limits of the non-Gaussian correlation functions. These soft limits are particularly clean detection channels, since in single-field inflation their momentum scalings are fixed by the symmetries of the inflationary background~\cite{Maldacena:2002vr, Creminelli:2004yq}.  The most straightforward interpretation of such non-analyticity in the correlation functions is therefore the presence of extra particles. 
Scalar fields with masses less than $\frac{3}{2}H$ give rise to monotonic scalings in the squeezed limit~\cite{Chen:2009zp, Baumann:2011nk}, while
those with masses greater than $\frac{3}{2}H$ 
 lead to oscillatory behavior~\cite{Noumi:2012vr, Arkani-Hamed:2015bza, Mirbabayi:2015hva, Chen:2015lza}. The effects of extra massive particles with spin have not been studied in as much detail. Such particles can naturally arise as massive Kaluza-Klein modes or as part of the tower of higher-spin states from string theory~\cite{Rindani:1985pi, Aragone:1988yx}. As was shown by 
AHM, the spins of new particles lead to a distinctive angular dependence of the soft limits of the non-Gaussian correlators.
The analysis of AHM was restricted to the squeezed limit of the bispectrum and interactions that maintained the approximate conformal invariance of the inflationary background.  While this assumption made their analysis particularly well controlled, it also implied that the amplitude of the signal is highly suppressed and only observable in the most optimistic and futuristic scenarios.  

\vskip 4pt
We will drop some of the restrictions of the analysis of AHM in our analysis. In particular, we will allow for a large breaking of conformal invariance within the framework of the effective field theory (EFT) of inflation~\cite{Cheung:2007st}. We will find that the signal due to massive spinning particles can be observable within the regime of validity of the EFT.
At the same time, the main spectroscopic features of particles with spin during inflation do not rely on conformal invariance and therefore still apply. On the other hand, couplings to particles with odd spins, which are disallowed in the conformally-invariant case, are permitted in the generic effective theory.  We also consider the breaking of special conformal invariance by giving the Goldstone fluctuations a nontrivial sound speed. 
In that case, we find a reduced exponential suppression in the particle production rate, and thus an enhanced level of non-Gaussianity. Finally, we also study the coupling to an external graviton~$\gamma_{ij}$.  
We demonstrate that the soft graviton limit of the correlator $\langle \gamma \zeta \zeta\rangle$ provides an interesting detection channel for extra particles.  Like in the case of massive scalar fields, there will be non-analytic scalings of non-Gaussianities close to the soft momentum limit, but this time only for particles with spin greater than or equal to two. 

 \paragraph{Outline}  In this paper, we analyze the allowed couplings of massive particles with spin to the Goldstone boson of broken time translations and the graviton, and discuss their observational signatures.  
 In Section~\ref{sec:dS}, we first collect the 
 equations of motion for massive fields with spin in de Sitter space, whose solutions are presented in Appendix~\ref{app:spindS}. In Section~\ref{sec:EFT}, we then construct the effective action for the leading interactions between the Goldstone boson $\pi$, the graviton $\gamma_{ij}$, and massive spinning fields $\sigma_{\mu_1 \ldots \mu_s}$. We analyze under what conditions the theory is under perturbative control and discuss various constraints on the sizes of the couplings.  In Section~\ref{sec:correlators}, we compute the correlation functions associated with the interactions of Section~\ref{sec:EFT}. We estimate the maximal amount of non-Gaussianity that is consistent with the constraints on the couplings of the effective theory. Details of the in-in computation are relegated to Appendix~\ref{app:inin}, and analytic results for soft limits are given in Appendix~\ref{app:squeezed}.
Our conclusions are presented in Section~\ref{sec:conclusions}. 

 \paragraph{Notation and conventions} 
We will use natural units, $c=\hbar=1$, with reduced Planck mass $\Mp^2=1/8\pi G$. Our metric signature is ($-++\hskip 1pt +$).  We will use Greek letters for spacetime indices, $\mu, \nu, \ldots =0,1,2,3$, and Latin letters for spatial indices, $i,j,\ldots=1,2,3$. Three-dimensional vectors are written in boldface, $\k$, and unit vectors are hatted, $\hat \k$. A shorthand for the symmetrization of tensor indices is $a_{(\mu} b_{\nu)} \equiv \frac{1}{2} (a_\mu b_\nu + a_\nu b_\mu)$.  Overdots and primes will denote derivatives with respect to physical time $t$ and conformal time $\eta$, respectively. The letter $\pi$ will refer both to $3.141\ldots$ and the Goldstone boson of broken time translations. 
The dimensionless power spectrum of a Fourier mode $f_\k$ is defined as 
\beq
\Delta_f^2(k) \equiv \frac{k^3}{2\pi^2} \langle f_\k f_{-\k}\rangle' \, , 
\eeq
where the prime on the expectation value indicates that the overall momentum-conserving delta function has been dropped.

\section{Spin in de Sitter Space}
\label{sec:dS}

We begin by reviewing a few elementary facts about massive fields with spin in four-dimensional de Sitter space, ${\rm dS}_4$. 
\paragraph{Spin-1}
The quadratic action of a massive spin-1 field $\sigma_\mu$ in de Sitter space is  
\begin{align}
S_1&= \int\d^4x\sqrt{-g}\left[-\frac{1}{2}\nabla_\mu\sigma_\nu\nabla^\mu\sigma^\nu+\frac{1}{2}(\nabla^\mu\sigma_\mu)^2 - \frac{1}{2}m_1^2\sigma^\mu\sigma_\mu\right] ,\label{spin1action}
\end{align}
where $m_1^2\equiv m^2+3H^2$, with $m$ being the mass of the field.\footnote{We define the mass parameter in such a way that the action acquires a gauge invariance in the massless limit, $m=0$. This is required in order for massless spinning fields to propagate the right number of degrees of freedom.  The mass defined in this way can also be identified as the mass of the field in the flat space limit \cite{Garidi:2003ys}.} 
The structure of the action (\ref{spin1action}) is uniquely fixed by requiring the absence of ghost degrees of freedom.\footnote{The ghost-free structure of the quadratic action will remain valid as long as nonlinear interactions can be treated perturbatively. } Up to integration by parts,
this is equivalent to the Proca action.
Variation of the action yields the equation of motion, $\Box\sigma_\mu - \nabla_\mu\nabla^\nu\sigma_\nu - m_1^2 \sigma_\mu = 0$, where $\Box\equiv \nabla^\mu\nabla_\mu$ denotes the Laplace-Beltrami operator on dS$_4$. Taking the divergence of this equation gives the constraint $\nabla^\mu\sigma_\mu = 0$. The on-shell equation of motion then becomes
\beq
\big(\Box-m_1^2\big)\hskip 1pt \sigma_\mu=0\, .\label{spin1eom}
\eeq
In Appendix~\ref{app:spindS}, we derive the solutions to this equation for the different helicity components of the field.

\paragraph{Spin-2}
The unique ghost-free quadratic action of a massive spin-2 field $\sigma_{\mu\nu}$ in de Sitter space is \cite{Higuchi:1986py} 
\begin{align}
S_2 = \int\d^4x\sqrt{-g}\,\bigg[&-\frac{1}{2}\nabla^\alpha \sigma^{\mu\nu}\nabla_\alpha \sigma_{\mu\nu} + \nabla^\mu \sigma_{\mu\nu} \nabla_\alpha \sigma^{\alpha\nu} - \nabla^\mu \sigma_{\mu\nu} \nabla^\nu \tilde\sigma + \frac{1}{2}\nabla^\mu \tilde\sigma \nabla_\mu \tilde\sigma \nonumber \\
& - \frac{1}{2}m_2^2(\sigma^{\mu\nu}\sigma_{\mu\nu} - \tilde\sigma^2)-\frac{3}{2}H^2\tilde\sigma^2\bigg]\, ,
\end{align}
where $m_2^2\equiv m^2+2H^2$ and $\tilde\sigma\equiv{\sigma^\mu}_\mu$ denotes the trace. Varying the action with respect to $\sigma_{\mu\nu}$, we obtain
\begin{align}
\Box \sigma_{\mu\nu}-2\nabla_{(\mu}\nabla^\alpha \sigma_{\nu)\alpha}+\nabla_\mu\nabla_\nu\tilde\sigma+g_{\mu\nu}(\nabla^\alpha\nabla^\beta\sigma_{\alpha\beta}-\Box\tilde\sigma)-m_2^2\sigma_{\mu\nu}+(m_2^2-3H^2)g_{\mu\nu}\tilde\sigma=0\, .
\end{align}
Taking the divergence gives $\nabla^\mu\sigma_{\mu\nu}=\nabla_\nu\tilde\sigma$, and plugging this back into the equation yields $(m^2-2H^2)\tilde\sigma=0$. For $m^2\ne 2H^2$, the equation of motion and the constraints satisfied by the field $\sigma_{\mu\nu}$ are\hskip 1pt\footnote{For $m^2=2H^2$, the system enjoys a (partial) gauge invariance $\sigma_{\mu\nu}\to \sigma_{\mu\nu}+\nabla_{(\mu}\nabla_{\nu)}\xi$, and the longitudinal (helicity-0) mode becomes non-dynamical \cite{Deser:1983mm}.}
\begin{align}
\big(\Box-m^2_2\big)\hskip 1pt\sigma_{\mu\nu}=0\, ,\quad \nabla^\mu\sigma_{\mu\nu} = 0 \, , \quad \tilde\sigma=0\, .\label{spin2onshell}
\end{align}
In Appendix~\ref{app:spindS}, we derive the solutions to the on-shell conditions (\ref{spin2onshell}).

\paragraph{Spin-${\boldsymbol{s}}$}

The Lagrangian for massive fields with arbitrary spin in flat space was constructed in~\cite{Singh:1974qz, Singh:1974rc}, and generalized to (A)dS spaces in~\cite{Zinoviev:2001dt}. 
For massive fields with spin greater than~2, the action is rather complex and requires introducing auxiliary fields of lower spins. 
An alternative, which we will follow, is to use a group theoretical approach to find the equations of motion directly~\cite{Deser:2003gw}.
A massive bosonic spin-$s$ field is described by a totally symmetric rank-$s$ tensor, $\sigma_{\mu_1\cdots\mu_s}$, 
subject to the constraints
\begin{align}
\nabla^{\mu_1}\sigma_{\mu_1\cdots\mu_s}=0 \, , \quad {\sigma^{\mu_1}}_{\mu_1\cdots\mu_s}=0\, .\label{spinscond}
\end{align}
The conditions in (\ref{spinscond}) project out the components of the tensor which transform as fields with lower spins. The Casimir eigenvalue equation of the de Sitter group then gives the wave equation satisfied by these fields:
\begin{align}
\left(\Box - m_s^2\right)\sigma_{\mu_1\cdots\mu_s} = 0\, ,\label{eom}
\end{align}
where $m_s^2 \equiv m^2-(s^2-2s-2)H^2$. The shift in the mass arises from the mismatch between the Casimir and Laplace-Beltrami operators in de Sitter space and is necessary to describe the correct representations for massless fields.
 Equivalently, it is required by imposing gauge invariance in the massless limit, $m=0$. Solutions to equation (\ref{eom}) are obtained in Appendix~\ref{app:spindS}.

\vskip 4pt
Following Wigner \cite{Wigner}, we identify the spectrum of particles by the unitary irreducible representations of the spacetime isometry group. For the de Sitter group ${\rm SO}(1,4)$, these representations fall into three distinct categories~\cite{Thomas, Newton}: 
\begin{center}
\begin{tabular}{ccc}
principal series & complementary series & discrete series \\[4pt]
\ \ $\displaystyle \frac{m^2}{H^2} \ge \left(s-\frac{1}{2}\right)^2$ \quad & \quad $\displaystyle s(s-1) < \frac{m^2}{H^2} < \left(s-\frac{1}{2}\right)^2$ \quad & \quad $\displaystyle \frac{m^2}{H^2} = s(s-1)-t(t-1)$\, ,
\end{tabular}
\end{center}
for $s,t=0,1,2,...$, with $t\le s$.
Masses that are not associated with one of the above categories are forbidden and correspond to non-unitary representations. At the specific mass values corresponding to the discrete series, the system gains an additional gauge invariance and some of the lowest helicity modes become pure gauge modes; this phenomenon is called partial masslessness~\cite{Deser:2001pe}. The spectrum of massive particles is contained in the principal and complementary series. We see that unitarity demands the existence of a lower bound, $m^2>s(s-1)H^2$, on the masses of fields that belong to this spectrum. For $s=2$, this is known as the Higuchi bound \cite{Higuchi:1986py}.

\vskip 4pt
In the late-time limit, the generators of the de Sitter isometries form the 3-dimensional conformal group. The asymptotic scaling of a spin-$s$ field is
\begin{align}
\lim_{\eta\to 0}\sigma_{i_1\cdots i_s}(\eta,\x) = \sigma_{i_1\cdots i_s}^+(\x)\,\eta^{\Delta_s^+-s}+\sigma_{i_1\cdots i_s}^-(\x)\,\eta^{\Delta_s^- -s}\, ,
\end{align}
where the conformal weight of the field  is defined as\hskip 1pt\footnote{Notice that for $s=0$, the case $m=0$  corresponds to a conformally coupled scalar field. For a minimally-coupled massless scalar, one should instead use $m^2\to m^2-2H^2$ in~\eqref{equ:mudef}.}
\begin{align}
\Delta_s^\pm = \frac{3}{2} \pm i\mu_s\, ,\quad {\rm with}\quad \mu_s \equiv \sqrt{\frac{m^2}{H^2}-\left(s-\frac{1}{2}\right)^2}\, .\label{equ:mudef}
\end{align}
In this paper, we will deal mostly with 
particles belonging to the principal series which covers the largest mass range and corresponds to $\mu_s\ge 0$. For real $\mu_s$, the asymptotic scaling is given by a complex-conjugate pair, resulting in a wavefunction that oscillates logarithmically in conformal time.
The complementary series has imaginary $\mu_s$ and corresponds to the interval $-i\mu_s\in (0,1/2)$. In that case, only the growing mode survives in the late-time limit.

\section[Spin in the EFT of Inflation]{Spin in the Effective Theory of Inflation}
\label{sec:EFT}

In this section, we will construct the leading interactions between the Goldstone boson of broken time translations $\pi$, the graviton $\gamma_{ij}$, and massive spinning fields $\sigma_{\mu_1\ldots \mu_s}$. 
We start, in \S\ref{ssec:Goldstone} and \S\ref{ssec:graviton}, by reviewing the effective actions for the Goldstone boson and the graviton.  In \S\ref{ssec:mixing}, we introduce the couplings to massive particles with spin;
first for the special cases $s=1$ and $2$, and then for arbitrary spin. We close,  in \S\ref{ssec:bounds}, by discussing how large the mixing interactions can be made while keeping the effective theory under theoretical control.

\subsection{Goldstone Action}
\label{ssec:Goldstone}

A time-dependent cosmological background induces a ``clock", i.e.~a preferred time slicing~
$\tilde t(t,\x)$ of the spacetime. In the inflationary context, surfaces of constant $\tilde t$ may be associated with the homogeneous energy density of the background. 
The slicing has a timelike gradient, and the unit vector perpendicular to the surface  of constant $\tilde t$ is
\beq
n_\mu \equiv \frac{\partial_\mu \tilde t}{\sqrt{-g^{\alpha\beta} \partial_\alpha \tilde t \partial_\beta \tilde t}}\, . \label{equ:nnu}
\eeq
The induced spatial metric on the slicing is $h_{\mu \nu} \equiv g_{\mu \nu} + n_\mu n_\nu$.  The metric $h_{\mu \nu}$ also serves to project spacetime tensors onto the spatial hypersurfaces. Geometric objects living on the hypersurfaces can be constructed from $h_{\mu \nu}$ and $n_\mu$. Examples are the intrinsic curvature, ${}^{(3)}R_{\mu \nu \rho \sigma}[h]$, and the extrinsic curvature, $K_{\mu \nu} \equiv {h_{(\mu}}^\rho \nabla_\rho n_{\nu)}$. Using the Gauss-Codazzi relation, the intrinsic curvature can be written in terms of (the projection of) the four-dimensional Riemann tensor $R_{\mu \nu \rho \sigma}$ and the extrinsic curvature $K_{\mu \nu}$. Higher-derivative objects can be constructed using the covariant derivative $\nabla_\mu$, defined with respect to $h_{\mu \nu}$. 

\vskip 4pt
In unitary gauge, the time coordinate $t$ is chosen to coincide with $\tilde t$. Fluctuations in the clock have been eaten by the metric, and the effective action for adiabatic fluctuations only contains metric perturbations. 
The action does not have to respect full diffeomorphism invariance, but only has to be invariant under time-dependent
spatial diffeomorphisms, 
$x^i \to x^i + \xi^i(t,\x)$. Besides terms that are invariant
under all diffeomorphisms (such as curvature invariants like $R$ and $R_{\mu \nu \rho \sigma} R^{\mu \nu \rho \sigma}$), the reduced symmetry of the system now allows many new terms in the action.
 The normal vector in (\ref{equ:nnu}) becomes $n_\mu = -  \delta^{0}_\mu/(-g^{00})^{1/2}$  in unitary gauge.
By contracting covariant tensors with $n_\mu$, we produce objects with uncontracted
upper $0$ indices, such as $g^{00}$ and $R^{00}$. It is easy to check that these are scalars under spatial diffeomorphisms.
Functions of $g^{00}$, $R^{00}$, etc.~are therefore allowed
in the effective action. In general, products of any four-dimensional covariant tensors
with free upper $0$ indices are allowed
operators.  In addition, we can have operators made out of the three-dimensional quantities describing the geometry of the spatial hypersurfaces (e.g.~$K_{\mu\nu}$).
The most general action constructed from these ingredients is~\cite{Cheung:2007st}
\beq
S = \int \d^4 x \sqrt{-g}\, {\cal L}(g^{00}, K_{\mu \nu}, R_{\mu \nu\rho \sigma}, \nabla_\mu, \ldots, t)\, , \label{equ:action}
\eeq
where the only free indices entering the functional ${\cal L}$ are upper $0$'s. The spacetime indices in (\ref{equ:action}) are contracted with the four-dimensional metric $g_{\mu \nu}$. Terms involving explicit contractions of the induced metric $h_{\mu \nu}$ do not lead to new operators.

\vskip 4pt
At leading order in derivatives, the action can be written in terms of $g^{00}$ alone,
\begin{align}
S = \int \d^4x\sqrt{-g}\left[\frac{1}{2}\Mp^2 R+\Mp^2\dot Hg^{00}-\Mp^2(3H^2+\dot H)+\sum_{n=2}^\infty \frac{M_n^4(t)}{n!}(\delta g^{00})^n+\cdots\right] , \label{equ:action2}
\end{align}
where $\delta g^{00} \equiv g^{00} + 1$.
The coefficients of the operators $1$ and $g^{00}$ have been fixed by the requirement that we are expanding around the correct FRW background with a given expansion rate $H(t)$.  This removes all tadpoles and the action starts quadratic in fluctuations.  Because time diffeomorphisms are broken, all operators are allowed to have time-dependent coefficients.
The limit of slow-roll inflation corresponds to $M_n \to 0$.  

\vskip 4pt
To make the dynamics of the theory defined by (\ref{equ:action2}) more transparent, we introduce the Goldstone boson associated with the spontaneous breaking of time-translation invariance. Through the St\"uckelberg trick, the field $\pi$ also restores the full diffeomorphism invariance of the theory. Specifically, we perform a spacetime-dependent
time reparameterization, $t\to \tilde t = t+\pi(t,\x)$.  The metric transforms in the usual way: e.g. 
\begin{align}
g^{00} &\to g^{00}+2\partial_\mu \pi g^{0\mu} + \partial_\mu \pi \partial_\nu \pi g^{\mu \nu}\, . \label{equ:g00}
\end{align}
Substituting this into (\ref{equ:action2}) gives the action for the Goldstone boson. In general, this action contains a complicated mixing between the Goldstone mode and metric fluctuations.  However, for most applications of interest, we can take the so-called decoupling limit, and evaluate the Goldstone action in the unperturbed background~\cite{Cheung:2007st}, $g_{\mu \nu} \to \bar g_{\mu \nu}$.
In this case, the transformation~(\ref{equ:g00}) reduces to $g^{00} \to -1 -2 \dot \pi +(\partial_\mu \pi)^2$, and the Goldstone Lagrangian becomes
\beq
{\cal L}_\pi = \Mp^2 \dot H (\partial_\mu \pi)^2 + 2 M_2^4 \left[\dot \pi^2 - \frac{\dot \pi (\partial_i\pi)^2}{a^2} \right] +\left(2M_2^4-\frac{4}{3} M_3^4\right) \dot \pi^3 + \cdots\, . \label{Lpi}
\eeq
We see that $M_2 \ne 0$ induces a nontrivial sound speed for the Goldstone boson, 
\beq
\cs^2 \equiv \frac{\Mp^2 \dot H}{\Mp^2\dot H-2M_2^4}\, .
\eeq
A small value of $\cs$ (large value of $M_2$) is correlated with an enhanced cubic interaction $\dot \pi (\partial_i \pi)^2$.
The Planck constraints on primordial non-Gaussianity imply $\cs \ge 0.024$~\cite{Ade:2015lrj}. For purely adiabatic fluctuations, the relationship between the comoving curvature perturbation $\zeta$ and the Goldstone boson is $\zeta=-H\pi +{\cal O}(\pi^2)$. The dimensionless power spectrum of $\zeta$ is found to be
\beq
\Delta_\zeta^2 \equiv \frac{k^3}{2\pi^2}P_\zeta(k)=\frac{1}{4\pi^2} \left(\frac{H}{f_\pi}\right)^4\, , \label{equ:Deltaz}
\eeq
where $f_\pi^4 \equiv 2 \Mp^2 |\dot H| \cs$ is the symmetry breaking scale~\cite{Baumann:2011su}. The observed amplitude of the power spectrum is $\Delta_\zeta^2=(2.14\pm 0.05)\times 10^{-9}$~\cite{Ade:2015xua}.

\subsection{Graviton Action}
\label{ssec:graviton}
The tensor sector of inflation is harder to modify~\cite{Creminelli:2014wna}. 
The leading correction to the Einstein-Hilbert action can be written as
\beq 
S = \int \d^4x\sqrt{-g}\left[\frac{1}{2}\Mp^2 R+ \hat M_2^2 \left(\delta K^{\mu \nu} \delta K_{\mu \nu}- \delta K^2\right)\right]  ,
\eeq
where the combination of extrinsic curvature tensors was chosen in a way that doesn't induce a scalar sound speed.
Inserting the transverse and traceless tensor perturbation of the metric, $g_{ij}=a^2(\delta_{ij}+\gamma_{ij})$, we find
\beq
{\cal L}_\gamma = \frac{\Mp^2}{8} \frac{1}{\ct^2}\left[\dot \gamma_{ij}^2 - \ct^2 \frac{(\partial_k \gamma_{ij})^2}{a^2} \right] + \cdots\, ,
\eeq
where we have defined the tensor sound speed
\beq
\ct^2 \equiv \frac{\Mp^2}{\Mp^2 + 2 \hat M_2^2}\, .
\eeq
As discussed in detail in~\cite{Creminelli:2014wna}, the tensor sound speed can be set to unity by a disformal transformation.  
This transformation makes the tensor sector canonical, and moves all the corrections to the scalar sector.  In this paper, we will make the same choice of frame and work with $\ct = 1$ throughout. The dimensionless power spectrum of $\gamma_{ij}$ is then given by
\beq
\Delta_\gamma^2 \equiv  \frac{k^3}{2\pi^2}P_\gamma(k)= \frac{2}{\pi^2} \frac{H^2}{\Mp^2}\, . \label{equ:Deltag}
\eeq
The current contraint on the tensor-to-scalar ratio, $r \equiv \Delta_\gamma^2/\Delta_\zeta^2 <0.07$~\cite{Array:2015xqh}, implies that $\Delta_\gamma^2 \lesssim 1.5\times 10^{-10}$.

\subsection{Mixing Interactions}
\label{ssec:mixing}
Next, we construct the effective action for interactions between the Goldstone boson, the graviton, and massive spinning fields. We will also consider self-interactions of the massive spinning fields, and focus on terms which contribute to the correlation functions $\langle\zeta\zeta\zeta\rangle$ and $\langle\gamma\zeta\zeta\rangle$ at tree level and at leading order in derivatives. Moreover, we will restrict our presentation to the subset of interactions which give rise to a distinctive angular dependence due to the exchange of the spinning fields. 

\subsubsection{Couplings to the Goldstone}

The construction of the effective action proceeds as above.  We first write down all operators consistent with the symmetries. Amongst them will be tadpole terms, which must add up to zero.
In unitary gauge, the basic building blocks involving spinning fields are $\sigma^{0\cdots 0}$ and all Lorentz-invariant self-interactions, e.g.~$\sigma^{\mu_1\cdots\mu_s}\sigma_{\mu_1\cdots\mu_s}$. 
The latter are invariant under all diffeomorphisms, so they don't lead to a coupling to $\pi$ after the St\"{u}ckelberg trick, whereas the former transform~as
\begin{align}
\sigma^{0\cdots 0} \to (\delta^0_{\mu_1}+\partial_{\mu_1}\pi)\cdots(\delta^0_{\mu_s}+\partial_{\mu_s}\pi)\,\sigma^{\mu_1\cdots\mu_s}\, . \label{equ:ST}
\end{align}
We may also have contractions with the curvature tensors, which appear at higher order in derivatives.

\paragraph{Spin-1}
We first analyze the  couplings between a massive spin-1 field $\sigma_\mu$ and the Goldstone boson~$\pi$. In unitary gauge, the operators of the effective action involve $g^{00}$ and $\sigma^0$. In order not to alter the background solution, these operators have to start at quadratic order in fluctuations. 
\begin{itemize}
\item At leading order in derivatives and to linear order in $\sigma_\mu$, the mixing Lagrangian is\hskip 1pt\footnote{Note that there are no terms involving $\delta g^{0\mu}\sigma_\mu$ in the effective action. This is because this operator does not satisfy the symmetries of the EFT, since the background value $\bar g^{0\mu}\sigma_\mu=-\sigma_0$ transforms nontrivially under spatial diffeomorphisms (and so does the fluctuation). } 
\begin{align}
{\cal L}_{\pi \sigma}^{(1)} &= \omega_1^3 \hskip 1pt \delta g^{00}\sigma^0  + \omega_2^3  \hskip 1pt  (\delta g^{00})^2\sigma^0 \,  .\label{spin1mixaction1}
\end{align}
Introducing $\pi$ using (\ref{equ:g00}) and (\ref{equ:ST}), we get 
\begin{align}
{\cal L}_{\pi \sigma}^{(1)} &= \omega_1^3  \hskip 1pt a^{-2}\big( 2\partial_i\pi\sigma_i -(\partial_i\pi)^2\sigma_0 - 2\dot\pi\partial_i\pi\sigma_i  \big) +(3\omega_1^3+4\omega_2^3)  \hskip 1pt \dot\pi^2\sigma_0 + \cdots\, ,
\end{align}
where we have taken the decoupling limit so that couplings to metric fluctuations become irrelevant.\footnote{The decoupling limit is not affected by the inclusion of mixing interactions, provided that we are in the perturbative regime. This can be shown by an ADM analysis of the metric perturbations~\cite{Noumi:2012vr, Delacretaz:2015edn}.} We also used the constraint $\nabla^\mu\sigma_\mu=0$, which we assume to hold at the background level, to replace $\dot\pi\sigma_0$ by $\partial_i\pi\sigma_i$.  Since only the cubic mixing $\dot\pi\partial_i\pi\sigma_i$ will lead to the characteristic angular structure in the resulting correlation functions (see \S\ref{ssec:ZZZ}), we will focus on the bispectrum created created by the combination of $\dot\pi\partial_i\pi\sigma_i$ and $\partial_i\pi\sigma_i$. Note that there is a single parameter $\omega_1$ controlling the size of these two interactions. This is a consequence of the nonlinearly-realized time translation symmetry. 

\item At quadratic order in $\sigma_\mu$, the mixing Lagrangian is
\begin{align}
{\cal L}_{\pi \sigma^2}^{(1)} &\ =\ \omega_3^2  \hskip 1pt \delta g^{00}(\sigma^0)^2+\omega_4^2  \hskip 1pt \delta g^{00}\sigma^\mu\sigma_\mu \label{spin1mixaction1} \\[6pt]
&\ \to\  -2(\omega_3^2-\omega_4^2)\dot\pi\sigma_0\sigma_0 -2\omega_4^2 a^{-2}\dot\pi\sigma_i\sigma_i\, ,
\end{align}
where in the second line we have introduced the Goldstone and taken the decoupling limit. We see that, this time, the size of the cubic interaction $\dot\pi\sigma_i\sigma_i$ is independent from the quadratic mixing term.  
\end{itemize}
Combining the above, we can write
\begin{align}
{\cal L}^{(1)}_{\rm mix} =\frac{1}{a^2}\left(\rho_1\hskip 1pt\partial_i\pi_c\sigma_i + \frac{1}{\Lambda_1}\dot\pi_c\partial_i\pi_c\sigma_i +\lambda_1\hskip 1pt \dot\pi_c\sigma_i\sigma_i\right) ,\label{Lmix1}
\end{align}
where $\pi_c \equiv f_\pi^2\hskip 1pt \pi$ is the canonically normalized Goldstone boson, and we defined 
\begin{align}
\rho_1\equiv \frac{2\omega_1^3}{f_\pi^2}\, , \quad \Lambda_1 \equiv -\frac{f_\pi^2}{\rho_1}\, ,\quad \lambda_1\equiv -\frac{2\omega_4^2}{f_\pi^2}
\, . \label{equ:coup}
\end{align}
We note that $\rho_1$ and $\Lambda_1$ are correlated, since they are both determined by the parameter $\omega_1$.

\paragraph{Spin-2}
Next, we consider the mixing between a massive spin-2 field and the Goldstone boson.
\begin{itemize}
\item At linear order in $\sigma_{\mu\nu}$, the mixing Lagrangian is 
\begin{align}
{\cal L}_{\pi \sigma}^{(2)} &=  \tilde \omega_1^3 \hskip 1pt \delta g^{00}\sigma^{00} + \tilde \omega_2^3 \hskip 1pt (\delta g^{00})^2\sigma^{00} + \tilde \omega_3^2 \hskip 1pt \delta K_{\mu\nu}\sigma^{\mu\nu}+ \tilde \omega_4^2 \hskip 1pt \delta g^{00}\delta K_{\mu\nu}\sigma^{\mu\nu}\, , \label{equ:320}
\end{align}
where it was necessary to include higher-derivative operators to get the relevant interactions for the spatial components $\sigma_{ij}$. In the decoupling limit, the mixing with the Goldstone boson~is 
\begin{align}
{\cal L}_{\pi \sigma}^{(2)} &=   \tilde \omega_1^3\left[ -2\dot\pi\sigma_{00}+a^{-2}(\partial_i\pi)^2\sigma_{00}+4a^{-2}\dot\pi\partial_i\pi\sigma_{0i}\right]  -(5 \tilde \omega_1^3-4 \tilde \omega_2^3) \hskip 1pt \dot\pi^2\sigma_{00}\nonumber\\[4pt]
&\quad - \tilde \omega_3^2\hskip 1pt a^{-4}\partial_i\partial_j\pi\sigma_{ij} +2 \tilde \omega_4^2\hskip 1pt a^{-4}\dot\pi\partial_i\partial_j\pi\sigma_{ij}+\cdots \, .\label{spin2mixlag1}
\end{align}
We will focus on the traceless part of $\sigma_{ij}$, which we denote by $\hat\sigma_{ij}$. Only the cubic mixing $\dot\pi\partial_i\partial_j\pi\hat\sigma_{ij}$ will lead to the characteristic angular structure in the bispectrum. 
Since the quadratic mixing does not affect the angular structure, we will simply choose $\partial_i\partial_j\pi\sigma_{ij}$ as a representative example.  Unlike the spin-1 case, the sizes of the quadratic and cubic mixing operators are controlled by two independent parameters, $\tilde\omega_3$ and $\tilde\omega_4$. 
\item At quadratic order in  $\sigma_{\mu\nu}$, the mixing Lagrangian is
\begin{align}
{\cal L}_{\pi \sigma^2}^{(2)} &\ =\   \tilde \omega_5^2\hskip 1pt\delta g^{00}(\sigma^{00})^2+ \tilde \omega_6^2\hskip 1pt\delta g^{00}\sigma^{\mu\nu}\sigma_{\mu\nu} \, , \\[6pt]
&\ \to\ -(2 \tilde \omega_5^2+2\tilde \omega_6^2)\hskip 1pt \dot\pi\sigma_{00}^2 - 2 \tilde \omega_6^2\left[2a^{-2}\dot\pi\sigma_{0i}\sigma_{0i}+a^{-4}\dot\pi\sigma_{ij}\sigma_{ij}\right]  + \cdots\, , \label{equ:322}
\end{align}
where the last term in (\ref{equ:322}) will lead to the angular structure that we are interested in. 
\end{itemize}
We will study the following mixing Lagrangian
\begin{align}
{\cal L}^{(2)}_{\rm mix} =\frac{1}{a^4}\left(\rho_2\hskip 1pt\partial_i\partial_j\pi_c\hat\sigma_{ij} + \frac{1}{\Lambda^2_2}\, \dot\pi_c\partial_i\partial_j\pi_c\hat\sigma_{ij} +\lambda_2\hskip 1pt \dot\pi_c\hat\sigma_{ij}\hat\sigma_{ij}\right) ,\label{Lmix2} 
\end{align}
where we defined 
\begin{align}
\rho_2\equiv -\frac{\tilde \omega_3^2 }{f_\pi^2}\, , \quad \Lambda_2 \equiv \frac{f_\pi^{2}}{\sqrt{2} \tilde \omega_4}\, ,\quad \lambda_2\equiv  -\frac{2 \tilde \omega_6^2}{f_\pi^2} \, .\label{spin2param}
\end{align}
This is similar to the spin-1 mixing Lagrangian (\ref{Lmix1}), except that the quadratic and cubic mixing parameters, $\rho_2$ and $\Lambda_2$, are now independent.
\paragraph{Spin-${\boldsymbol s}$} 
Performing the same analysis for a field with arbitrary spin $s> 2$, we find the following mixing Lagrangian 
\begin{align}
{\cal L}^{(s)}_{\rm mix} = \frac{1}{a^{2s}}\left(\rho_s \hskip 1pt\partial_{i_1\cdots i_s}\pi_c\hat\sigma_{i_1\cdots i_s} + \frac{1}{\Lambda_s^s}\dot\pi_c\partial_{i_1\cdots i_s}\pi_c\hat\sigma_{i_1\cdots i_s}+\lambda_s\hskip 1pt\dot\pi_c\hat\sigma_{i_1\cdots i_s}^2\right) ,\label{spinsLmix} 
\end{align}
where $\partial_{i_1\cdots i_s}\equiv \partial_{i_1}\cdots\partial_{i_s}$.  As in the case of spin-2, these interactions generically arise from independent operators, i.e.~$\rho_s$, $\Lambda_s$, and $\lambda_s$ are independent parameters. 

\vskip 4pt
The mixing in (\ref{spinsLmix}) can convert hidden non-Gaussianity in the $\sigma$-sector into visible non-Gaussianity in the $\pi$-sector. To allow for this possibility, we add cubic self-interactions to the action for $\sigma$, which schematically we can write as
\begin{align}
a^{3s}{\cal L}_{\sigma^3}^{(s)} \equiv \begin{cases} \xi_s\hskip 1pt\hat\sigma\cdot\hat\sigma\cdot\hat\sigma & s \ {\rm even},\\ \xi_s\hskip 1pt\hat\sigma\cdot\hat\sigma\cdot (\partial\hat\sigma) & s\ {\rm odd},\end{cases} \label{spinsLmixX} 
\end{align}
with suitable symmetric contractions of spatial indices. 

\subsubsection{Couplings to the Graviton}

We will also be interested in the couplings between massive spinning fields and the graviton, $\gamma_{ij}$. For simplicity, we will only consider linear couplings to $\gamma_{ij}$, but the generalization to higher orders will essentially be straightforward.

\paragraph{Spin-1}
The leading couplings to the graviton arise from
\begin{align}
{\cal L}_{ \gamma \pi \sigma}^{(1)} &= \omega_5^2 \hskip 1pt \delta g^{00} g^{\mu\nu}\nabla_\mu\sigma_\nu -\frac{m_1^2}{2}\sigma^{\mu\nu}\sigma_{\mu\nu}\, .
\label{Lgamma1}
\end{align}
Note that, in our perturbative treatment, the on-shell conditions for $\sigma_\mu$ hold at the background level, so that $\bar g^{\mu\nu}\nabla_\mu\sigma_\nu=0$. 
The first term in (\ref{Lgamma1}) is therefore proportional to $g^{\mu\nu}\nabla_\mu\sigma_\nu=\delta g^{\mu\nu}\nabla_\mu\sigma_\nu$ and starts at cubic order in fluctuations. In terms of $\pi$ and $\gamma_{ij}$, the mixing Lagrangian becomes
\begin{align}
{\cal L}_{\gamma \pi \sigma}^{(1)} = \frac{1}{a^2}\frac{1}{\Mp}\big(\tau_1\hskip 1pt \dot\pi_c\gamma_{ij}^c\partial_i\sigma_j +m_1^2\hskip 1pt\gamma_{ij}^c\sigma_i\sigma_j\big)\ ,\label{Lmix1gamma}
\end{align}
where $\tau_1\equiv 4\omega_5^2/f_\pi^2$ and $\gamma_{ij}^c\equiv \frac{1}{2}\Mp\gamma_{ij}$ denotes the canonically normalized graviton. The first term in (\ref{Lmix1gamma}) is higher order in derivatives than the operator $\gamma_{ij}\partial_i\pi\sigma_j$. However, the latter only arises from the tadpole $\sigma^0$, and is therefore required to have a vanishing coefficient. Moreover, a quadratic mixing between the spin-1 field and the graviton is forbidden by kinematics: any such mixing will involve spatial gradients and hence must vanish because the graviton is transverse, $\partial_i\gamma_{ij}=0$.

\paragraph{Spin-2}
The couplings between a massive spin-2 field and the graviton follow from
\begin{align}
{\cal L}_{ \gamma \pi \sigma}^{(2)} &= \tilde\omega_3^2\hskip 1pt \delta K_{\mu\nu}\sigma^{\mu\nu} + \tilde \omega_7^3 \hskip 1pt \delta g^{00} g^{\mu\nu}\sigma_{\mu\nu} -\frac{m_2^2}{2}\sigma^{\mu\nu}\sigma_{\mu\nu}\, . 
\label{Lgamma2}
\end{align}
Note that we have already encountered the operator $\delta K_{\mu\nu}\sigma^{\mu\nu}$ in (\ref{equ:320}). In our perturbative treatment, the on-shell traceless condition holds at the background level, $\bar g^{\mu\nu}\sigma_{\mu\nu}=0$, which implies that $g^{\mu\nu}\sigma_{\mu\nu} =  \delta g^{\mu\nu}\sigma_{\mu\nu}$, so that
the second term in (\ref{Lgamma2}) starts at cubic order in fluctuations. The cubic operator $\delta g^{00}\delta g^{\mu\nu}\nabla_\mu\sigma_{\nu 0}$  will not be considered, since its effects are indistinguishable from those of the first term in (\ref{Lmix1gamma}). Introducing $\pi$ and $\gamma_{ij}$, the mixing Lagrangian becomes 
\begin{align}
{\cal L}_{\gamma \pi \sigma}^{(2)} = \frac{1}{a^2}\frac{1}{\Mp}\left(\tilde\rho_2\hskip 1pt \dot\gamma_{ij}^c\hat\sigma_{ij} + \tau_2\hskip 1pt\dot\pi_c\gamma_{ij}^c\hat\sigma_{ij} +\frac{m_2^2}{a^2}\hskip 1pt\gamma_{ij}^c\hat\sigma_{ik}\hat\sigma_{kj}\right) ,\label{Lmix2gamma}
\end{align}
where 
$\tilde\rho_2\equiv -\rho_2 f_\pi^2$ and $\tau_2 \equiv 4\tilde \omega_7^3/ f_\pi^2$. Note that we have only kept the spatial components in the coupling to the mass term. Unlike in the spin-1 case, there is now a quadratic mixing between the spin-2 field and the graviton, whose size is correlated with the $\pi$-$\sigma$ mixing in (\ref{Lmix2}). The other possible form of mixing $\gamma_{ij}\sigma_{ij}$ comes from the tadpole $\tilde\sigma$ and is thus absent. 

\paragraph{Spin-$\boldsymbol s$}
For arbitrary spin $s>2$, the leading interactions with the graviton and the Goldstone take the following form 
\begin{align}
\hskip -5pt{\cal L}_{ \gamma \pi \sigma}^{(s)}= \frac{1}{a^{2s-2}}\frac{1}{\Mp}\hskip -2pt\left(\tilde\rho_s\hskip 1pt \partial_{i_3\cdots i_s}\dot\gamma^c_{i_1i_2}\hat\sigma_{i_1\cdots i_s}+ \tau_s\hskip 1pt\gamma_{i_1i_2}^c\partial_{i_3\cdots i_s}\pi_c\hat\sigma_{i_1\cdots i_s} + \frac{m_s^2}{a^2}\gamma_{i_1j_1}^c\hat\sigma_{i_1\cdots i_s}\hat\sigma_{j_1\cdots i_s}\right) , \label{Lmixsgamma}
\end{align}
where $\tilde\rho_s\equiv -\rho_s f_\pi^2$.
Again, we have only kept interactions that involve the purely spatial components of the field. In practice, there are other low-dimensional operators that can also contribute to the correlator $\langle\gamma\zeta\zeta\rangle$ with the same angular structure, such as $\dot\gamma_{ij}\sigma_{ij0\cdots 0}$. 

\subsection{Bounds on Mixing Coefficients}
\label{ssec:bounds}

It is important to determine how large the mixing interactions of the previous section can be made while keeping the effective theory under theoretical control. 
In this section, we will discuss bounds arising from {\it i}\hskip 1pt)~the requirement that the mixing interactions can be treated perturbatively and {\it ii}\hskip 1pt) the absence of superluminal propagation.\footnote{Additional bounds could arise from the analyticity of the $S$-matrix, which is a requirement for a weakly-coupled, Lorentz-invariant ultraviolet completion \cite{Adams:2006sv} (see e.g.~\cite{Baumann:2015nta} for an application in the context of inflation).}  
Finally, we also consider what range of coefficients yields a technically natural effective field theory, in the sense of stability under radiative corrections. In Section~\ref{sec:correlators}, we will consider the implications of these constraints on the size of non-Gaussianities.

\subsubsection{Perturbativity}
\label{ssec:perturbative}

We wish to treat the mixing interactions as perturbative corrections to the free-field actions for the Goldstone boson and the massive spinning fields.
Since massive particles decay outside the horizon and oscillate rapidly inside the horizon, the dominant contributions to correlation functions will occur at horizon crossing of the Goldstone boson, corresponding to frequencies of order $H$. 
Consistency of the perturbative description therefore requires that the sizes of the mixing interactions at $\omega \sim H$ are smaller than the terms in the free-field actions. 
This puts constraints on the couplings in the mixing Lagrangians discussed in the previous section. 
For $\cs=1$, the criteria for a consistent perturbative treatment require little more than dimensional analysis.  The dimensionful couplings of relevant interactions have to be less than $H$, while those of irrelevant interactions have to be greater than $H$. The dimensionless couplings of marginal interactions have to be less than unity. For example,  for the couplings appearing in (\ref{Lmix2}), we require $\{\rho_2,\hskip 1pt \lambda_2\} <1$ and $\Lambda_2>H$. Similar considerations apply for the couplings in (\ref{Lmix1}) and (\ref{spinsLmix}). For $\cs \ne 1$, determining the perturbativity constraints on the mixing parameters requires a more careful analysis. Spatial gradients of the Goldstone mode are enhanced and the correlation functions can receive contributions from a second time scale, the time of crossing of the sound horizon. We will return to this complication in Section~\ref{sec:correlators}.

\subsubsection{Superluminality}

The breaking of time diffeomorphism invariance can modify the actions for spinning fields of Section~\ref{sec:dS} by introducing additional non-Lorentz-invariant interactions. For concreteness, we will confine our discussion in this subsection  to the case of spin one, but we expect similar results to hold for higher spins. In unitary gauge, the most general quadratic action for a spin-1 field is
\begin{align}
S_\sigma = \int\d^4x\sqrt{-g}\left[-\frac{1}{4} F^{\mu\nu}F_{\mu\nu} + \frac{a_1}{2} F^{0\mu}{F^0}_\mu -\frac{1}{2}m^2(\sigma^\mu\sigma_\mu-a_0\, \sigma^0\sigma^0)\right] ,\label{spin1EFTaction}
\end{align}
where $F_{\mu\nu}\equiv \partial_\mu\sigma_\nu-\partial_\nu\sigma_\mu$, and the structure of the kinetic part is enforced by gauge invariance in the massless limit.  
The departure from the Lorentz-invariant action is characterized by the parameters $a_0$ and~$a_1$, which lead to nontrivial sound speeds for the longitudinal mode, $c_0\equiv 1/\sqrt{1+a_0}$, and for the transverse mode, $c_1\equiv 1/\sqrt{1+a_1}$.
To see this, we consider the on-shell equations of motion in the flat-space limit,
\begin{align}
\ddot\sigma_0-c_0^2\hskip 1pt\nabla^2\sigma_0 + m^2\sigma_0 &=0\, ,\\
\ddot\sigma_i^{\rm t}-c_1^2\hskip 1pt\nabla^2\sigma_i^{\rm t} + m^2\sigma_i^{\rm t} &= 0\, ,
\end{align}
where $\sigma^{\rm t}_i$ denotes the transverse mode, $\partial_i\sigma^{\rm t}_i=0$. The components of the spin-1 field propagate subluminally with no gradient instability as long as $\{a_0,a_1\}\ge 0$. A tachyonic instability is avoided for $m^2>0$.

\vskip 4pt
Even if the spin-1 field propagates subluminally, the mixing with the Goldstone boson $\pi$ can lead to superluminal propagation in the coupled system. Requiring the absence of superluminality imposes a constraint on the size of the quadratic mixing term in (\ref{Lmix1}). 
To derive this constraint, we consider the on-shell equations of motion for the coupled system,
\begin{align}
\ddot\sigma_0-c_0^2\hskip 1pt\nabla^2\sigma_0 + m^2\sigma_0 &=-c_0^2\hskip 1pt\rho_1\hskip 1pt m^{-2}\nabla^2\dot{\pi}_c\, ,\\
\ddot\sigma_i^{\rm t}-c_1^2\hskip 1pt\nabla^2\sigma_i^{\rm t}+m^2\sigma_i^{\rm t} &= 0\, ,\\
\ddot\pi_c -\cs^2\hskip 1pt\nabla^2\pi_c &=-\cs^3\rho_1 \hskip 1pt(c_0^{-2}\dot\sigma_0-\rho_1\hskip 1pt m^{-2}\nabla^2\pi_c)\, .
\end{align}
We see that the transverse mode does not mix with $\pi$, and hence its dispersion relation is unmodified. After diagonalizing the coupled $\pi$-$\sigma$ system, the dispersion relations obeyed by the normal modes are
\begin{align}
\omega^2_\pm = \frac{1}{2}\left[\big(c_0^2+\cs^2(1+2\delta^2)\big)k^2+m^2 \pm\sqrt{\big[(c_0^2-\cs^2)k^2+m^2\big]^2+4\cs^4k^4\delta^2(1+\delta^2)}\, \right] ,
\end{align}
where $\delta^2\equiv \cs^3\rho_1^2/m^2$. For large $k$, subluminality implies the following constraint
\begin{align}
\frac{\rho^2_1}{m^2}\, \le\, \frac{1-c_0^2}{2-c_0^2} \,\frac{1-\cs^2}{\cs^3}\, .\label{SubluminalityCond}
\end{align}
Note that the mixing is required to vanish if either $\cs$ or $c_0$ are equal to $1$. (A similar result for the mixing with a scalar field was found in \cite{Baumann:2011nk}.)
However, even a relatively small deviation of $\cs$ and $c_0$ from $1$ is sufficient to allow $\rho_1$ to be of order $H$ (i.e.~of order the maximal size allowed by perturbativity). For simplicity, we will therefore work with $c_0=c_1 \approx 1$, but incorporating nontrivial sound speeds for the spin-1 field could be done straightforwardly using modifications of the mode functions given in Appendix~\ref{app:spindS}. Similarly, we will assume that all spinning fields obey a relativistic dispersion relation.

\subsubsection{Naturalness}
\label{sec:naturalness}

Finally, we will consider constraints arising from the radiative stability of the effective theory, e.g.~we require that the masses of spinning fields do not receive large loop corrections. This is more of a philosophic criterion rather than a strict consistency condition.

\begin{itemize}
\item Let us consider the interaction
$\dot\pi_c\partial_{i_1\cdots i_s}\pi_c\sigma_{i_1\cdots i_s}$ in (\ref{spinsLmix}), suppressed by the scale~$\Lambda_s$. At one loop, this term generates the following correction to the non-Lorentz-invariant mass term,
\begin{align}
\delta m_{\sigma_{i_1\cdots i_s}}^2 \sim \frac{1}{\Lambda_s^{2s}}\int\d\omega\hskip 1pt \d^3k\, \frac{\omega^2k^{2s}}{[\cs^{-3}(\omega^2-\cs^2k^2)]^2} \,\sim\, \cs^{3-2s}\frac{\Lambda^{2s+2}}{\Lambda_s^{2s}}\, .\label{equ:loop}
\end{align}
Naturalness of the mass of the spinning field requires $\delta m_{\sigma_{i_1\cdots i_s}}^2  \lesssim m_s^2 \sim H^2$. To estimate the size of (\ref{equ:loop}), we take the cutoff of the $\pi$-loop to be of order the strong coupling scale of the Goldstone sector.  For $\cs=1$, we can, in principle, extend the $\pi$-loop up to the symmetry breaking scale, i.e.~$\Lambda \sim f_\pi$, while, for $\cs \ll 1$, the effective theory of the Goldstone becomes strongly coupled at $\Lambda \sim f_\pi \cs$. We will therefore use $\Lambda \sim f_\pi \cs$ for all values of $\cs$. The condition for radiative stability then becomes
\begin{align}
\left(\frac{H}{\Lambda_s}\right)^s\ \lesssim\ \left( \frac{(2\pi \Delta_\zeta)^{s+1}}{\cs^{5}}\right)^{1/2}\, .
\label{spinsloop}
\end{align}
Typically, this constraint requires $\Lambda_s$ to be slightly larger than $f_\pi$. 
\item Next, we consider the interaction $\lambda_s\hskip 1pt\dot\pi_c\sigma_{i_1\cdots i_s}^2$  in (\ref{spinsLmix}). This leads to the following radiative correction to the non-Lorentz-invariant mass term,
\begin{align}
\delta m^2_{\sigma_{i_1\cdots i_s}}\sim \lambda_s^2 \int\d\omega\hskip 1pt\d^3k\,\frac{\omega^2}{\cs^{-3}(\omega^2-\cs^2k^2)(\omega^2-k^2-m^2)} \sim \cs^2\hskip 1pt\lambda_s^2\hskip 1pt\Lambda^2\, .\label{loop2}
\end{align}
Cutting off the loop at $\Lambda\sim f_\pi \cs$, we obtain the following constraint for radiative stability:
\begin{align}
\lambda_s\ \lesssim\ \frac{(2\pi\Delta_\zeta)^{1/2}}{\cs^2}\, . \label{loop2X}
\end{align} 
The interaction $\lambda_s\hskip 1pt\dot\pi_c\sigma_{i_1\cdots i_s}^2$ can also give a correction to the kinetic term for the Goldstone.  However, on dimensional grounds, it is easy to see that this interaction only contributes a negligible correction to the sound speed of $\pi$.
\item Lastly, the radiative correction generated by the cubic self-interaction of the spinning fields~is
\begin{align}
\delta m^2_{\sigma_{i_1\cdots i_s}}\sim \begin{cases} \xi_s^2 & s\ \text{even},\\ \xi_s^2\Lambda^2 & s\ \text{odd},\end{cases}
\end{align}
where the couplings $\xi_s$ are of dimensions zero and one for odd and even spins, respectively, cf.~(\ref{spinsLmixX}).
For even spins, we only get a fixed finite correction to the mass term.  Since we require $\xi_s<H$ for perturbative control, the loop contribution from this interaction is guaranteed to be small. For odd spins, it is natural to take the cutoff for the $\sigma$-loop to be the strong coupling scale $\Lambda_s$. We then get
\begin{align}
\xi_s^2\ \lesssim\ \frac{H^2}{\Lambda_s^2}\ \lesssim\ \left(\frac{(2\pi\Delta_\zeta)^{s+1}}{\cs^5}\right)^{1/s}\, ,
\end{align}
where we have used the naturalness constraint (\ref{spinsloop}) on $\Lambda_s$ in the second inequality.
\end{itemize}

\section{Cosmological Correlators}
\label{sec:correlators}

We will now compute the effects of massive particles with spin on the correlation functions of the Goldstone boson and the graviton. Following~\cite{Arkani-Hamed:2015bza}, we will study separately the contributions from local and non-local processes. 
Local processes are, by definition, those whose imprint can be mimicked by adding a local operator in the low-energy effective theory of the light fields alone. Non-local processes, on the other hand, capture particle production effects which cannot be mimicked by additional local operators. While the latter are the distinctive signature of extra particles during inflation, the amplitude of such effects is exponentially suppressed for masses above the Hubble scale. We will discover that the sound speed of the Goldstone boson plays a crucial role in controlling the relative size of the local and non-local processes.

\begin{figure}[t!]
\centering
\includegraphics[scale=0.5]{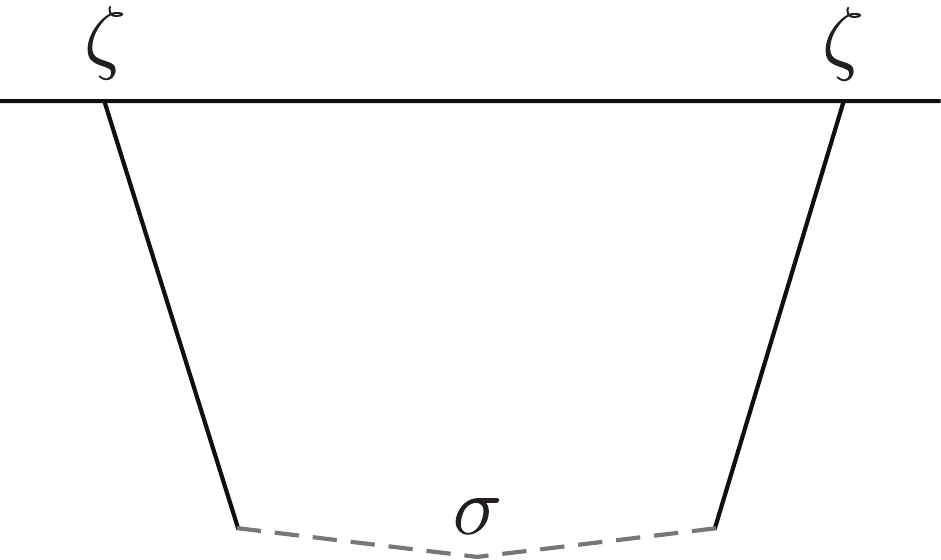}\vskip 3pt
\caption{Tree-level diagram contributing to the two-point function $\langle\zeta\zeta\rangle$. The solid and dashed lines represent the curvature perturbation $\zeta$  
and a massive spin-$s$ field $\sigma_{i_1\cdots i_s}$, respectively. \label{fig:2pt}}
\end{figure}

\subsection[${\langle \zeta\zeta\rangle}$]{$\boldsymbol{\langle \zeta\zeta\rangle}$}
\label{sec:PS}

Before discussing a potentially richer structure in the bispectra, we will gain some useful insights by first examining the effect of massive particles on the power spectrum $\langle \zeta \zeta \rangle$ (see Fig.~\ref{fig:2pt}). We will separate the correlation function into distinct contributions coming from  local and non-local processes. 
Spin doesn't play a big role in the correction to the power spectrum, so for simplicity we will consider a minimally-coupled massive scalar field~$\sigma$, whose two-point function in de Sitter space is 
\begin{align}
\langle\sigma_\k(\eta)\sigma_{-\k}(\eta')\rangle' = \frac{\pi}{4}H^2(\eta\eta')^{3/2}e^{-\pi\mu}H_{i\mu}(-k\eta)H^*_{i\mu}(-k\eta')\, ,\label{scalar2pt0}
\end{align}
where $H_{i\mu}\equiv H_{i\mu}^{(1)}$ is the Hankel function of the first kind and $\mu\equiv \sqrt{m^2/H^2-9/4}$. 
We will focus on massive particles belonging to the principal series, so that $\mu\ge 0$.  
The local part of the two-point function has support only at coincident points in position space, while the non-local part describes correlations over long distances. 
In Fourier space, the local and non-local parts of the two-point function are analytic and non-analytic in the momentum~$k$, respectively. In the late-time limit, we have
\begin{align}
\lim_{\eta, \eta'\to 0}\langle\sigma_\k(\eta)\sigma_{-\k}(\eta')\rangle'_{\rm local}&= \frac{H^2(\eta\eta')^{3/2}}{4\pi}\Gamma(-i\mu)\Gamma(i\mu)\left[e^{\pi\mu}\Big(\frac{\eta}{\eta'}\Big)^{i\mu}+e^{-\pi\mu}\Big(\frac{\eta}{\eta'}\Big)^{-i\mu}\right] ,\label{scalar2ptlocal}\\
\lim_{\eta, \eta'\to 0}\langle\sigma_\k(\eta)\sigma_{-\k}(\eta')\rangle'_{\text{non-local}}&= \frac{H^2(\eta\eta')^{3/2}}{4\pi}\left[\Gamma(-i\mu)^2\Big(\frac{k^2\eta\eta'}{4}\Big)^{i\mu}+\Gamma(i\mu)^2\Big(\frac{k^2\eta\eta'}{4}\Big)^{-i\mu}\right] .\label{scalar2ptnonlocal}
\end{align}
Away from the late-time limit, we use a series expansion of the Hankel function,
\begin{align}
H_{i\mu}(x) &= \sum_{n=0}^\infty\sum_{\pm} c_n^\pm(\mu,x)\, ,\quad  c_n^\pm(\mu,x) \equiv \pm\frac{(-1)^n}{n!}\frac{e^{\pi\mu(1\pm 1)/2}}{\sinh\pi\mu}\frac{\left(x/2\right)^{2n\pm i\mu}}{\Gamma(n+1+i\mu)} \, ,
\end{align}
to decompose the two-point function (\ref{scalar2pt0}) into its local and non-local pieces.
Summing over the set of local and non-local contributions, the two-point function can be split into\footnote{Away from the late-time limit, we are summing an infinite series of local/non-local elements for the propagator, in which case the distinction between the local and non-local parts is not as sharp. Nevertheless, we will see that this decomposition still leads to some useful insights.}
\begin{align}
\langle\sigma_\k(\eta)\sigma_{-\k}(\eta')\rangle'_{\text{local}} &= \frac{\pi}{4}\frac{H^2(\eta\eta')^{3/2}}{\sinh^2\pi\mu}\left[e^{\pi\mu}J_{i\mu}(-k\eta)J_{i\mu}^*(-k\eta')+e^{-\pi\mu}J^*_{i\mu}(-k\eta)J_{i\mu}(-k\eta')\right] ,\\
\hskip -3pt\langle\sigma_\k(\eta)\sigma_{-\k}(\eta')\rangle'_{\text{non-local}} &= \frac{\pi}{4}\frac{H^2(\eta\eta')^{3/2}}{\sinh^2\pi\mu}\left[J_{i\mu}(-k\eta)J_{i\mu}(-k\eta')+J^*_{i\mu}(-k\eta)J^*_{i\mu}(-k\eta')\right] ,
\end{align}
where $J_{i\mu}$ denotes the Bessel function of the first kind.

\vskip 4pt
To illustrate the distinct roles played by local and non-local parts, let us consider a coupling between $\pi$ and $\sigma$ of the form $\int\d^4x\,  a^3\hskip 1pt \rho\hskip 1pt \dot\pi_c\hskip 1pt\sigma$~\cite{Chen:2012ge, Pi:2012gf}. 
At tree level, this produces the following correction to the power spectrum of $\zeta$:
\begin{align}
\langle\zeta_{\k}\zeta_{-\k}\rangle' = P_\zeta(k) \left[1+ \frac{\cs^2\rho^2}{H^2}\big( {\cal C}_1+{\cal C}_2\big)\right] ,\label{scalar2pt}
\end{align}
where 
\begin{align}
{\cal C}_1&\equiv \frac{\pi}{4}e^{-\pi\mu}\,\bigg| \int_0^\infty \frac{\d x}{x}\,\sqrt{x} H_{i\mu}(x)\hskip 1pt e^{i\cs x} \bigg|^2\, , \label{C1}\\
{\cal C}_2&\equiv  -\frac{\pi}{2} \hskip 1pt  e^{-\pi\mu}\, {\rm Re} \left[ \int_0^\infty \frac{\d x}{\sqrt{x}}\, H_{i\mu}(x)e^{-i \cs x}\int_x^\infty \frac{\d y}{\sqrt{y}}\, H_{i\mu}^*(y) \hskip 1pt e^{-i\cs y}\right] . \label{C2}
\end{align}
The integral in~(\ref{C1}) can be evaluated analytically to give
\begin{align}
{\cal C}_1= \frac{\pi^2}{2\cosh^2\pi\mu}\, {}_2F_1\hskip -1pt\left(\frac{1}{2}-i\mu\,,\frac{1}{2}+i\mu\,,1\,,\frac{1-\cs}{2}\right)^2\, ,\label{C1cs}
\end{align}
where ${}_2F_1$ is the hypergeometric function. It is instructive to consider the $\cs \to 1$ and $\cs\to 0$ limits of the result (\ref{C1cs}):
\begin{figure}[t!]
\centering
\vskip 5pt
\includegraphics[scale=1]{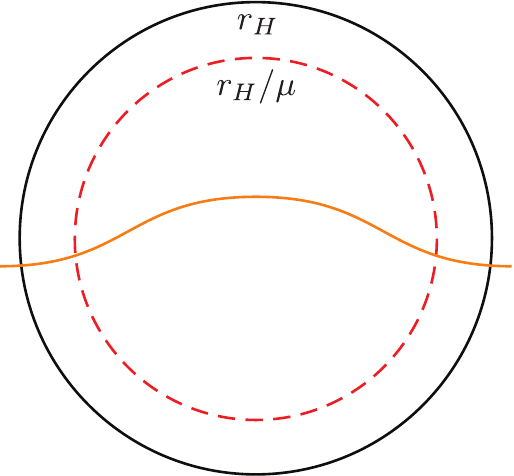}\qquad\qquad\includegraphics[scale=1]{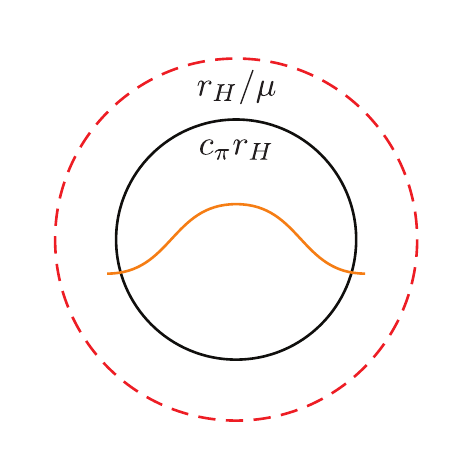}
\caption{\label{fig:horizon} Pictorial representations of the horizon crossing scale of the Goldstone boson (solid) and the scale associated with the turning point in the dynamics of a massive particle (dashed), with the left (right) diagram corresponding to $\cs =1$ ($\cs <\mu^{-1}$). The Hubble radius is denoted by $r_H\equiv H^{-1}$. We see that for $\cs  <\mu^{-1}$ the horizon crossing of the Goldstone boson occurs before the turning point of the massive particles, while for $\cs =1$ it occurs after.}
\end{figure} 
\begin{itemize}
\item For $\cs=1$, the hypergeometric function becomes unity, and (\ref{C1cs}) scales as $e^{-2\pi\mu}$ for large~$\mu$, as expected for the pair-production of massive particles. 
\item In the limit $c_\pi\to 0$, we instead get
\begin{align}
	\lim_{\cs\to 0}{\cal C}_1 = \frac{\pi^2}{2\cosh^2\pi\mu}\times \frac{\pi}{\Gamma(\frac{3}{4}+\frac{i\mu}{2})^2\hskip 2pt \Gamma(\frac{3}{4}-\frac{i\mu}{2})^2}\, ,
\end{align}
which scales as $e^{-\pi\mu}$ for large $\mu$ instead of the usual Boltzmann factor $e^{-2\pi\mu}$. 
\end{itemize}

\begin{figure}[t!]
\centering
\vskip 5pt
\includegraphics[scale=0.85]{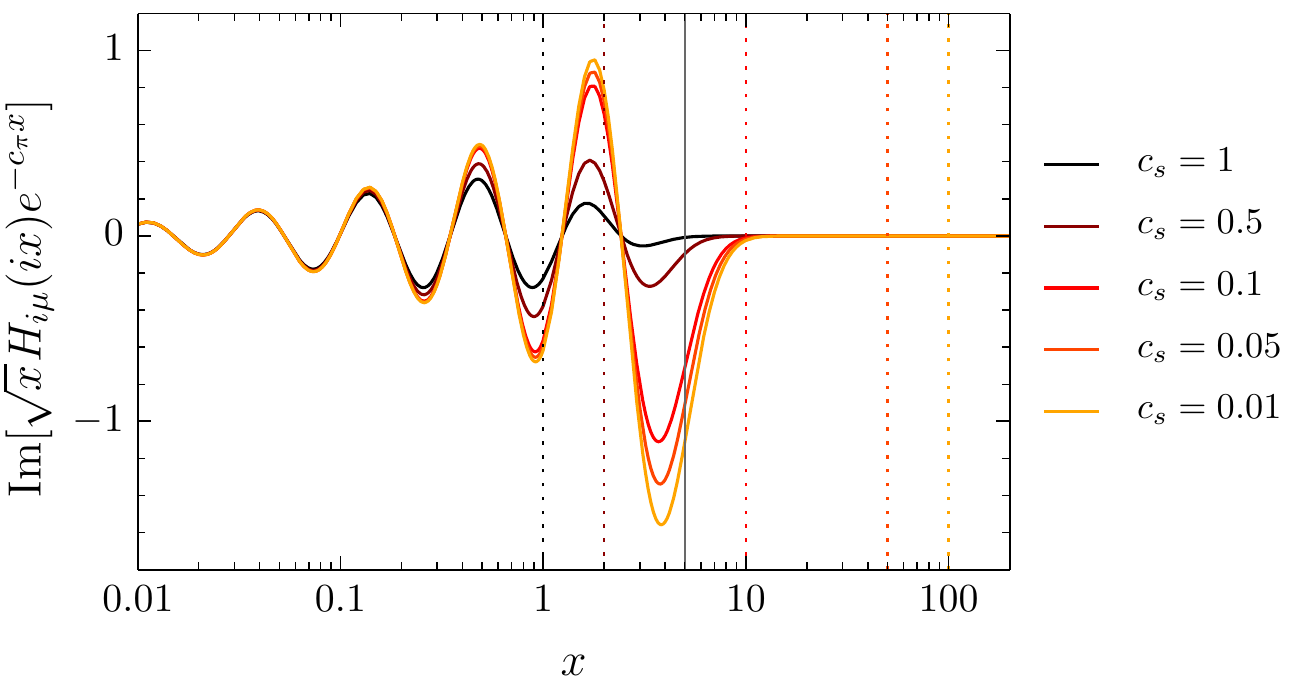}
\caption{\label{fig:integrand} 
Wick-rotated integrand of the integral in (\ref{C1}) as a function of $x=|k\eta|$ and for $\mu=5$. The vertical dotted lines indicate the times of sound horizon crossing of $\pi$, i.e.~$x = \cs^{-1}$, for each value of $\cs$. The solid vertical line marks the turning point of $\sigma$, i.e.~$x=\mu$. }
\end{figure}

To see why the exponential suppression of ${\cal C}_1$ changes for $\cs \ll 1$, we need to consider the change in the dynamics of $\sigma$ and $\pi$.
There are two relevant timescales in the problem:  
\vskip 4pt
{\it i}\hskip 1pt) at the {\it turning point}, $|k\eta| \sim \mu$, the mode function of the massive particle starts to decay, 
\vskip 2pt
{\it ii}\hskip 1pt)  at the {\it sound horizon crossing}, $|k\eta| \sim \cs^{-1}$, the Goldstone boson freezes. 
\vskip 4pt
\noindent
For $\cs=1$, event~{\it i}\hskip 1pt) occurs before {\it ii}\hskip 1pt), while for $\cs< \mu^{-1}$, the order is reversed (see Fig.~\ref{fig:horizon}). 
As a consequence, the integral in (\ref{C1}) is dominated at the horizon crossing of $\pi$ for $\cs=1$, while it is dominated by the turning point of $\sigma$ for $\cs < \mu^{-1}$. This is illustrated in Fig.~\ref{fig:integrand}, where we show the Wick-rotated integrand of the integral in (\ref{C1}) as a function of $x=|k\eta|$. 
A notable feature is the peak at $x \sim \mu$, which increases for small $\cs$.
For $\cs = 1$, the turning point occurs before horizon crossing and the overlap between $\pi$ and $\sigma$ is suppressed. For $\cs < \mu^{-1}$, on the other hand, the turning point occurs after the freeze-out of the Goldstone, which enhances the feature at $x\sim \mu$. This qualitatively explains the boost in the amplitude of ${\cal C}_1$ for small $\cs$.

\vskip 4pt
Let us now consider the time-ordered integral ${\cal C}_2$ in (\ref{C2}). For general $\cs$, it cannot be evaluated analytically, but some insights can be obtained by taking the limits $\cs \to 1$ and $\cs \to 0$:
\begin{itemize}
\item For $\cs=1$, the above decomposition of the $\sigma$-propagator into local and non-local pieces leads to \cite{Chen:2012ge}
\begin{align}
{\cal C}_2|_{\rm local} &=\frac{e^{\pi\mu}}{8\sinh\pi\mu}{\rm Re}\left[\psi^{(1)}\hskip -1pt\left(\frac{3}{4}+\frac{i\mu}{2}\right)-\psi^{(1)}\hskip -1pt\left(\frac{1}{4}+\frac{i\mu}{2}\right)\right]-e^{-2\pi\mu}\, (i\mu\leftrightarrow -i\mu)\, , \label{C2local}\\
{\cal C}_2|_{\text{non-local}} &= 0 \, , 
\end{align}
where $\psi^{(1)}(z)=\partial_z^2\ln \Gamma(z)$ is the polygamma function of order 1.\footnote{There are also logarithmically divergent 
terms within the separate integrals for the local and non-local
parts. These are the result of an imperfect decomposition between the two terms away from the late-time limit and the fact that we are integrating over time. However, these terms exactly cancel in the sum over all contributions, so that the final result remains finite.} For large $\mu$, the first term in (\ref{C2local}) scales as $\mu^{-2}$, which has a simple interpretation: a heavy field contributes to non-renormalizable interactions in the low-energy effective theory of the light fields with coefficients given by inverse powers of the mass of the heavy field. The second term is instead suppressed by $e^{-2\pi\mu}$, describing an effect which cannot be captured by a local Lagrangian of the light fields alone. Finally, we see that the non-local part of the $\sigma$-propagator does not contribute to the correction to the power spectrum. 

\item In the limit $\cs \to 0$, we find 
\begin{align}
\lim_{\cs \to 0}{\cal C}_2|_{\rm local}&=0\, ,\\
\lim_{\cs \to 0}{\cal C}_2|_{\text{non-local}}&= - \frac{\pi^2}{2\cosh^2\pi\mu}\times \frac{\pi}{\Gamma(\frac{3}{4}+\frac{i\mu}{2})^2\hskip 2pt \Gamma(\frac{3}{4}-\frac{i\mu}{2})^2}\, . \label{C2cs0}
\end{align}
We wish to highlight several features of this result. First, the local contribution to ${\cal C}_2$ vanishes.
This follows from the simple fact that the Goldstone bosons become non-propagating when $\cs=0$; hence, they can only communicate to each other through non-local effects. 
Second, the non-local contributions to ${\cal C}_1$ and ${\cal C}_2$ precisely cancel each other, implying that the correction to the two-point function (\ref{scalar2pt}) vanishes faster than $\cs^2$ in the limit $\cs \to 0$. This is the result of the cancellation between the contributions from the forward and backward branches of the integration contour. A way to see this is to drop the exponentials in $\cs$ in \eqref{C1} and \eqref{C2}, and notice that ${\cal C}_1+{\cal C}_2$ is now proportional to the sum of all Schwinger-Keldysh propagators for the $\sigma$ field; these propagators add up to zero. Of course, for small (but finite) $\cs$, we do not expect this cancellation to be exact. 
\end{itemize}
\begin{figure}[h!]
\centering
\vskip 5pt
\includegraphics[scale=.9]{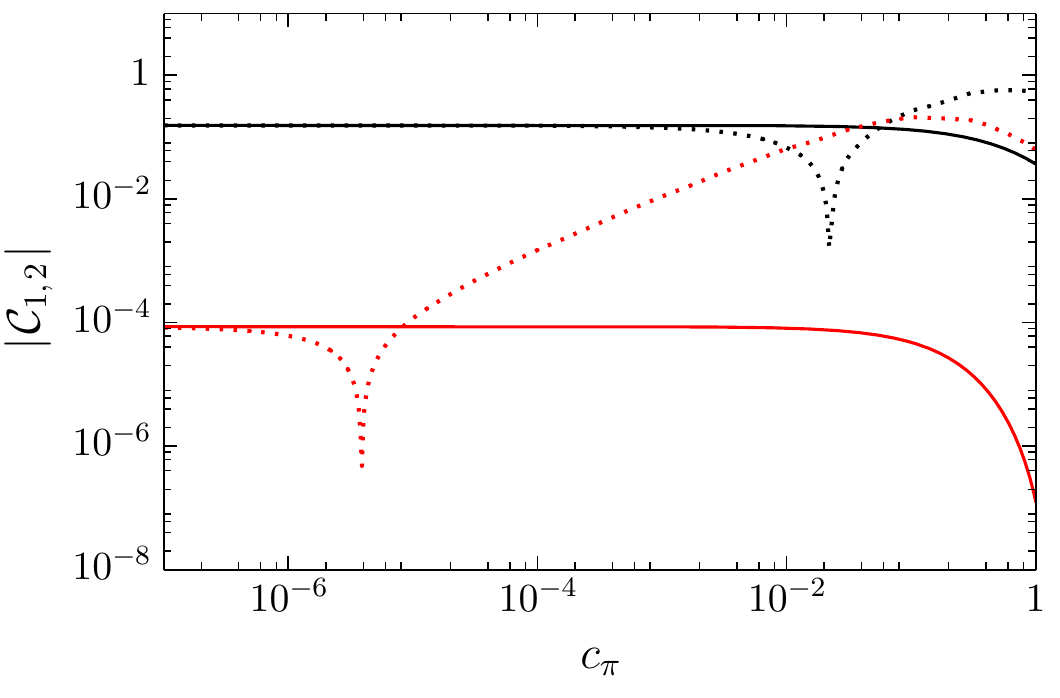}
\caption{\label{fig:PS}
${\cal C}_1$ and ${\cal C}_2$ as functions of $\cs$ for $\mu=1$ (black) and $\mu=3$ (red). The solid and dotted lines denote ${\cal C}_1$ and ${\cal C}_2$, respectively.}
\end{figure} 
To understand how the result for general $\cs$ interpolates between these two limiting behaviours, we evaluate ${\cal C}_2$ numerically. Figure~\ref{fig:PS} shows the analytical result (\ref{C1cs}) for ${\cal C}_1$ and a numerical computation of ${\cal C}_2$, both as functions of $\cs$. As $\cs$ is lowered, the exponential dependence on $\mu$ for both of the integrals changes. For ${\cal C}_1$, this happens relatively quickly when $\cs \lesssim \mu^{-1}$, agreeing with the intuition that reversing the ordering of the turning point of $\sigma$ and the horizon exit of $\pi$ changes the solution qualitatively. 
On the other hand, the transition in the exponential behavior for ${\cal C}_2$ only occurs for very small $\cs$, typically much smaller than the lower limit required for perturbative control of the non-renormalizable interaction $\dot\pi(\partial_i\pi)^2$ associated with $\cs$. This implies that, while for ${\cal C}_2$ the dependence on $\mu > 1$ will not change much within the allowed range of $\cs > 10^{-2}$, the exponential suppression $e^{-2\pi\mu}$ of ${\cal C}_1$ can be reduced to $e^{-\pi\mu}$ when $\cs<\mu^{-1}$.

\subsection[${\langle \zeta\zeta\zeta\rangle}$]{$\boldsymbol{\langle \zeta\zeta\zeta\rangle}$}
\label{ssec:ZZZ}

Next, we consider the imprints of massive spinning particles on the three-point function $\langle\zeta\zeta\zeta\rangle$. In single-field inflation, a long-wavelength curvature perturbation locally corresponds to a rescaling of the background experienced by short-wavelength fluctuations. As a result, the bispectrum $\langle \zeta \zeta \zeta\rangle$ satisfies a consistency relation for the squeezed limit~\cite{Maldacena:2002vr, Creminelli:2004yq, Cheung:2007sv}.
In particular, we can write a Taylor expansion around the squeezed limit, 
\begin{align} \label{eq:conscond}
\lim_{k_1 \ll k_3}\langle \zeta_{\k_1}\zeta_{\k_2}\zeta_{\k_3}\rangle' = P_\zeta(k_1)P_\zeta(k_3)\sum_{n=0}^\infty b_n \left(\frac{k_1}{k_3} \right)^n\, ,
\end{align}
where the leading coefficient is determined by the tilt of the scalar power spectrum, $b_0=-(n_s-1)$.  
The consistency condition furthermore fixes the coefficient of the linear term, $b_1$, and partially constrains higher-order coefficients~\cite{Creminelli:2012ed, Assassi:2012zq, Hinterbichler:2013dpa,Pimentel:2013gza,Berezhiani:2013ewa}. 
Since the contributions coming from $b_0$ and $b_1$ cannot be measured by a local observer \cite{Tanaka:2011aj, Pajer:2013ana}, any physical effect will only appear at order $(k_1/k_3)^2$~\cite{Creminelli:2011rh}. A crucial consequence of the consistency relation is the existence of the Taylor expansion \eqref{eq:conscond} with only integer powers of $k_1/k_3$. Interesting non-analytic deviations from (\ref{eq:conscond}), however, are known to arise in the presence of additional fields. For example, fractional powers $(k_1/k_3)^{\nu}$ can be present in quasi-single-field inflation~\cite{Chen:2009zp}, with scaling $0 <\nu\le3/2 $ in between the fully constrained $(k_1/k_3)^0$ term and the physical $(k_1/k_3)^2$ term. In this section, we will study such deviations for additional fields that carry spin.

\begin{figure}[h!]
\centering
\subfloat[\label{fig:3pt_a}]{\includegraphics*[scale=0.42]{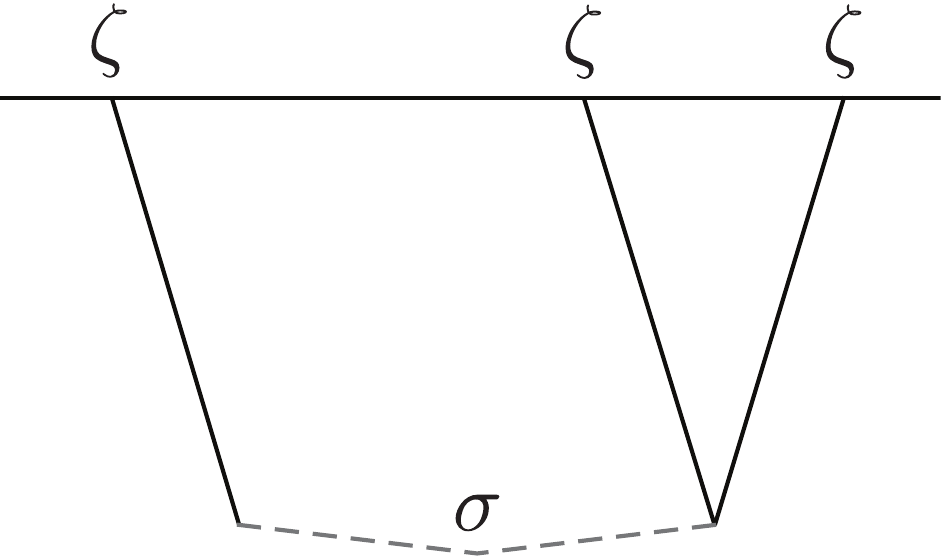} }\qquad\subfloat[\label{fig:3pt_b}]{\includegraphics*[scale=0.42]{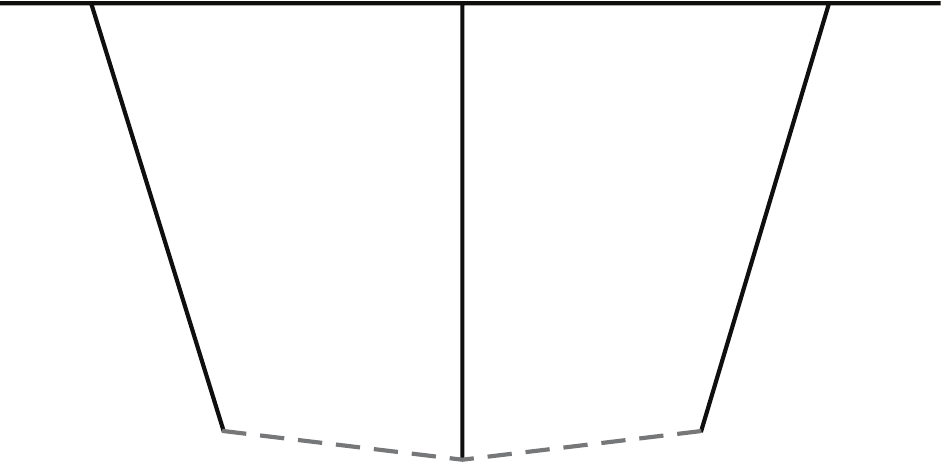}}\qquad
\subfloat[\label{fig:3pt_c}]{\includegraphics*[scale=0.42]{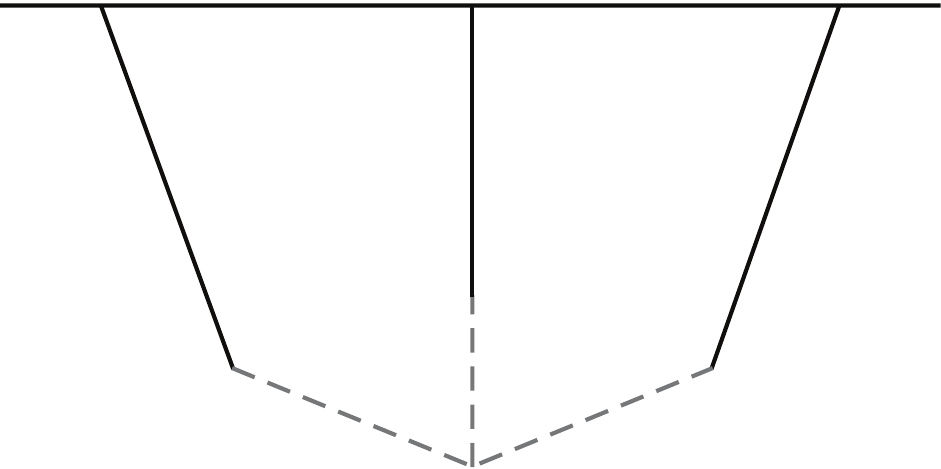}}\vskip -3pt
\caption{Tree-level diagrams contributing to $\langle\zeta\zeta\zeta\rangle$. The solid and dashed lines represent the curvature perturbation $\zeta$ and a spinning field $\sigma_{i_1\cdots i_s}$, respectively. \label{fig:3pt}}
\end{figure}

\vskip 4pt
Figure~\ref{fig:3pt} shows all possible tree-level contributions to $\langle\zeta\zeta\zeta\rangle$. The three diagrams share many qualitative features, so to avoid repetition we will mostly concentrate on the analysis of the single-exchange diagram [(a)], and only highlight the differences that arise for the other two diagrams~[(b,c)].  We will split the contributions to the bispectrum into its local and non-local parts. To avoid confusion with the alternative usage of ``local non-Gaussianity", we will refer to these contributions as  {\it analytic} and {\it non-analytic}, respectively. (This terminology highlights the distinctive scaling behavior in the squeezed limit.) Although we will ultimately be interested in the behavior of the latter, the observability of the signal will depend on the full bispectrum, so we will present the results for both types of contributions. 
As before, we will mostly restrict our analysis to particles in the principal series, with $\mu_s\ge 0$.

\paragraph{Single-exchange diagram}

We will first compute the bispectrum associated with the exchange of a single spinning field (Fig.~\ref{fig:3pt_a}). The relevant interaction Lagrangian is [cf.~eq.~(\ref{spinsLmix})]
\begin{align}
	{\cal L}_I = \frac{1}{a^{2s}}\left(\rho_s \hskip 1pt\partial_{i_1\cdots i_s}\pi_c\hat\sigma_{i_1\cdots i_s} + \frac{1}{\Lambda_s^s}\dot\pi_c\partial_{i_1\cdots i_s}\pi_c\hat\sigma_{i_1\cdots i_s}\right) . \label{equ:LintX}
\end{align}
We obtain the following bispectrum
\begin{align}
\frac{\langle\zeta_{\k_1}\zeta_{\k_2}\zeta_{\k_3}\rangle'}{\Delta_\zeta^4}= \alpha_s\Delta_\zeta^{-1} \times P_s(\hat \k_1\cdot\hat\k_3) \times {\cal I}^{(s)}(\mu_s,\cs,k_1,k_2,k_3) + \text{5 perms.}\, ,\label{zeta3spinssingle}
\end{align}
where an integral representation of the function ${\cal I}^{(s)}$ is given in Appendix~\ref{app:inin}. The dimensionless parameters $\alpha_s$ are 
\begin{align}
\alpha_s\equiv \frac{1}{\cs^{s-3/2}}\frac{\rho_s}{H^{2-s}}\,\left(\frac{H}{\Lambda_s}\right)^s  \, .\label{alpha1}
\end{align}
where we have included powers of $\cs$ in $\alpha_s$, so that the function ${\cal I}^{(s)}$ does not scale parametrically with $\cs$. By this we mean that ${\cal I}^{(s)}$ saturates to a constant value in the limit of small $\cs$, similar to the behavior of the integrals (\ref{C1}) and (\ref{C2}). The requirement of a perturbative treatment of non-Gaussianity implies that 
\beq
\alpha_s< 1\, \, .
\eeq
Notice that we have a stronger perturbativity condition on the bare parameters $\rho_s$ and $\Lambda_s$ for subluminal $\cs$, which takes into account the fact that the dispersion relation, $\omega=\cs k$, is non-relativistic. 

\vskip 4pt
\noindent
{\it Size of NG.}---It is customary to quantify the size of non-Gaussianity by the parameter 
\begin{align}
f_{\rm NL} \equiv\frac{5}{18}\frac{\langle\zeta_{\k_1}\zeta_{\k_2}\zeta_{\k_3}\rangle'}{P^2_\zeta(k)}\, ,\label{fnldef}
\end{align}
where the bispectrum is evaluated in the equilateral configuration, $k_1=k_2=k_3 \equiv k$. The overall size of the non-Gaussianity can only partially be read off from the prefactor in (\ref{zeta3spinssingle}), since there is a hidden dependence on  $\mu_s$ in the function ${\cal I}^{(s)}$. An estimate for the size of non-Gaussian signal~is 
\beq
f_{\rm NL}\, \sim\, f(\mu_s) \hskip 1pt \alpha_s \Delta_\zeta^{-1}\, , \label{eq:fnlsingle}
\eeq
where $f(\mu_s)$ gives the appropriate mass suppressions for the analytic and non-analytic parts\hskip 1pt\footnote{The displayed $\mu_s$ scalings are the asymptotic behaviors for large $\mu_s$. There is also a polynomial dependence in $\mu_s$ for the non-analytic part which competes with the exponential suppression for intermediate values of $\mu_s$.}${}^{,\hskip 1pt}$\footnote{It is more useful to consider this separation of the signal in the squeezed limit, where the distinction between the analytic and non-analytic parts becomes sharp, as we will show below.}
\begin{align}
	f(\mu_s)\ \equiv\ \begin{cases}
	\, \mu_s^{-2}\quad\ & \text{analytic},\\
		\, e^{-\pi\mu_s}\quad\ & \text{non-analytic},\quad \cs=1, \\
		\, e^{-\pi\mu_s/2}\quad\ &  \text{non-analytic},\quad \cs<\mu_s^{-1}.
	\end{cases}
\end{align}
We see that there are two sources of suppression in the signal: the mass suppression as a function of $\mu_s$ and the mixing efficiency parameterized by $\alpha_s$. At the same time, there is a $\Delta_\zeta^{-1}\approx 10^5$ enhancement in the signal. It is this large factor that can, in principle, allow for observable non-Gaussianity even in the presence of the above suppressions. 
The size of the analytic part is only power-law suppressed and thus dominates for large mass, whereas the non-analytic part is always accompanied by an exponential Boltzmann suppression. For $\cs =1$, the dominant non-analytic term is suppressed by $e^{-\pi\mu_s}$. 
As explained in \cite{Arkani-Hamed:2015bza,Mirbabayi:2015hva}, this arises from the quantum interference of two wavefunctions: $\Psi[2\sigma]\propto e^{-\pi\mu_s}$ for pair-produced massive particles and $\Psi[0\sigma]$ for the wavefunction involving no spontaneously created massive particles. This interference contribution is larger than the probability of pair-producing massive particles, which is $|\Psi[2\sigma]|^2\propto e^{-2\pi\mu_s}$.  For $\cs <\mu_s^{-1}\ll 1$, the exponential suppression of the non-analytic part changes to $e^{-\pi\mu_s/2}$. We have already encountered this phenomenon in \S\ref{sec:PS}: for small $\cs$ the horizon crossing  of the Goldstone boson occurs {\it before} the turning point in the mode function of the massive particle. In this case, we are picking out the contribution of the wavefunction for a pair of massive particles not in the late-time limit, but at the turning point, which comes with a different exponential factor.

\vskip 4pt
In \S\ref{sec:naturalness}, we derived naturalness constraints on the mixing parameters of the effective theory. For the parameter $\alpha_s$ in (\ref{alpha1}), the radiative stability of the mass (\ref{spinsloop}) implies
\begin{align}
\alpha_s \lesssim \ \bigg(\frac{2\pi\Delta_\zeta}{\cs^2}\bigg)^{(s+1)/2}\, .
\label{naturalsingle}
\end{align}
For $\cs=1$, this naturalness constraint is rather strong, implying that large non-Gaussianity, $f_{\rm NL} > 1$, is only possible if additional physics, such as supersymmetry, 
stabilizes the mass of the spinning 
particle, or if the mass term is fine-tuned. For $\cs \ne 1$, the current observational constraint $\cs \ge 0.024$~\cite{Ade:2015lrj} still allows for naturally large non-Gaussianity, although within a rather narrow range in the small $\cs$ regime.

\vskip 4pt 
Some comments are in order concerning the observability of particles with odd spins. In~\cite{Arkani-Hamed:2015bza}, it was shown that the diagram due to the exchange of an odd-spin  
particle vanish exactly at leading order in the weak breaking of conformal symmetry. 
At subleading orders, however, there are non-zero contributions from odd-spin  
particles.\footnote{When the approximate conformal invariance is valid, we can think of this in terms of correlation functions of the inflaton $\Phi(t,\x)=\phi(t)+\varphi(t,\x)$, where $\dot\phi \ne 0$ characterizes the weak breaking of conformal symmetry. The leading three-point function for the inflaton perturbation $\varphi$ will be given by the four-point function of $\Phi$ with one external leg set to $\dot\phi$:
\begin{align}
\langle\varphi\varphi\varphi\rangle'\propto \langle \varphi\varphi\sigma\rangle'\langle\sigma\varphi\dot\phi\rangle' \propto \dot\phi\langle \varphi\varphi\sigma\rangle' \langle\sigma\varphi\rangle_{\rm inf}'	\, ,
\end{align}
where $\langle\cdots\rangle_{\rm inf}$ denotes an inflationary correlation function which breaks conformal symmetry \cite{Kundu:2014gxa}. However, in the conformally symmetric case, $\langle \varphi\varphi\sigma\rangle$ vanishes when $\sigma$ has odd spin \cite{Giombi:2011rz}. The next-to-leading order result is given by the six-point function with three insertions of $\dot\phi$,
\begin{align}
\langle\varphi\varphi\varphi\rangle' \propto \langle \dot\phi\varphi\varphi\sigma\rangle' \langle\sigma\varphi\dot\phi^2\rangle'	\propto \dot\phi^3\langle \varphi\varphi\sigma\rangle_{\rm inf}'\langle\sigma\varphi\rangle_{\rm inf}' \, .
\end{align}
This is suppressed by an additional factor of $\dot\phi^2$, but notice that the correlator $\langle\varphi\varphi\sigma\rangle_{\rm inf}$, not being constrained by conformal symmetry, does not have to vanish for odd-spin $\sigma$.} When conformal symmetry is strongly broken, these terms become as important as the leading ones, and odd-spin  
particles can leave an equally relevant imprint on the correlation function
$\langle \zeta \zeta \zeta \rangle$. 
Nevertheless, the amplitude of the bispectrum with an intermediate spin-1  
particle is 
\begin{align}
f_{\rm NL} \sim f(\mu_1)\hskip 1pt \sqrt{\cs}\,\frac{\rho_1^2}{H^2}\, .
\end{align}
As long as the mixing is perturbative, $\rho_1 < H$,  this non-Gaussianity is constrained to be less than unity.  We see that a spin-1  
particle cannot lead to large non-Gaussianity because the size of the cubic vertex in  (\ref{Lmix1}) is tied to the quadratic mixing coefficient. In fact, the same reasoning applies to the coupling to scalar fields, which is why the single-exchange diagram has been neglected in the context of quasi-single-field inflation \cite{Chen:2009zp, Baumann:2011nk}.
This fact, however, is only tied to spins zero and one, and the bispectrum does not have to be suppressed for higher odd-spin 
particles. Moreover, we will see that the diagrams involving more than a single exchange can allow for observable non-Gaussianity, even for spin one. 

\vskip 4pt 
\noindent
{\it Shape of NG.}---Before considering the general shape of the bispectrum, we will first analyze the singular behavior of the bispectrum in the squeezed limit, mainly concentrating on  
particles with even spins. We will quote results whose derivations can be found in Appendix~\ref{app:squeezed}.
\begin{itemize}
\item For the analytic part of the bispectrum, we get 
\begin{align}
\lim_{k_1 \ll k_3}\langle\zeta_{\k_1} \zeta_{\k_2}\zeta_{\k_3}\rangle' \propto  \frac{1}{k_1^3 k_3^3}\left(\frac{k_1}{k_3}\right)^2\, .\label{analyticSq}
\end{align}
We see that the local effects of massive particles lead to the same squeezed limit behavior as for single-field inflation, cf.~(\ref{eq:conscond}). This is expected, since the massive particle can be integrated out for large $\mu_s$, producing an effective cubic vertex of the form $\dot\pi(\hat\partial_{i_1\cdots i_s}\pi)^2$. The presence of extra particles therefore cannot be inferred from this part of the signal. Although the analytic part of the non-Gaussianity is itself interesting and more information can be gained by analyzing its shape for general momentum configurations, we have to treat it as an effective noise in the squeezed limit as far as the detection of extra particles is concerned.

\item  
For the non-analytic part, we find
\begin{align}
\lim_{k_1 \ll k_3} \langle\zeta_{\k_1}\zeta_{\k_2}\zeta_{\k_3}\rangle' \,\propto\, \frac{1}{k_1^3k_3^3} \left(\frac{k_1}{k_3}\right)^{3/2}P_s(\hat\k_1\cdot\hat\k_3)\cos\left[\mu_s\ln\left(\frac{k_1}{k_3}\right)+\phi_s\right] ,\label{singlenonanalytic}
\end{align}
where the phase $\phi_s$ is uniquely fixed in terms of $\mu_s$ and $\cs$ (see Appendix~\ref{app:squeezed}). 
The suppression factor $(k_1/k_3)^{3/2}$ represents the dilution of the physical particle number density due to the volume expansion. 
This non-analytic scaling in the squeezed limit, corresponding to an intrinsically non-local process, cannot be mimicked by a local interaction within the effective theory of a single field.
The signal contains oscillations in $\ln(k_1/k_3)$, with a  frequency  set by the mass of the spinning particle. This is due to the fact that the wavefunctions of massive particles oscillate logarithmically in time on superhorizon scales. The spin of the extra particle is reflected in the angular dependence, which is given by a Legendre polynomial of the angle between the short and long momenta. 

\vskip 4pt
The above behavior applies for  
particles in the principal series, for which $\mu_s\ge 0$. For  
particles in the complementary series, $\mu_s$ becomes imaginary and the scaling of the squeezed bispectrum changes to 
\begin{align}
\lim_{k_1 \ll k_3}  \langle\zeta_{\k_1}\zeta_{\k_2}\zeta_{\k_3}\rangle'&\,\propto\, \frac{1}{k_1^3k_3^3}\left(\frac{k_1}{k_3}\right)^{3/2-\nu_s} P_s(\hat\k_1\cdot\hat\k_3)\, ,\label{tsssqueezed4}
\end{align}
with $\nu_s\equiv - i\mu_s$ real. For $s\ge 2$, unitarity implies $\nu_s\in [0,1/2)$, and the singular behavior in the squeezed limit is suppressed by at least $k_1/k_3$ compared to the leading term in the consistency relation (\ref{eq:conscond}).

\vskip 4pt
The fact that the polarization tensors corresponding to odd-spin particles are odd under the exchange of two short momenta, together with momentum conservation, implies that the signal will gain an extra suppression factor of $k_1/k_3$ in the squeezed limit compared to the case of even spin. This means that the non-analytic part due to odd-spin  
particles scales as $(k_1/k_3)^{5/2}$ in the squeezed limit, which is more suppressed than the analytic part that scales as $(k_1/k_3)^2$. The latter, however, have an analytic dependence on momenta and correspond to local correlations in position space. Thus, the presence of odd-spin particles could still be inferred from long-distance correlations, although it might be subdominant compared to other non-local effects. 
\end{itemize}
It is possible to understand the different behaviors in the squeezed limit intuitively. For concreteness, let us consider the exchange of a spin-2 field involving the interactions  $\partial_i\partial_j\pi\hat\sigma_{ij}$ and $\dot\pi\partial_i\partial_j\pi\hat\sigma_{ij}$.
The bispectrum in the isosceles-triangle configuration, $k_2=k_3$, consists of three different permutations of the external legs:
\begin{align}
\langle\zeta_{\k_1}\zeta_{\k_2}\zeta_{\k_3}\rangle' \, \propto\,\underbrace{\includegraphicsbox[scale=.38]{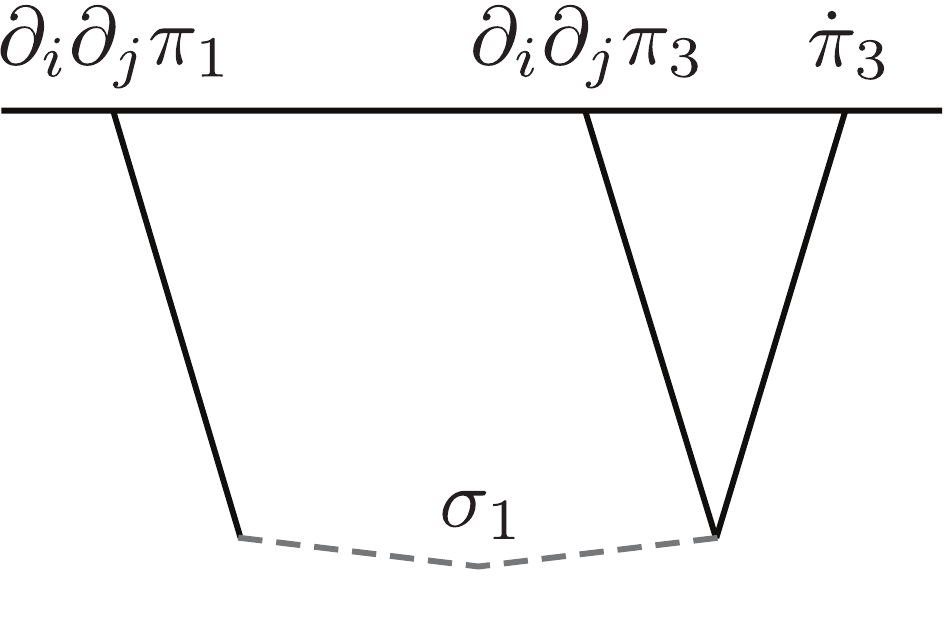}}_{I_1\, \equiv\,  {\cal I}(k_1,\hskip 1pt k_3,\hskip 1pt k_3)} +\underbrace{\includegraphicsbox[scale=.38]{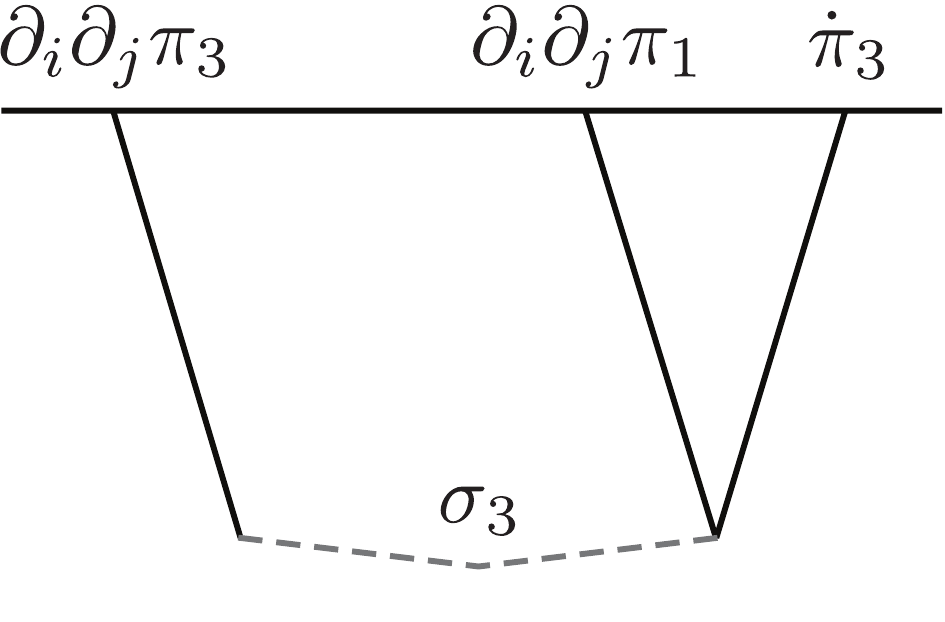}}_{I_2\, \equiv\, {\cal I}(k_3,\hskip 1pt k_1,\hskip 1pt k_3)} +\underbrace{\includegraphicsbox[scale=.38]{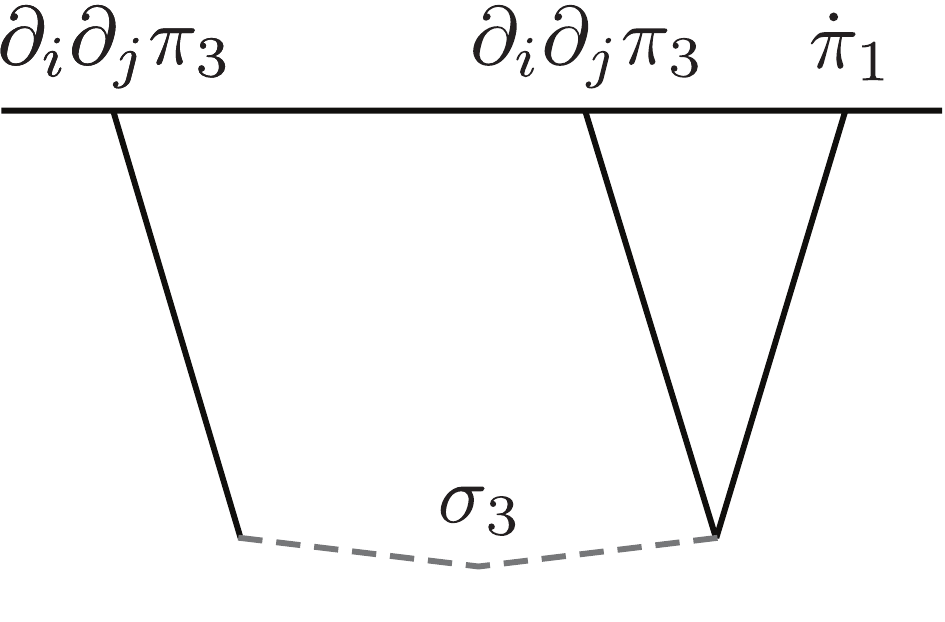}}_{I_3\, \equiv\, {\cal I}(k_3,\hskip 1pt k_3,\hskip 1pt k_1)}  \ ,\label{eq:diagrams}
\end{align}
where $\pi_n\equiv \pi(\k_n)$, $\sigma_n\equiv \sigma_{ij}(\k_n)$ and ${\cal I}(k_1,k_2,k_3)\propto P_2(\hat\k_1\cdot\hat\k_3)\, {\cal I}^{(2)}(\mu_2,\cs,k_1,k_2,k_3)$. The non-analytic squeezed limit (\ref{singlenonanalytic}) arises if the massive exchange particle carries the soft momentum, corresponding to the contribution $I_1$ in (\ref{eq:diagrams}). 
This describes a non-local conversion process between the massive particle and the Goldstone boson between the horizon crossing times of the long and short modes. However, when the mass of the extra particle becomes large, it can be integrated out and the same effect will be captured by a local vertex. In that case, the bispectrum should become indistinguishable from that produced by a self-interaction of $\pi$, namely $\dot\pi(\hat\partial_{ij}\pi)^2$. Note, in particular, that this interaction is symmetric under the exchange of the momenta associated with the two external legs with spatial gradients. This allows us to gauge how well the interaction is approximated by a local vertex by looking at how similar the terms $I_1$ and $I_2$ are. Both $I_2$ and $I_3$ will lead to analytic scalings in the squeezed limit, where the latter produces (\ref{analyticSq}).

\begin{figure}[h!]
\centering
\includegraphics[scale=0.75]{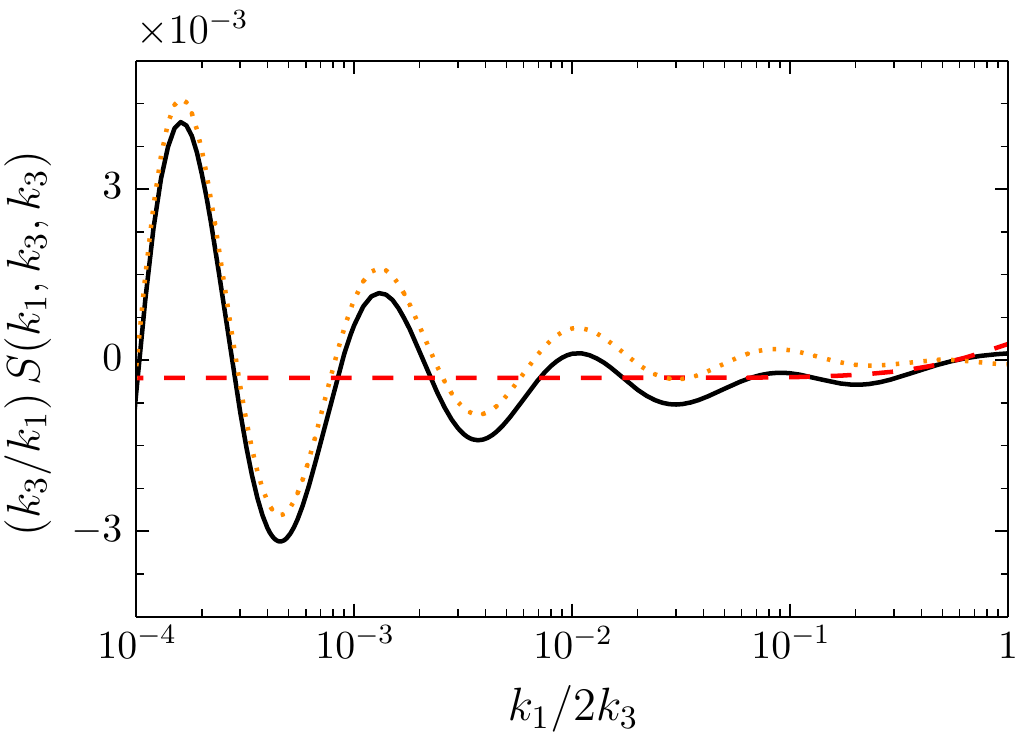}\quad\includegraphics[scale=0.75]{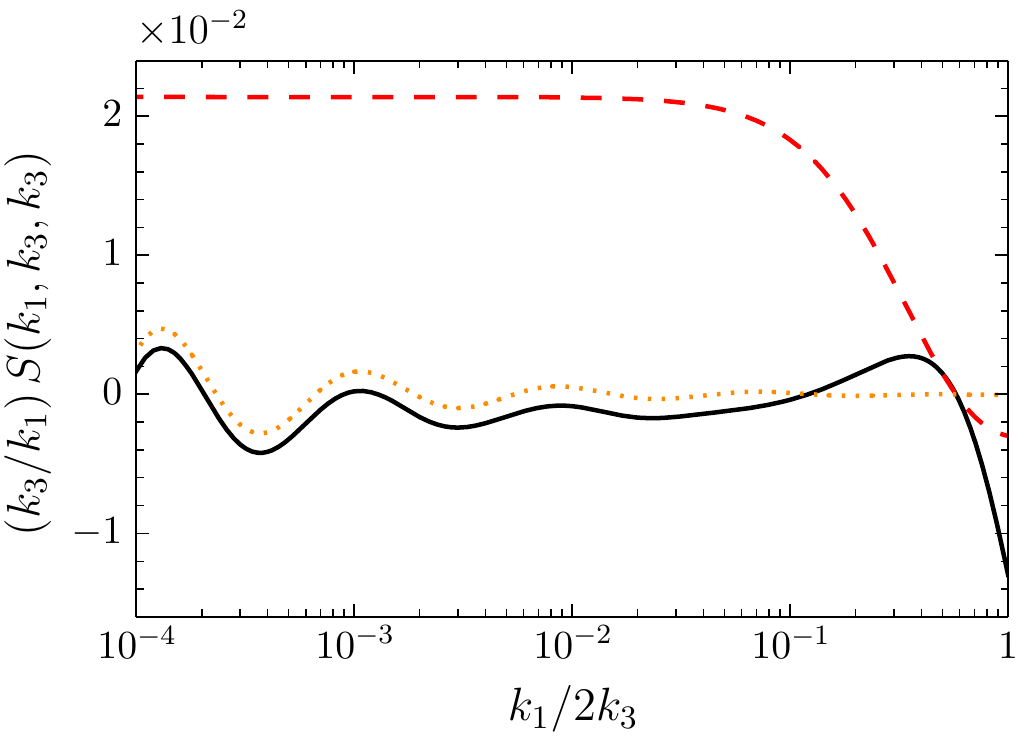}
\includegraphics[scale=0.75]{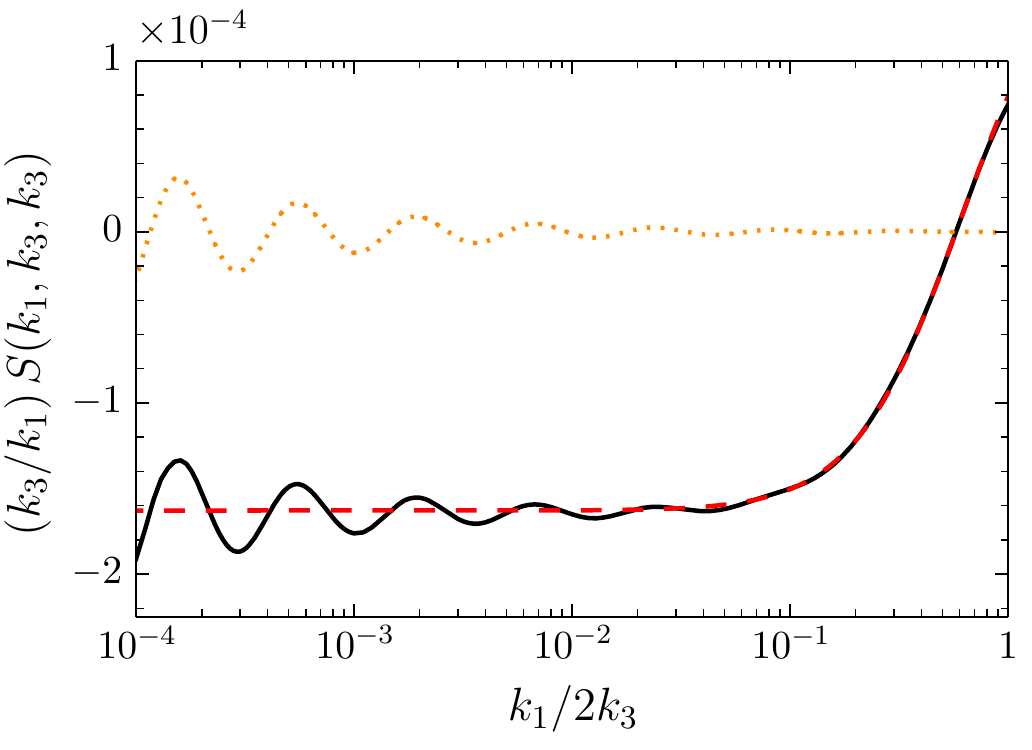}\quad\includegraphics[scale=0.75]{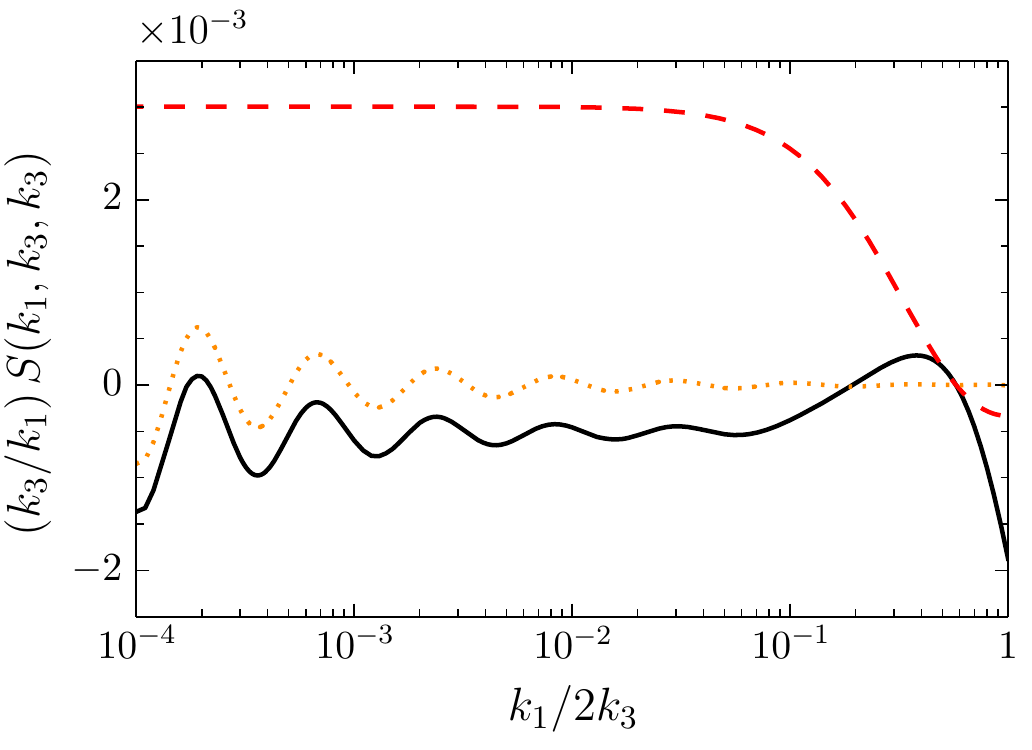}
\includegraphics[scale=0.75]{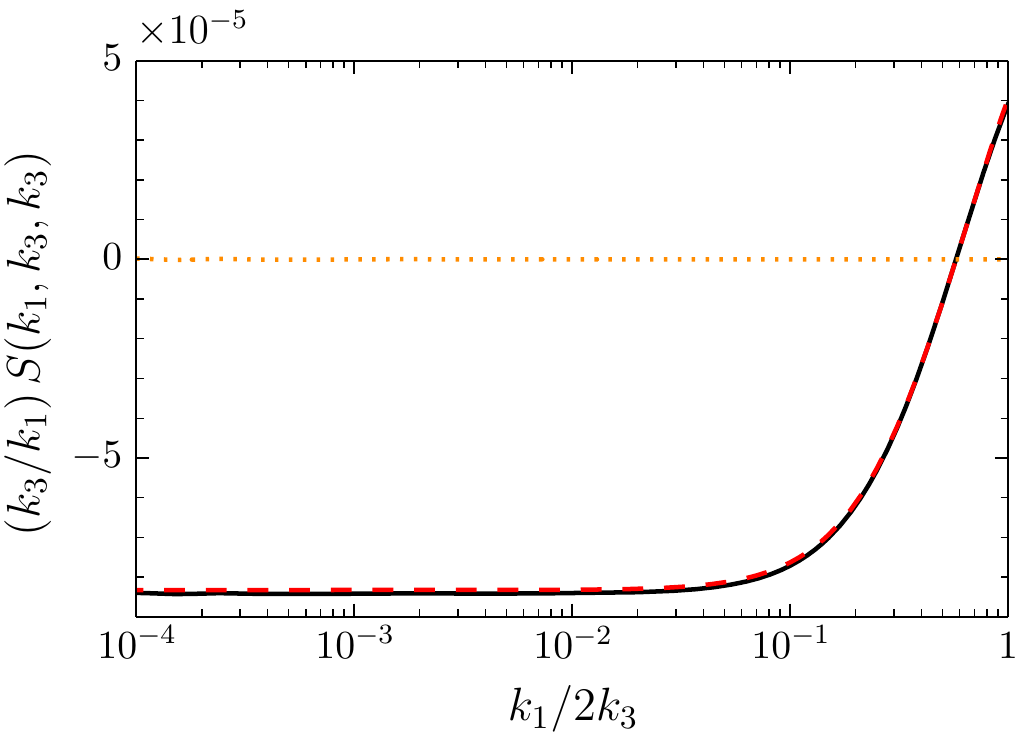}\quad\includegraphics[scale=0.75]{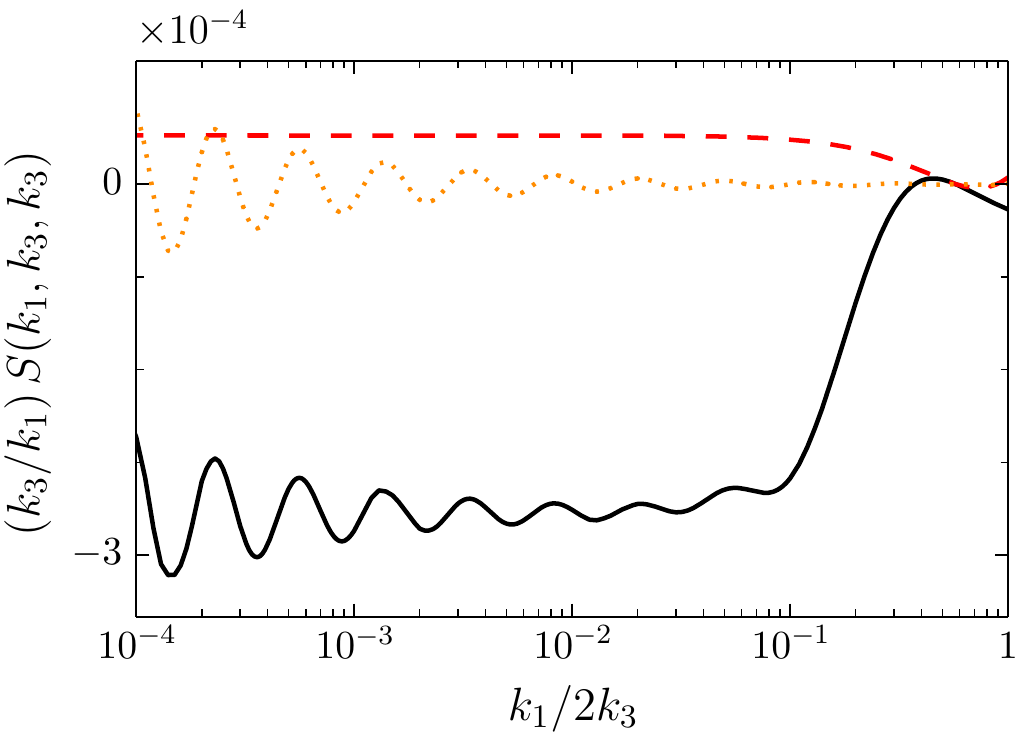}
\caption{Shape functions (in units of $\alpha_2\Delta_\zeta^{-1}$) for the spin-2 single-exchange diagram in the isosceles-triangle configuration, $k_2=k_3$, with $\mu_2=3$ ({\it top}), $\mu_2=5$ ({\it middle}), and $\mu_2=7$ ({\it bottom}) for $\cs=1$ ({\it left}) and $\cs=0.1$ ({\it right}).  The solid and dashed lines correspond to the numerical results for the parts of the signal corresponding to   
the terms $I_1$ and $I_2$ in (\ref{eq:diagrams}), respectively.
Not shown in the figure is the term $I_3$, which produces an analytic scaling in the squeezed limit and is needed to obtain the full bispectrum. Convergence of the solid and dashed lines indicates that the same effect can be captured by a local vertex $\dot\pi(\hat\partial_{ij}\pi)^2$ in the single-field EFT. The dotted lines show the analytical predictions for the non-analytic part.  
\label{fig:Spin2Single}}
\end{figure}

\vskip 4pt
To analyze the  
shape of the bispectrum for general momentum configurations, we proceed numerically. For this purpose, it is convenient to define a dimensionless shape function 
\begin{align}
S(k_1,k_2,k_3) \equiv \frac{k_1^2k_2^2k_3^2}{(2\pi)^4}\frac{\langle\zeta_{\k_1}\zeta_{\k_2}\zeta_{\k_3}\rangle'}{\Delta_\zeta^4} \, .\label{shapefuncsingle}
\end{align}
Figure~\ref{fig:Spin2Single} shows two-dimensional projections of the shape function for spin 2 with $\mu_2= 3,\hskip 1pt 5,\hskip 1pt 7$ and $\cs=1,\hskip 1pt 0.1$ in the isosceles-triangle configuration, $k_2=k_3$. 
For the reasons explained in the previous paragraph, in Fig.~\ref{fig:Spin2Single} we have shown separately the shape functions corresponding to the contributions $I_1$ and $I_2$ in (\ref{eq:diagrams}).\footnote{We have omitted $I_3$ in the plots, which has the same analytic scaling as in~(\ref{analyticSq}) and thus shows no interesting features. Of course, this contribution should be added in order to obtain the full bispectrum.} As anticipated, these contributions exhibit different scalings in the squeezed limit. The plots show $(k_3/k_1)\times S$, so that the analytic part is expected to approach a constant in the squeezed limit, while the non-analytic part grows as $(k_1/k_3)^{-1/2}$ for small $k_1$. 
We see that the shape of the bispectrum is mostly governed by the non-analytic part for small mass, giving almost pure oscillations. The amplitude of this effect, however, goes as $e^{-\pi\mu_2}$ for large $\mu_2$. The analytic part, being power-law suppressed, therefore takes over in size as the mass increases, and the shape approaches the equilateral form in the limit of large mass. 
For large mass, it is clear that the non-Gaussianity is dominated by the analytic piece, with small oscillations coming from the non-analytic piece indicating the presence of a heavy mode. For $\cs =1$, the contributions $I_1$ and $I_2$ lead to the same shape of the bispectrum for $\mu_2=7$, indicating that the $\pi$-$\sigma$ conversion process has become local. Indeed, in this case the bispectrum precisely overlaps with that of the local interaction $\dot\pi(\hat\partial_{ij}\pi)^2$. For small $\cs$, we have argued that the exponential suppression is instead $e^{-\pi\mu_2/2}$. The fact that we see more pronounced oscillations for $\cs=0.1$ is a consequence of this. Moreover, for small $\cs$, the shapes of the contributions $I_1$ and $I_2$ are no longer identical. Note that, in order for the massive particle to be integrated out, the time of its turning point should be much earlier than the time at which the Goldstone boson crosses its sound horizon, which translates into the condition $\cs>\mu_2^{-1}$. For $\cs=0.1$, this condition is not satisfied for the list of mass parameters used in the figure, which is the reason why we do not see the convergence to the local behavior. We have checked that the convergence does indeed happen for sufficiently large $\mu_2 > \cs^{-1}$. 

\begin{figure}[t!]
\centering
\qquad\qquad
\includegraphics[scale=0.8]{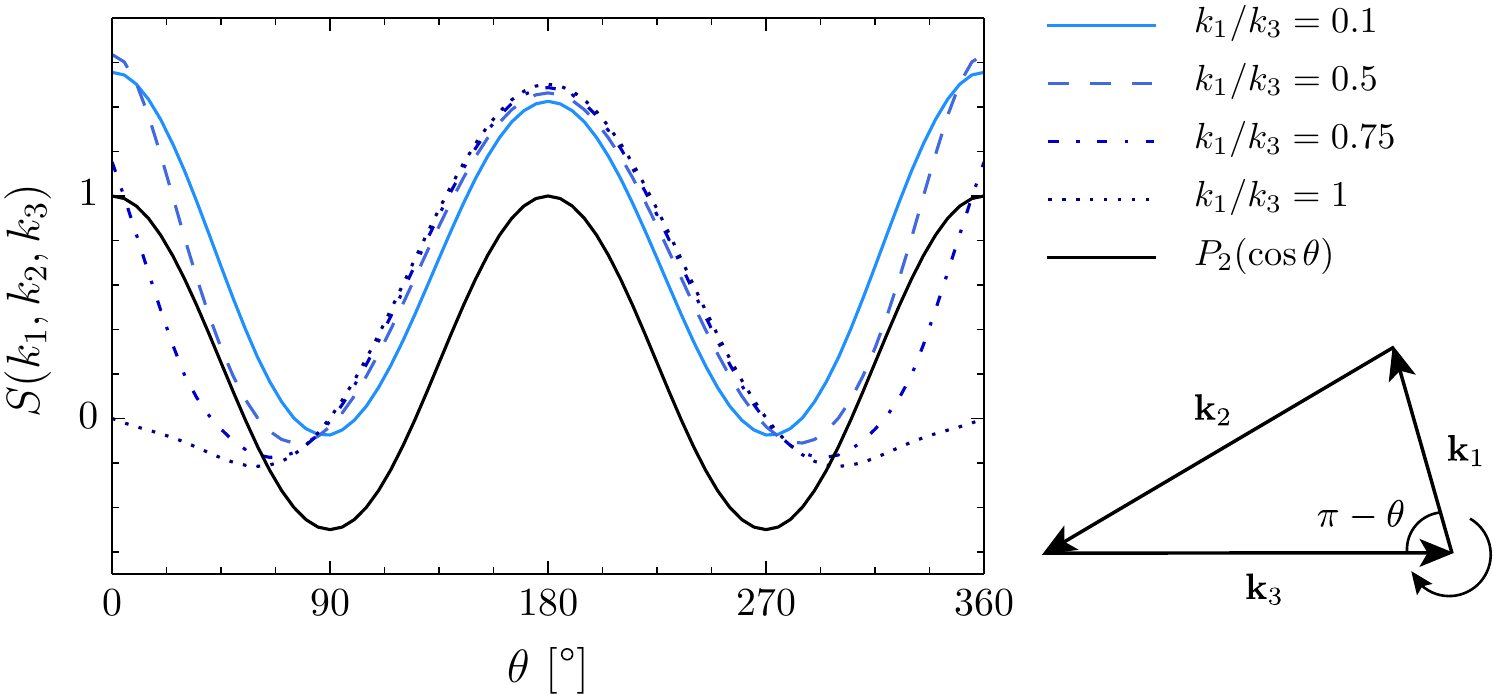}
\caption{\label{fig:Spin2SingleAngle} Shape functions of the spin-2 single-exchange diagram with $\mu_2=5$ and $\cs=1$ as a function of the base angle $\theta=\cos^{-1}(\hat\k_1\cdot\hat\k_3)$ for fixed ratios of $k_1/k_3$. For easy comparison, the plot has been normalized such that the height difference between $\theta=90^\circ$ and $\theta=180^\circ$ of each curve is fixed to 3/2.}
\end{figure}

\vskip 4pt
Another characteristic of the signal due to spinning particles is its angular dependence. Figure~\ref{fig:Spin2SingleAngle} shows the shape function of the total signal as a function of the angle between the long and short momenta, $\theta \equiv \cos^{-1}(\hat\k_1\cdot \hat \k_3)$, for a range of momentum configurations with fixed $k_1/k_3$. For visualization purposes, the plot has been rescaled so that it can be compared more easily to the Legendre polynomial $P_2(\cos\theta)$. As expected, the angular dependence converges to the pure Legendre behavior as the triangle becomes squeezed, $k_1/k_3 \ll 1$. 
The non-zero offset is due to the analytic part which doesn't carry any angular dependence. We also see that the angular dependence deviates from the pure Legendre behavior as the triangle approaches the equilateral shape. Still, the peak around the flat triangle ($\theta=180^\circ$) remains prominent regardless of the momentum configuration. This suggests that the information about a particle's spin can still be inferred without necessarily going to very squeezed momentum configurations, since the width of the peak is still fixed by the polarization tensor of the spinning 
particle. This property can serve as an important tool for detecting odd-spin particles, whose signal in the squeezed limit necessarily gains an extra suppression in the soft momentum.

\paragraph{Double-exchange diagram}
The bispectrum for the double-exchange diagram~(Fig.~\ref{fig:3pt_b})~is  
\begin{align}\label{zeta3spin1double}
\frac{\langle \zeta_{\k_1}\zeta_{\k_2}\zeta_{\k_3}\rangle'}{\Delta_\zeta^4} =  \tilde\alpha_s\Delta_\zeta^{-1} \times P_s(\hat\k_2\cdot\hat\k_3)  \times {\cal J}^{(s)}(\mu_s,\cs,k_1,k_2,k_3) +\text{5 perms.}\ ,
\end{align}
where the function ${\cal J}^{(s)}$ is given explicitly in Appendix~\ref{app:inin}, and the dimensionless parameters $\tilde \alpha_s$ are  
\begin{align}
\tilde\alpha_s\equiv \lambda_s\hskip 1pt\left(\frac{\rho_s}{H^{2-s}}\right)^2<1\, ,
\end{align}
with $\lambda_s$ and $\rho_s$ defined in~(\ref{spinsLmix}).

\vskip 4pt
The size of the non-Gaussianity associated with the double-exchange diagram can be read off from (\ref{eq:fnlsingle}) after replacing $\alpha_s$ by $\tilde\alpha_s$, but with an extra suppression of $\mu_s^{-2}$, because this diagram involves another particle exchange. The condition for radiative stability~(\ref{loop2X}) imposes the following upper limit on the size of the mixing parameter: 
\begin{align}
\tilde\alpha_s \ \lesssim \ \frac{(2\pi\Delta_\zeta)^{1/2}}{\cs^2}\, .
\end{align}
Notice that this is a much weaker constraint than the corresponding constraint for the single-exchange diagram (\ref{naturalsingle}). Depending on the values of $\cs$, this may or may not be stronger than the requirement for perturbativity, $f_{\rm NL} < \Delta_\zeta^{-1}$. This diagram can thus naturally produce detectable levels of non-Gaussianity, even for $\cs=1$.

\vskip 4pt
Note that this diagram involves two $\pi$-$\sigma$ conversion processes. When one of these processes becomes local, the double-exchange diagram becomes essentially equivalent to the single-exchange diagram. This can be seen by replacing one of the $\hat\sigma_{i_1\cdots i_s}$  
legs in the cubic vertex $\dot\pi\hat\sigma^2_{i_1\cdots i_s}$ by $\partial_{i_1\cdots i_s}\pi$, after which the interaction becomes the same as the cubic vertex for the single-exchange diagram. As a result, the squeezed-limit behavior for this diagram is essentially the same as that of the single-exchange diagram. Hence, the analysis we have presented for the single-exchange diagram applies also to the double-exchange diagram.

\paragraph{Triple-exchange diagram}
As indicated in (\ref{spinsLmix}), there is a slight difference between the form of the cubic self-interaction of spinning fields for even and odd spins. For concreteness, we will present the results for the former. The bispectrum for the triple-exchange diagram  (Fig.~\ref{fig:3pt_c}) is
\begin{align}\label{zeta3spin1triple}
\frac{\langle \zeta_{\k_1}\zeta_{\k_2}\zeta_{\k_3}\rangle'}{\Delta_\zeta^4} =  \hat\alpha_s\Delta_\zeta^{-1} \times P(\hat\k_1,\hat\k_2,\hat\k_3) \times {\cal K}^{(s)}(\mu_s,\cs, k_1,  k_2,  k_3) + \text{5 perms.}\, ,
\end{align}
where $P(\hat\k_1,\hat\k_2,\hat\k_3) \equiv \varepsilon^0(\hat\k_1)\hskip -1pt\cdot\hskip -1pt\varepsilon^0(\hat\k_2)\hskip -1pt\cdot\hskip -1pt\varepsilon^0(\hat\k_3)$ is a symmetric contraction of the longitudinal polarization tensors $\varepsilon^0_{i_1\cdots i_s}$ (see Appendix~\ref{app:spindS} for the precise definition of the polarization tensor) that reduces to $P_s(\hat\k_1\cdot\hat\k_3)$ in the squeezed limit. The couplings $\hat \alpha_s$ are
\begin{align}
\hat\alpha_s\equiv \xi_s \left(\frac{\rho_s}{H^{2-s}}\right)^2<1\, ,
\end{align}
where $\xi_s$ was introduced in (\ref{spinsLmix}).
The function ${\cal K}^{(s)}$ can be found in Appendix~\ref{app:inin}. 

\vskip 4pt
The size of the non-Gaussianity associated with this diagram can, again, be read off from~(\ref{eq:fnlsingle}), with $\alpha_s$ replaced by $\hat\alpha_s$, and taking into account an extra suppression of $\mu_s^{-4}$. Although the qualitative features of the non-analytic signal will be similar to that of the other diagrams, there are some relevant differences. First, as shown in \S\ref{sec:naturalness}, naturalness does not constrain the size of the coupling $\xi_s$, so the triple-exchange diagram allows for a naturally large non-Gaussianity. This is to be contrasted especially with the single-exchange diagram, where the naturalness criterion imposed a strong constraint on the size of the corresponding non-Gaussianity.
Second, when the mass of the particle becomes large, the bispectrum is well-captured by a local vertex, namely $(\hat\partial_{i_1\cdots i_s}\pi)^3$ with symmetric contraction of indices. Notice that, due to the number of spatial gradients, for $s>2$ the squeezed-limit bispectrum  is suppressed by more than $(k_1/k_3)^2$ for small $k_1$. This makes the non-analytic part, scaling as $(k_1/k_3)^{3/2}$, a rather clean signal in the squeezed limit. 

\paragraph{Summary}

All diagrams in Fig.~\ref{fig:3pt}, except for the single-exchange diagram for spin one, can yield sizable non-Gaussianities within the perturbative regime. In order for this to be natural, the single-exchange diagram requires new physics or fine-tuning to stabilize the mass of the spinning 
particle, whereas both the double- and triple-exchange diagrams can naturally produce large non-Gaussianities. The non-analytic part of the bispectrum is suppressed by $e^{-\pi\mu_s}$ for $\cs=1$, but only by $e^{-\pi\mu_s/2}$ for small $\cs$. Typically, we find that $f_{\rm NL}\gtrsim {\cal O}(1)$ from the non-analytic part is possible if $\mu_s\lesssim 5$ for $\cs=1$ and $\mu_s\lesssim 10$ for $\cs \ll 1$.

\newpage
\subsection[${\langle \gamma\zeta\zeta\rangle}$]{$\boldsymbol{\langle \gamma\zeta\zeta\rangle}$}

Lastly, we consider the tensor-scalar-scalar correlation function $\langle \gamma \zeta \zeta\rangle$. In single-field inflation,  a long-wavelength tensor fluctuation is locally equivalent to a spatially anisotropic coordinate transformation. Again, we can Taylor expand the expectation value around the squeezed limit, thus obtaining
\begin{align}
\lim_{k_1\ll k_3}\langle\gamma_{\k_1}^{\lambda}\zeta_{\k_2}^{\phantom\lambda}\zeta_{\k_3}^{\phantom\lambda}\rangle' = P_\gamma(k_1) P_\zeta(k_3) \sum_{n=0}^\infty d_n \left(\frac{k_1}{k_3}\right)^n \, ,\label{tensorcc}
\end{align}
where $\gamma^{\lambda}$, with $\lambda=\pm 2$, denotes the positive or negative helicity components of the graviton. As in the case of the scalar bispectrum, the leading coefficients are determined by the single-field consistency relation~\cite{Maldacena:2002vr} (see also \cite{Creminelli:2012ed, Hinterbichler:2013dpa}). In particular, $d_0$ in (\ref{tensorcc}) is given by 
\beq
d_0=\frac{1}{16}{\cal E}_2^{\lambda}(\hat\k_1 \cdot\hat\k_3) \big[3-(n_s-1)\big]\, , \label{equ:d0}
\eeq
where 
${\cal E}_2^{\lambda}(\hat\k_1\cdot\hat\k_3)\equiv \hat k_3^i\hskip 1pt\hat k_3^j\hskip 1pt\varepsilon^{\lambda}_{ij}(\hat\k_1)$, with $\varepsilon_{ij}^\lambda\varepsilon_{ij}^{\lambda *}=4$. 
When the consistency relation holds, it also completely fixes the linear term $d_1$ in (\ref{tensorcc}), and physical effects appear at order $(k_1/k_3)^2$. The presence of new particles during inflation invalidates the Taylor expansion and leads to non-analytic scalings in (\ref{tensorcc}). 
Our goal in this section is to study these characteristic signatures of massive spinning 
particles.
\begin{figure}[h!]
\centering
\subfloat[\label{fig:3pt_tss_a}]{\includegraphics[scale=0.42]{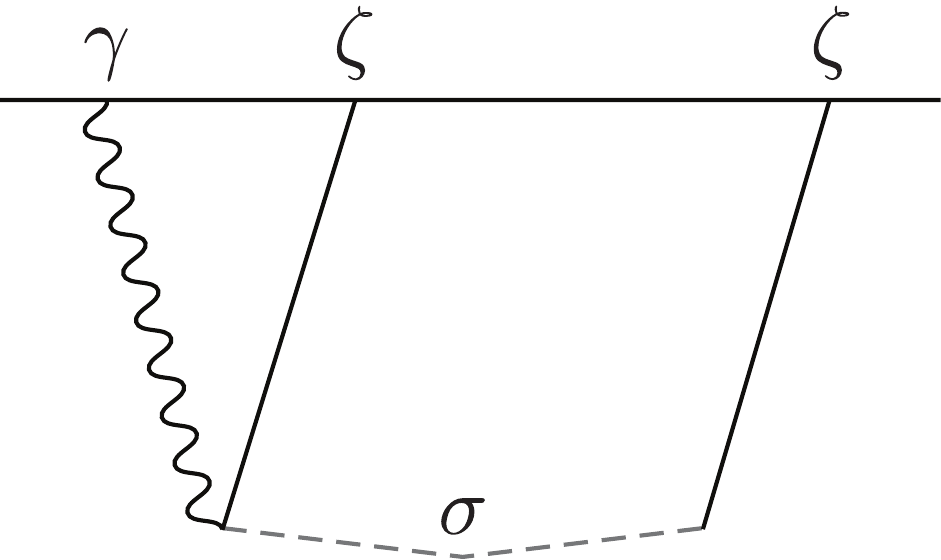}}\qquad\subfloat[\label{fig:3pt_tss_b}]{\includegraphics[scale=0.42]{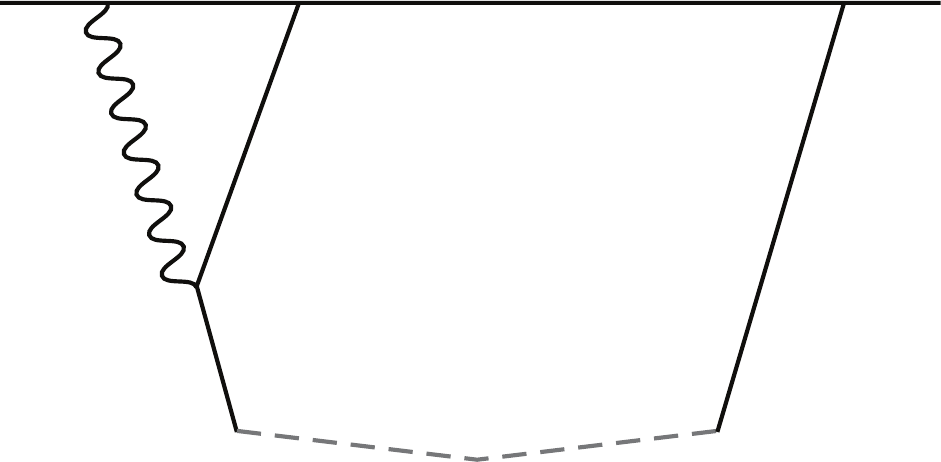}}
\qquad\subfloat[\label{fig:3pt_tss_c}]{\includegraphics[scale=0.45]{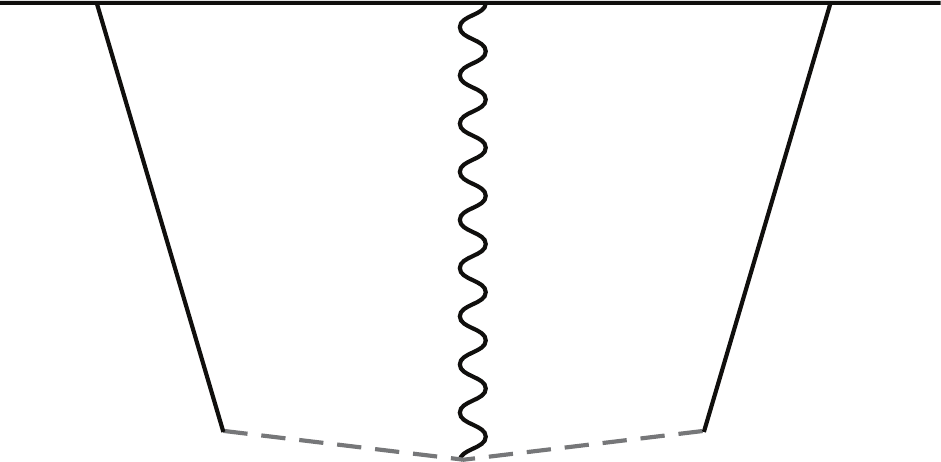}}\\ \vskip -3pt
\subfloat[\label{fig:3pt_tss_d}]{\includegraphics[scale=0.42]{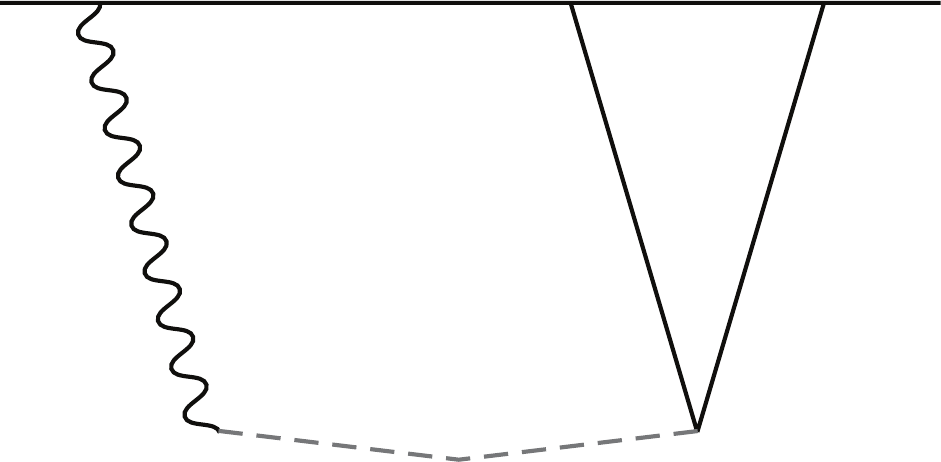}}\qquad\subfloat[\label{fig:3pt_tss_e}]{\includegraphics[scale=0.42]{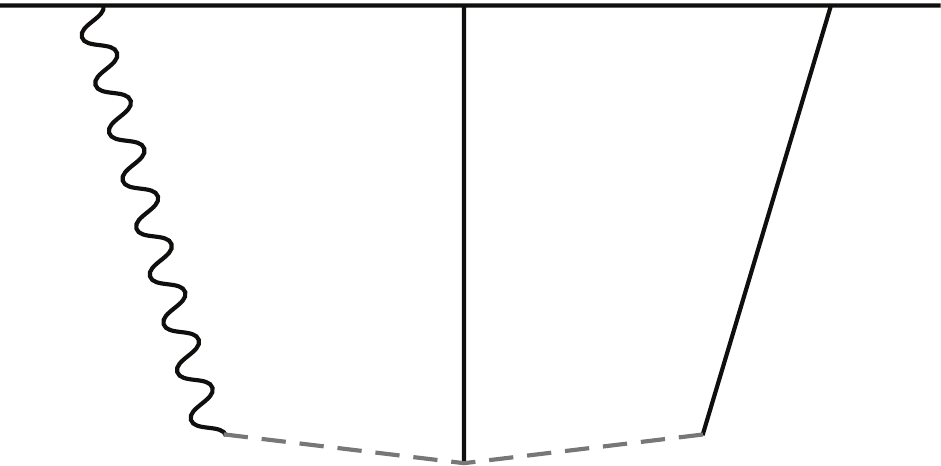}}
\qquad\subfloat[\label{fig:3pt_tss_f}]{\includegraphics[scale=0.42]{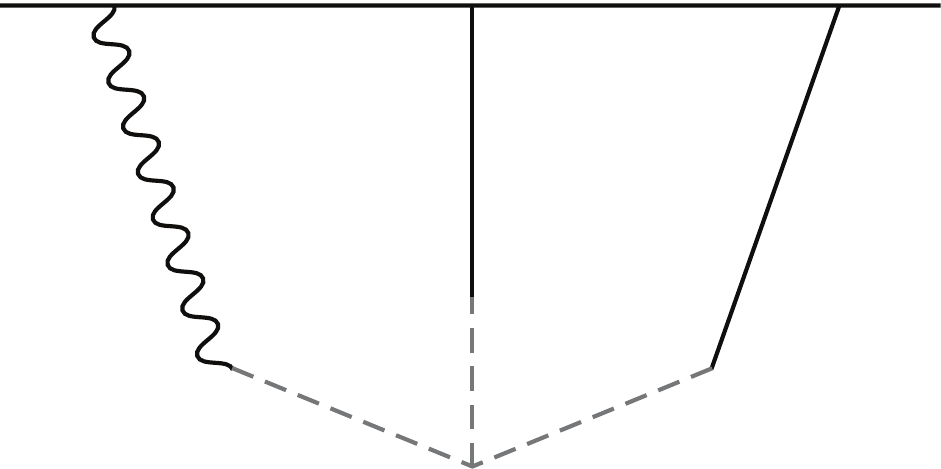}}\vskip -3pt
\caption{Tree-level diagrams contributing to $\langle\gamma\zeta\zeta\rangle$. The solid, dashed, and wavy lines represent the curvature perturbation $\zeta$, a spinning field $\sigma_{i_1\cdots i_s}$, and the graviton $\gamma_{ij}$, respectively.  
\label{fig:3pt2}} 
\end{figure}

\vskip 4pt
All tree-level diagrams contributing to $\langle \gamma \zeta \zeta\rangle$ are shown in Fig.~\ref{fig:3pt2}. Not all of these diagrams can lead to a nontrivial deviation from the consistency relation. 
For the diagrams [(a-c)] the same symmetry that generates the tensor consistency relation enforces corrections to the power spectrum, so that the relation in (\ref{tensorcc}) and (\ref{equ:d0}) is preserved~\cite{Dimastrogiovanni:2015pla}. 
Only the diagrams [(d-f)], which involve a quadratic mixing between the graviton and the intermediate particle, can lead to such a deviation. These diagrams have the same structure as those in Fig.~\ref{fig:3pt}, except that one of the legs in the quadratic mixing is replaced by an external graviton, so that the exchanging particle must carry the same helicity as the graviton. In the following, we will present results for the diagrams [(d-f)], mostly focusing on the single-exchange diagram [(d)] to avoid repetition. 
The quadratic $\gamma$-$\sigma$ mixing vanishes for spins 0 and 1, so only particles with $s\ge 2$ will contribute. 

\paragraph{Single-exchange diagram}
We first consider the single-exchange diagram (Fig.~\ref{fig:3pt_tss_d}). The relevant interaction Lagrangian is [c.f.~eqs.~(\ref{spinsLmix}) and (\ref{Lmixsgamma})]
\begin{align}
{\cal L}_I = \frac{1}{a^{2s}}\left(-\frac{f_\pi^2}{\Mp}\rho_s\hskip 1pt a^2\hskip 1pt \partial_{i_3\cdots i_s}\dot\gamma_{i_1i_2}^c\hat\sigma_{i_1\cdots i_s} + \frac{1}{\Lambda^s_s}\, \dot\pi_c\partial_{i_1\cdots i_s}\pi_c\hat\sigma_{i_1\cdots i_s} \right) .\label{Lmix2gamma2}
\end{align}
Using (\ref{equ:Deltaz}) and (\ref{equ:Deltag}), we can write the coefficient of the quadratic mixing term as $- \rho_{s}\sqrt{r/8} \hskip 1pt H$. The perturbativity condition on the $\pi$-$\sigma$ mixing, $\rho_{s}<1$, implies that the $\gamma$-$\sigma$ mixing carries an extra suppression factor of $\sqrt{r/8}$. The bispectrum corresponding to the single-exchange diagram~is 
\begin{align}
\frac{\langle\gamma_{\k_1}^{\lambda}\zeta_{\k_2}^{\phantom\lambda}\zeta_{\k_3}^{\phantom\lambda}\rangle'}{\Delta_\gamma\Delta_\zeta^3}=\alpha_2 \hskip1pt \sqrt{r}\hskip 1pt\Delta_\zeta^{-1} \times {\cal E}_2^{\lambda}(\hat\k_1\cdot\hat\k_3)\hskip 1pt \hat P_s^{\lambda}(\hat\k_1\cdot\hat\k_3) \times {\cal B}^{(s)}(\mu_s,\cs,k_1,k_2,k_3)  + (\k_2\leftrightarrow \k_3)\, ,\label{tss3ptspinssingle}
\end{align}
where $\hat{P}_s^\lambda \equiv  (1-x^2)^{-\lambda/2} P_s^\lambda$,  with $P_s^\lambda$ the associated Legendre polynomial. The function ${\cal B}^{(s)}$ is given explicitly in Appendix~\ref{app:inin}. 

\vskip 4pt
\noindent
{\it Size of NG.}---We quantify the size of the tensor-scalar-scalar bispectrum by 
\begin{align}
f_{\rm NL}^{\gamma\zeta\zeta} \equiv \frac{6}{17} \sum_{\lambda=\pm2} \frac{\langle\gamma_{\k_1}^{\lambda}\zeta_{\k_2}^{\phantom\lambda}\zeta_{\k_3}^{\phantom\lambda}\rangle'}{P^{1/2}_\gamma(k)P_\zeta^{3/2}(k)}\, ,\label{fnltss}
\end{align}
where the bispectrum is evaluated in the equilateral configuration, $k_1=k_2=k_3 \equiv k$, with vectors maximally aligned with the polarization tensor. This choice of normalization agrees with that adopted in \cite{Meerburg:2016ecv} and implies $f_{\rm NL}^{\gamma\zeta\zeta}=\sqrt r/16$ for single-field slow-roll inflation \cite{Maldacena:2002vr}. An estimate of the size of the non-Gaussianity from the single-exchange diagram is
\begin{align}
f_{\rm NL}^{\gamma\zeta\zeta} \,\sim\, g(\mu_s)\hskip 1pt \alpha_s\sqrt{r} \hskip 1pt\Delta_\zeta^{-1}\, ,
\end{align}
where $g(\mu_s)$ denotes the appropriate mass suppressions for the analytic and non-analytic parts, which in the large $\mu_s$ limit scale as\hskip 1pt\footnote{The exponential suppression of the non-analytic part of the signal applies to  particles in the principal series. Unlike the scalar case, this exponential suppression cannot be reduced to $e^{-\pi\mu_s/2}$, since the graviton propagates with $c_\gamma=1$. For particles belonging to the complementary series, the non-analytic part of the signal would not be exponentially suppressed.}
\begin{align}
	g(\mu_s)\ \equiv\ \begin{cases}
	\, \mu_s^{-2}\quad\ & \text{analytic},\\
		\, e^{-\pi\mu_s}\quad\ & \text{non-analytic}.
	\end{cases}
\end{align}
The enhancement of $f_{\rm NL}^{\gamma\zeta\zeta}$ by the large factor $\Delta_\zeta^{-1}$ means that, in principle, the signal could be significantly larger than the one predicted from single-field slow-roll inflation, $f_{\rm NL}^{\gamma\zeta\zeta}\gg \sqrt{r}/16$, even in the perturbative regime.  As in the scalar case, the condition for radiative stability gives a rather strong constraint on the naturally allowed size of the bispectrum associated with the single-exchange diagram [cf.~(\ref{naturalsingle})]. While the size of the single-exchange diagram is strongly constrained by naturalness, both the diagrams~[(e,f)] can lead to naturally large non-Gaussianity, as in the  case of the scalar bispectrum.
Future constraints on $f_{\rm NL}^{\gamma\zeta\zeta}$ from observations of the $\langle BTT\rangle$ correlator of CMB anisotropies were discussed in~\cite{Meerburg:2016ecv}. The proposed CMB Stage~IV experiments~\cite{Abazajian:2013oma} will have the sensitivity to reach $\sigma(\sqrt{r} f_{\rm NL}^{\gamma\zeta\zeta}) \sim 0.1$, which suggests that the tensor non-Gaussianity due to massive spinning particles would be detectable for $ r\gtrsim 10^{-4} \hskip 1pt [g(\mu_s)\alpha_s]^{-1}$.\footnote{Producing a large tensor contribution while keeping the scalar contribution small may require some fine-tuned cancellation between interactions in the scalar sector. This is because the interaction vertices in (\ref{Lmix2gamma2}) and (\ref{equ:LintX})  arise from the same operators in unitary gauge. Suppressing the effects of the interactions in (\ref{equ:LintX}) would require balancing them against additional interactions such as $\dot\pi\sigma_{0\cdots 0}$.} 

\vskip 4pt
\noindent
{\it Shape of NG.}---In the squeezed limit, $\langle \gamma \zeta \zeta \rangle$ behaves in the following ways: 
\begin{itemize}
\item 
The analytic part scales as 
\begin{align}
\lim_{k_1 \ll k_3} \langle\gamma_{\k_1}^{\lambda}\zeta_{\k_2}^{\phantom\lambda}\zeta_{\k_3}^{\phantom\lambda}\rangle' \propto  \frac{1}{k_1^3 k_3^3}\left(\frac{k_1}{k_3}\right)^{s} {\cal E}_2^{\lambda}(\hat\k_1\cdot\hat\k_3)\hskip 1pt \hat P_s^{\lambda}(\hat\k_1\cdot\hat\k_3)\, .\label{tssanalyticSq}
\end{align}
Notice that the suppression of the analytic part in the squeezed limit increases with spin. This can be understood by looking at the form of the local vertex after integrating out the massive particle, which becomes $\dot\pi\partial_{i_1\cdots i_s}\pi\partial_{i_3\cdots i_s}\dot\gamma_{i_1i_2}$. As we will see below, this means that the analytic part of the signal will be subdominant compared to its non-analytic counterpart in the soft graviton limit. 

\item For $\mu_s \ge 0$, the squeezed limit of the non-analytic part of the bispectrum scales as
\begin{align}
\lim_{k_1 \ll k_3} \langle\gamma^{\lambda}_{\k_1}\zeta_{\k_2}^{\phantom\lambda}\zeta_{\k_3}^{\phantom\lambda}\rangle' \propto \frac{1}{k_1^3k_3^3}\hskip -1.5pt\left(\frac{k_1}{k_3}\right)^{3/2}\hskip -1pt{\cal E}_2^{\lambda}(\hat\k_1\cdot\hat\k_3)\hskip 1pt \hat P_s^{\lambda}(\hat\k_1\cdot\hat\k_3) \cos\left[\mu_{s}\ln\hskip -1pt\left(\frac{k_1}{ k_3}\right)+\tilde\phi_{s}\right] ,\label{tsssqueezed}
\end{align}
where the phase $\tilde\phi_{s}$ is a function of $\mu_{s}$ and $\cs$ (see Appendix~\ref{app:squeezed}). Coupling to a particle with spin greater than two induces an extra angular structure. For imaginary $\mu_s$, we instead have
\begin{align}
\lim_{k_1 \ll k_3} \langle\gamma^{\lambda}_{\k_1}\zeta_{\k_2}^{\phantom\lambda}\zeta_{\k_3}^{\phantom\lambda}\rangle'&\,\propto\, \frac{1}{k_1^3k_3^3}\left(\frac{k_1}{k_3}\right)^{3/2-\nu_s} {\cal E}_2^{\lambda}(\hat\k_1\cdot\hat\k_3)\hskip 1pt \hat P_s^{\lambda}(\hat\k_1\cdot\hat\k_3)\, ,\label{tsssqueezed3}
\end{align}
with $\nu_{s}\equiv -i\mu_{s} \in [0,1/2)$. This gives a non-analytic $(k_1/k_3)^{3/2-\nu_{s}}$ correction to the leading term of the consistency relation (\ref{tensorcc}). Since unitarity implies $\nu_{s} < 1/2$, the squeezed-limit bispectrum due to massive spinning particles will be suppressed by at least $k_1/k_3$ compared to the 
leading term in the tensor consistency relation.\footnote{A deviation from the leading term of the consistency relation due to spinning particles can arise in a number of ways: First, the unitarity bound can be evaded  
if the de Sitter isometries are not fully respected in the quadratic action of the spinning field~\cite{Blas:2009my, Bordin:2016ruc}.  Another possibility involves partially massless fields with spin greater than two, since the late-time behavior of these fields does not obey the same restrictions as for the massive case. It would be interesting to explore these possibilities further.}
\end{itemize}

\paragraph{Other diagrams}
The extensions to the diagrams [(e,f)] are completely analogous to the scalar case. Similar to the scalar three-point function, these diagrams have the advantage that they are less constrained by naturalness considerations.

\section{Conclusions}
\label{sec:conclusions}

In this paper, we have studied the imprints of massive particles with spin on cosmological correlators using the framework of the effective field theory of inflation \cite{Cheung:2007st}. This generalizes the work of Arkani-Hamed and Maldacena (AHM)~\cite{Arkani-Hamed:2015bza} to cases where conformal symmetry is strongly broken. Let us summarize our results and contrast them with the conclusions of AHM:

\begin{itemize}
\item In AHM's more conservative analysis, the overall size of non-Gaussianity was too small to be observable even in the most optimistic experimental scenarios. Our results are cautiously more optimistic. Within the regime of validity of the effective field theory, we can accommodate observable non-Gaussianity as long as the masses of the new particles aren't too far above the Hubble scale during inflation.

\item The key spectroscopic features of massive particles with spin do not rely on conformal invariance and therefore continue to hold in our analysis. As explained in~\cite{Arkani-Hamed:2015bza}, the masses and spins of extra particles during inflation can be extracted by measuring the momentum dependence in the squeezed limit.

\item Our systematic effective field theory treatment of massive spinning  
particles during inflation allows for a complete characterization of their effects on non-Gaussian cosmological correlators, including their imprints beyond the squeezed limit.  We showed that the characteristic angular dependence resulting from the presence of particles with spin persists even for more general momentum configurations.  
Having access to the complete correlation functions will be valuable for future data analysis.

\item We also studied the effects of an explicit breaking of special conformal symmetry by introducing a sound speed $\cs$ for the Goldstone fluctuations.
We found that, for $\cs <\mu_s^{-1}$, the exponential suppression in the production of the massive particles, $e^{-\pi\mu_s}$, is changed to $e^{-\pi\mu_s/2}$. For a given mass, the size of non-Gaussianity is therefore enhanced (or less suppressed) for small~$\cs$.

\item Finally, we showed that particles with spin greater than or equal to two lead to a signature in the squeezed limit of $\langle\gamma\zeta\zeta\rangle$. This signal may be observable in the $\langle BTT\rangle$ correlator of CMB anisotropies~\cite{Meerburg:2016ecv}.
\end{itemize}

\begin{figure}[t!]
\centering
\quad \includegraphics[scale=0.9]{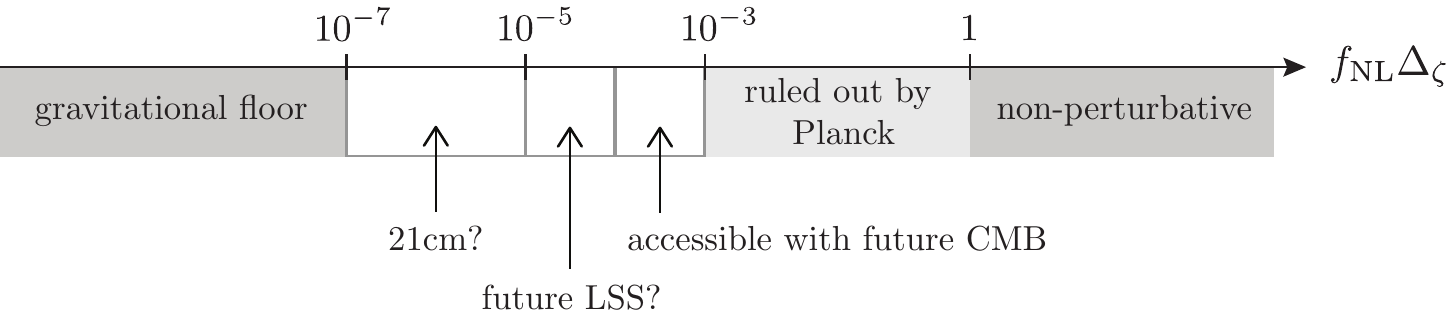}
\caption{\label{fig:fNLconstraints} Schematic illustration of current and future constraints on (scale-invariant) primordial non-Gaussianity. The ``gravitational floor" denotes the minimal level of non-Gaussianity created by purely gravitational interactions during inflation~\cite{Maldacena:2002vr}. }
\end{figure}

Figure~\ref{fig:fNLconstraints} is a schematic illustration of current and future constraints on (scale-invariant) primordial non-Gaussianities. We see that the perturbatively interesting regime spans about seven orders of magnitude in $f_{\rm NL}$. Of this regime, three orders of magnitude have been ruled out by current CMB observations, leaving a window of opportunity of about four orders of magnitude. Accessing these low levels of non-Gaussianity will be challenging.  Even optimistic projections for future CMB observations won't reduce the constraints by more than an order of magnitude. Digging deeper will require new cosmological probes, such as observations of the large-scale structure (LSS) of the universe~\cite{Alvarez:2014vva} and the tomography of the 21cm transition of neutral hydrogen gas~\cite{Loeb:2003ya}. Our results, together with~\cite{Chen:2009zp, Baumann:2011nk, Noumi:2012vr,Arkani-Hamed:2015bza,Flauger:2016idt}, will help to find optimal observational strategies for extracting the subtle imprints of extra particles during the inflationary era.

\paragraph{Acknowledgements} We thank Paolo Creminelli, Garrett Goon, Dan Green, Juan Maldacena, Daan Meerburg, Mehrdad Mirbabayi, Enrico Pajer, Rafael Porto, Eva Silverstein, and Marko Simonovi\'c for helpful discussions, and Maldacena and Meerburg for comments on a draft. H.L.~thanks the Institute of Physics at the University of Amsterdam for its hospitality. H.L~acknowledges support from the EPSRC and the Cambridge Overseas Trust.
D.B.~and G.P.~acknowledge support from a Starting Grant of the European Research Council (ERC STG Grant 279617).

\newpage
\appendix
\section{More on Spin in de Sitter Space}
\label{app:spindS}
In this appendix, we will derive various mathematical results that have been used in this work. In \S\ref{app:mode}, we obtain the mode functions for massive spinning fields in de Sitter space by solving their equations of motion. We then derive the formula for the two-point function in \S\ref{app:2pt}. 

\paragraph{Preliminaries}

We will work with the components of the spinning field $\sigma_{\mu_1\cdots\mu_s}$ projected onto spatial slices, i.e.~$\sigma_{i_1\cdots i_n \eta\cdots \eta}$.
We will find it convenient to write these as
 \beq
 \sigma_{i_1\cdots i_n \eta\cdots \eta} = \sum_\lambda\sigma_{n,s}^\lambda\hskip 1pt \varepsilon^\lambda_{i_1\cdots i_n}\, , \label{equ:A1}
 \eeq 
 where $\varepsilon^\lambda_{i_1\cdots i_n}$ is a suitably normalized polarization tensor (see insert below). The sub/superscripts on the mode functions $\sigma^\lambda_{n,s}$ label three ``quantum numbers'': $s$ is the spin (or the rank) of the spacetime tensor field, $n$ is its ``spatial" spin, and $\lambda$ is the helicity component of the spatial spin.

\begin{framed}
{\small
\noindent{\it Polarization tensors.}---In this insert, we will derive explicit expressions for the polarization tensors of arbitrary spin and helicity. The longitudinal polarization tensors are functions of $\hat \k$, while the transverse polarization tensors in addition depend on two polarization directions $\hat{\boldsymbol\varepsilon}^{\pm}$, with $\hat\k\cdot\hat{\boldsymbol\varepsilon}^\pm=0$. Since $\hat{\boldsymbol\varepsilon}^+$ and $\hat{\boldsymbol\varepsilon}^-$ are related to each other by the reality condition $\hat{\boldsymbol\varepsilon}^+ = (\hat{\boldsymbol\varepsilon}^{-})^*$, let us denote one of them by $\hat{\boldsymbol\varepsilon}$. 
The polarization tensors of helicity $\lambda$ satisfy the following conditions:
\begin{itemize}
\item[{\it i}\hskip 1pt)] symmetric: 
$\varepsilon_{i_1\cdots i_s}^\lambda=\varepsilon^\lambda_{(i_1\cdots i_s)}$.
\item[{\it ii}\hskip 1pt)] traceless: $\varepsilon^\lambda_{iii_3\cdots i_s}=0$.
\item[{\it iii}\hskip 1pt)] transverse: $\hat k_{i_1}\cdots\hat k_{i_n}\varepsilon^\lambda_{i_1\cdots i_s} = 0$, when $n > s- \lambda$.
\end{itemize}
The last condition implies that the polarization tensor is of the form 
\begin{align}
\varepsilon_{i_1\cdots i_s}^\lambda (\hat\k,\hat{\boldsymbol\varepsilon}) = \varepsilon^\lambda_{(i_1\cdots i_\lambda} (\hat{\boldsymbol\varepsilon})\, f_{i_{\lambda+1}\cdots i_s)}(\hat\k)\, ,
\end{align}
where $\hat k_{i_1}\varepsilon_{i_1\cdots i_\lambda}^\lambda(\hat{\boldsymbol\varepsilon}) = 0$ and $f_{i_1\cdots i_{s-\lambda}}$ is some tensor. Let us contract with vectors $\q$ and define
\begin{align}
F_s^\lambda(x,y,z) \equiv  q_{i_1}\cdots q_{i_s}\varepsilon_{i_1\cdots i_s}^\lambda(\hat\k,\hat{\boldsymbol\varepsilon})\, ,
\end{align}
where we have defined $x\equiv q^2$, $y\equiv \q\cdot\hat\k$, and $z\equiv q_{i_1}\cdots q_{i_\lambda} \varepsilon_{i_1\cdots i_\lambda}^\lambda$. The function $F_s^\lambda$ is a  homogeneous polynomial in $\q$, so that
\begin{align}
2x \hskip 1pt F^\lambda_{s,x} +  y \hskip 1pt  F^\lambda_{s,y} + \lambda z \hskip 1pt  F^\lambda_{s,z} = s \hskip 1pt F^\lambda_{s}\, .\label{app:homogeneous}
\end{align}
The transverse and traceless conditions translate into
\begin{align}
&z\hskip 1pt F^\lambda_{s,z} = F^\lambda_s\, ,\label{app:transverse} \\
&4x \hskip 1pt F^\lambda_{s,xx}+4y \hskip 1pt F^\lambda_{s,xy}+4\lambda z \hskip 1pt F^\lambda_{s,xz}+2d \hskip 1pt F^\lambda_{s,x}+F^\lambda_{s,yy} = 0\, , \label{app:traceless}
\end{align}
where $d$ is the number of spatial dimensions. 
Taking derivatives of (\ref{app:homogeneous}) and (\ref{app:transverse}), and substituting into (\ref{app:traceless}), we get
\begin{align}
(x-y^2)F^\lambda_{s,yy} -(2\lambda+d-1)y \hskip 1pt F^\lambda_{s,y} + (s-\lambda)(s+\lambda+d-2)F^\lambda_s = 0\, .\label{Feq}
\end{align}
Without loss of generality, we now set $x=q^2 \equiv 1$.  The solution to (\ref{app:transverse}) and (\ref{Feq}) is
\begin{align}
F^\lambda_s(y,z) \propto z \hat{P}^{\beta_\lambda}_{\beta_s} (y)\, ,
\end{align}
where $\hat{P}_{\beta_s}^{\beta_\lambda}$ is part of the associated Legendre polynomial $P_{\beta_s}^{\beta_\lambda}$ of degree $\beta_s \equiv \frac{1}{2}(2s+d-3)$ and order $\beta_\lambda\equiv \frac{1}{2}(2\lambda+d-3)$, defined by $ P^{\beta_\lambda}_{\beta_s} (y)= (1-y^2)^{\beta_\lambda/2}\hat{P}_{\beta_s}^{\beta_\lambda}$. We will set $P_s^{\beta_\lambda}\equiv P_s^{|\beta_\lambda|}$ and distinguish the opposite helicities only by the phase. For $d=3$, this reduces to
\begin{align}
F^\lambda_s(y,z) \propto z \hat{P}_s^\lambda (y)\, .\label{app:legendre}
\end{align}
This result includes longitudinal polarization tensors for $\lambda=0$ and $z=1$. It is straightforward to obtain explicit expressions for the polarization tensors by stripping off the contractions with $\q$ in (\ref{app:legendre}) and symmetrizing the indices:
\begin{align}
\varepsilon_{i_1\cdots i_s}^\lambda(\hat\k,\hat{\boldsymbol\varepsilon}) = \frac{1}{(2\lambda-1)!!}\, \sum_{n=0}^{s-\lambda} B_n\, \varepsilon^\lambda_{(i_1\cdots i_\lambda}(\hat{\boldsymbol\varepsilon})\hskip 1pt\hat k_{i_{\lambda+1}}\cdots \hat k_{i_{\lambda+n}}\delta_{i_{\lambda+n+1}\cdots i_s)}\, ,\label{PTformula}
\end{align}
where
\begin{align}
B_n &\equiv \frac{2^s}{n!(s-n-\lambda)!}\frac{\Gamma[\frac{1}{2}(n+\lambda+1+s)]}{\Gamma[\frac{1}{2}(n+\lambda+1-s)]}
\, , \quad\delta_{i_1\cdots i_n} \equiv \begin{cases} \delta_{i_1 i_2}\cdots\delta_{i_{n-1}i_n} & n\ \text{even} \\ 0 & n\ \text{odd}\end{cases}\, .
\end{align}
The self-contraction of the polarization tensors can be written as
\begin{align}
\varepsilon^{\lambda}_{i_1\cdots i_s}\varepsilon^{\lambda *}_{i_1\cdots i_s} &= \frac{(2s-1)!!(s+\lambda)!}{2^\lambda [(2\lambda-1)!!]^2s!(s-\lambda)!}\hskip 1pt \varepsilon^{\lambda}_{i_1\cdots i_\lambda}\varepsilon^{\lambda *}_{i_1\cdots i_\lambda}\, .\label{selfcontraction}
\end{align}
When choosing the orthogonal direction to be in, say, the $z$-direction, there will be in total of $2^s$ non-zero components for the polarized tensor $\varepsilon^s_{i_1\cdots i_s}$, which are $\pm 1$ or $\pm i$ up to a phase. This means that $\varepsilon^s_{i_1\cdots i_s}\varepsilon^{s *}_{i_1\cdots i_s} = 2^s$ with some overall normalization which we set to unity for convenience.}
\end{framed}

\subsection{Mode Functions}\label{app:mode}
In this section, we will derive the de Sitter mode functions for fields with spin. We will explicitly derive the mode functions for fields with spins 1 and 2, and present the results for arbitrary spin at the end.

\paragraph{Spin-1}
The equation of motion of a massive spin-1 field $\sigma_\mu$ is 
\begin{align} 
(\Box- m_1^2) \sigma_\mu = 0\, , 
\end{align}
with $\nabla^\mu \sigma_\mu=0$ and $m_1^2=m^2+3H^2$. The components $\sigma_\eta$ and $\sigma_i$ then satisfy 
\begin{align}
\sigma_\eta''-\left(\partial_j^2-\frac{m^2/H^2-2}{\eta^2}\right)\sigma_\eta &= \frac{2}{\eta}\partial_i\sigma_i\, ,\label{spin1eq1}\\
\sigma_i''-\left(\partial_j^2-\frac{m^2/H^2}{\eta^2}\right)\sigma_i&=\frac{2}{\eta} \partial_i\sigma_\eta\, ,\label{spin1eq2}
\end{align}
where a prime denotes a derivative with respect to conformal time, and
\begin{align}
\sigma_\eta'-\frac{2}{\eta}\sigma_\eta &=\partial_i\sigma_i\, .\label{spin1trans}
\end{align}
To decouple equations (\ref{spin1eq1}) and (\ref{spin1eq2}), we expand the field $\sigma_\mu$ into its different helicity components, 
\beq
\sigma_\mu =  \sum_{\lambda=-1}^1\sigma^{(\lambda)}_\mu\, , 
\eeq
where
\begin{align}
\sigma_\eta^{(0)}&= \sigma_{0,1}^{0} \, ,  \qquad \quad \, \,\, \sigma^{(\pm 1)}_\eta =0\, , \label{spin1helicity0} \\ 
\sigma^{(0)}_i&= \sigma^{0}_{1,1} \hskip 1pt\varepsilon^0_i \, , \quad  \quad\ \hskip 2pt \sigma^{(\pm 1)}_i = \sigma^{\pm 1}_{1,1} \varepsilon^{\pm 1}_i \, .\label{spin1helicity}
\end{align}
We demand that the polarization vectors $\varepsilon^{\lambda}_i(\hat\k)$ satisfy 
\begin{align}
\hat k_i\varepsilon_i^0=1 \, , \quad \hat k_i\varepsilon_i^{\pm 1} = 0\, , \quad \varepsilon_i^{\pm 1} = \varepsilon_i^{\mp 1 *}\, , \quad \varepsilon_i^{\pm 1} \varepsilon_i^{\pm 1*}=2\, .\label{spin1pol}
\end{align}
The choice of the normalization (\ref{spin1pol}) uniquely fixes the longitudinal polarization vector to be $\varepsilon_i^0(\hat \k)=\hat k_i$, and the transverse polarization vectors are fixed up to a phase. For momentum along the $z$-direction, they can be chosen to be $\varepsilon_i^{\pm 1}(\hat\z) = (1,\pm i,0)$.

\vskip 4pt
In terms of the mode functions defined in (\ref{spin1helicity0}) and (\ref{spin1helicity}), eqs.~(\ref{spin1eq1}) and (\ref{spin1eq2}) decouple 
\begin{align}
{\sigma_{0,1}^{0}}''-\frac{2}{\eta}{\sigma_{0,1}^{0}}'+\left(k^2+\frac{m^2/H^2+2}{\eta^2}\right)\sigma_{0,1}^{0} &=0\, ,\label{spin1u0}\\
{\sigma^{0}_{1,1}}''-\frac{k^2 \eta^2}{k^2\eta^2+m^2/H^2}\frac{2}{\eta}{\sigma^{0}_{1,1}}'+\left(k^2+\frac{m^2/H^2}{\eta^2}\right)\sigma^{0}_{1,1} &=0\, ,\label{spin1v0}\\
{\sigma^{\pm 1}_{1,1}}''+\left(k^2+\frac{m^2/H^2}{\eta^2}\right)\sigma^{\pm 1}_{1,1} &=0\, ,\label{spin1v1}
\end{align}
and the transverse condition (\ref{spin1trans}) becomes
\begin{align}
\sigma_{1,1}^{0}=-\frac{i}{k}\left({\sigma_{0,1}^{0}}'-\frac{2}{\eta}\sigma_{0,1}^{0}\right) .\label{spin1trans2}
\end{align}
The solutions to these equations with the Bunch-Davies initial condition are 
\begin{align}
\sigma_{0,1}^{0} &= {\cal A}_1 \hskip 1pt N_1 (-k\eta)^{3/2}H_{i\mu_1}\, ,\label{spin1phi0}\\
\sigma^{0}_{1,1} &= \frac{i}{2}  {\cal A}_1\hskip 1pt  N_1 (-k\eta)^{1/2}\Big[k\eta\big(H_{i\mu_1+1}-H_{i\mu_1-1}\big)-H_{i\mu_1}\Big]\, ,\label{spin1v1sol} \\
\sigma^{\pm 1}_{1,1} &= {\cal A} _1\hskip 1pt  Z_1^{\pm 1}(-k\eta)^{1/2}H_{i\mu_{1}}\, ,\label{spin1v1sol}
\end{align}
where ${\cal A}_1 \equiv  e^{i\pi/4}e^{-\pi\mu_1/2}$ and $Z_1^{\pm1}$ denotes the normalization constant for the helicity-$\pm 1$ mode of the spin-$1$ field. We have also suppressed the argument $-k\eta$ of the Hankel functions $H_{i\mu_1}\equiv H_{i\mu_1}^{(1)}$ for brevity. 

\vskip 4pt
A few comments are in order. First, note that for $m=0$ equation (\ref{spin1v1}) for the transverse mode becomes the flat space wave equation, whose solutions are simply plane waves. This is because the action of a massless spin-1 field is conformally invariant, so the mode in de Sitter space behaves as if it were in flat space. On the other hand, we do not see this behavior for the longitudinal mode. In particular, the longitudinal mode blows up relative to the transverse mode as we go to the infinite past $\eta\to -\infty$. We can understand this as follows. The mass term $m^2/H^2\eta^2$ in the action (\ref{spin1action}) is time dependent, so the spin-1 field is effectively massless in the infinite past, in which case the longitudinal mode turns into a pure gauge mode.

\vskip 4pt
We still need to determine
 the normalization constants $N_1$ and $Z_1^{\pm 1}$. This is done by imposing orthonormality of mode functions under the inner product
\begin{align}
\left\langle \sigma^{(\lambda)}_\mu(\k,\eta)e^{i\k\cdot\x}, \sigma^{(\lambda')}_\nu(\k',\eta)e^{i\k'\cdot\x}\right\rangle = \delta_{\lambda\lambda'}\delta(\k-\k')\, .\label{spin1ortho}
\end{align}
This orthonormality condition guarantees that we get the standard equal-time commutation relation upon canonical quantization. We have
\begin{align}
\left\langle \sigma^{(0)}_\mu(\k,\eta)e^{i\k\cdot\x}, \sigma^{(0)}_\nu(\k',\eta)e^{i\k'\cdot\x}\right\rangle &= -i\eta^{\mu\nu}\int\d^3x\left[\sigma_\mu^{(0)} {\sigma_\nu^{(0)*}}' - {\sigma_\mu^{(0)}}' \sigma_\nu^{(0)*}\right] e^{i(\k-\k')\cdot \x}\nonumber\\
&=-i\left[-{\cal W}(\sigma^{0}_{0,1},\sigma^{0*}_{0,1})+{\cal W}(\sigma^{0}_{1,1},\sigma^{0*}_{1,1})\right] \delta(\k-\k')\, , \label{spin1inner}
\end{align}
where ${\cal W}$ denotes the Wronskian. Substituting (\ref{spin1phi0}) and  (\ref{spin1v1sol}), we obtain
\begin{align}
{\cal W}(\sigma^{0}_{0,1},\sigma^{0*}_{0,1}) &= \frac{4ik^3\eta^2}{\pi} \times N_1^2\, , \label{XX}\\
{\cal W}(\sigma^{0}_{1,1},\sigma^{0*}_{1,1})&=\frac{4ik(k^2\eta^2+1/4+\mu_1^2)}{\pi} \times N_1^2\, . \label{YY}
\end{align}
Note that the time dependences in (\ref{XX}) and (\ref{YY}) cancel in (\ref{spin1inner}).
Imposing (\ref{spin1ortho}), we then get
\begin{align}
N_1 = \sqrt{\frac{\pi}{2}}\frac{1}{\sqrt{2k}}\frac{1}{(1/4+\mu_1^2)^{1/2}} = \sqrt{\frac{\pi}{2}}\frac{1}{\sqrt{2k}}\frac{H}{m}\, .\label{spin1norm}
\end{align}
The normalization for the transverse mode can be determined in a similar way. We get
\begin{align}
Z_1^{\pm 1}=\sqrt{\frac{\pi}{2}}\frac{1}{\sqrt{2k}}\, .
\end{align}
Notice that the normalization for the longitudinal mode blows up when $m=0$, which, again, does not signal any pathologies, since the longitudinal mode becomes a pure gauge mode in this limit. 

\paragraph{Spin-2}
The equations of motion and the constraints satisfied by a massive spin-2 field $\sigma_{\mu\nu}$ are
\begin{align}
(\Box-m^2-2H^2)\sigma_{\mu\nu}=0\ ,\quad \nabla^\mu\sigma_{\mu\nu} = 0 \ , \quad \tilde\sigma \equiv \sigma^\mu{}_\mu=0\ .
\end{align}
In terms of components, these are 
\begin{align}
\sigma_{\eta\eta}''+\frac{2}{\eta}\sigma_{\eta\eta}'-\left(\partial_k^2-\frac{m^2/H^2-6}{\eta^2}\right)\sigma_{\eta\eta} &= \frac{4}{\eta}\partial_i\sigma_{i\eta}+\frac{2}{\eta^2}\sigma_{ii}\, ,\label{spin2:00}\\
\sigma_{i\eta}''+\frac{2}{\eta}\sigma_{i\eta}'-\left(\partial_k^2-\frac{m^2/H^2-6}{\eta^2}\right)\sigma_{i\eta} &=\frac{2}{\eta}\partial_i\sigma_{\eta\eta}+\frac{2}{\eta}\partial_j\sigma_{ij}\, ,\\
\sigma_{ij}''+\frac{2}{\eta}\sigma_{ij}'-\left(\partial_k^2-\frac{m^2/H^2-2}{\eta^2}\right)\sigma_{ij}&=\frac{4}{\eta}\partial_{(i}\sigma_{j)\eta}+\frac{2}{\eta^2}\sigma_{\eta\eta}\delta_{ij}\, ,
\end{align}
and 
\begin{align}
\sigma_{\eta\eta}'-\partial_i\sigma_{i\eta}-\frac{1}{\eta}\sigma_{\eta\eta}-\frac{1}{\eta}\sigma_{ii}&=0\, ,\label{spin2trans}\\
\sigma_{i\eta}'-\partial_j\sigma_{ij}-\frac{2}{\eta}\sigma_{i\eta}&=0\, ,\label{spin2trans2}\\
\sigma_{\eta\eta}-\sigma_{ii}&=0\, .
\end{align}
As before, we expand the Fourier modes into helicity eigenstates 
\begin{align}
\sigma_{\mu\nu} = \sum_{\lambda=-2}^2 \sigma^{(\lambda)}_{\mu\nu}\, .
\end{align}
Let us denote the traceless part of the spatial tensor by $\hat\sigma_{ij}$, so that $\sigma_{ij}=\hat\sigma_{ij}+\frac{1}{3}\sigma_{\eta\eta}\delta_{ij}$, and  decompose the mode functions into different helicities: 
\begin{align}
\sigma_{\eta\eta}^{(0)} &= \sigma^{0}_{0,2}\, ,  \quad \ \, \,\,\,\, \, \hskip 1pt \sigma_{\eta\eta}^{(\pm 1)}=0\, ,  \qquad \quad \, \,  \,\,\sigma^{(\pm 2)}_{\eta\eta}=0\, , \\
\sigma^{(0)}_{i\eta} &=  \sigma^{0}_{1,2}\hskip 1pt  \varepsilon^0_i\, , \quad\hskip 1pt  \   \sigma^{(\pm 1)}_{i\eta}= \sigma^{\pm 1}_{1,2}\hskip 1pt \varepsilon^{\pm 1 }_i  \, ,\quad \sigma^{(\pm 2)}_{i\eta}= 0\, , \\
\hat \sigma^{(0)}_{ij} &= \sigma^{0}_{2,2} \hskip 1pt\varepsilon_{ij}^0 \ ,\quad \hat \sigma^{(\pm 1)}_{ij}= \sigma^{\pm 1}_{2,2}\hskip 1pt \varepsilon_{ij}^{\pm 1}\, , \quad  \hat\sigma^{(\pm 2)}_{ij}=  \sigma^{\pm 2}_{2,2}\hskip 1pt \varepsilon_{ij}^{\pm 2}\, .
\end{align}
Demanding that the polarization tensors satisfy 
\begin{align}
\hat k_i\hskip 1pt \varepsilon_{ij}^{0}= \varepsilon_j^{0}\, , \quad \hat k_i\hskip 1pt \varepsilon_{ij}^{\pm 1}= \frac{3}{2}\varepsilon_j^{\pm 1} \, ,\quad k_i\hskip 1pt\varepsilon_{ij}^{\pm 2}= 0\, ,\quad \varepsilon_{ij}^{\pm 2} = \varepsilon_{ij}^{\mp 2* }\, , \quad \varepsilon_{ij}^{\pm 2}\varepsilon_{ij}^{\pm 2*}=4\, ,
\end{align}
 leads to
\begin{align}
\varepsilon_{ij}^0= \frac{3}{2}\left(\hat k_i\hat k_j-\frac{1}{3}\delta_{ij}\right) ,\quad \varepsilon_{ij}^{\pm 1} = \frac{3}{2}\left(\hat k_i\varepsilon_j^{\pm 1} + \hat k_j\varepsilon_i^{\pm 1}\right)\, ,
\end{align}
and fixes $\varepsilon_{ij}^{\pm 2}$ up to a phase. For $\hat \k$ along the $z$-direction, this can be chosen to be 
\begin{align}
\varepsilon^{\pm 2}_{ij}(\hat\z) = \begin{pmatrix} 1 & \pm i & 0\\ \pm i & -1 & 0 \\ 0 & 0 & 0\end{pmatrix} .
\end{align}
The equations satisfied by the different helicity modes are 
\begin{align}
{\sigma_{0,2}^{0}}''-\frac{2}{\eta}{\sigma_{0,2}^{0}}'+\left(k^2+\frac{m^2/H^2}{\eta^2}\right)\sigma_{0,2}^{0} =0\, ,\\
{\sigma_{1,2}^{\pm 1}}''+\left(k^2+\frac{m^2/H^2-2}{\eta^2}\right)\sigma^{\pm 1}_{1,2} =0\, ,\\
{\sigma^{\pm 2}_{2,2}}''+\frac{2}{\eta}{\sigma^{\pm 2}_{2,2}}'+\left(k^2+\frac{m^2/H^2-2}{\eta^2}\right)\sigma^{\pm 2}_{2,2} =0\, ,
\end{align}
subject to the transverse conditions
\begin{align}
\sigma^0_{1,2} =-\frac{i}{k}\left({\sigma^0_{0,2}}'-\frac{2}{\eta}\sigma^0_{0,2}\right) ,\quad  \sigma^0_{2,2} &= -\frac{i}{k}\left({\sigma^0_{1,2}}'-\frac{2}{\eta}\sigma^0_{1,2} \right)-\frac{1}{3}\sigma^0_{0,2} \, , \\
\sigma^{\pm 1}_{2,2} &=-\frac{i}{k}\left({\sigma^{\pm 1}_{1,2}}'-\frac{2}{\eta}\sigma^{\pm 1}_{1,2}\right) .
\end{align}
The solutions with Bunch-Davies initial conditions are
\begin{align}
\sigma^0_{0,2} &= {\cal A}_2\hskip 1pt  N_2 (-k\eta)^{3/2}H_{i\mu_2}\, ,\label{spin2phi0}\\
\sigma^0_{1,2} &= \frac{i}{2} {\cal A}_2\hskip 1pt N_2(-k\eta)^{1/2}\Big[k\eta\big(H_{i\mu_2+1}-H_{i\mu_2-1}\big)-H_{i\mu_2}\Big]\, ,\label{spin2v1sol}\\
\sigma^0_{2,2} &=\frac{1}{12} {\cal A}_2\hskip 1pt  N_2 (-k\eta)^{-1/2}\Big[6k\eta\big((2+i\mu_2)H_{i\mu_2+1}-(2-i\mu)H_{i\mu_2-1}\big)-(9-8k^2\eta^2)H_{i\mu_2}\Big]\, ,
\end{align}
for the longitudinal modes, and
\begin{align}
\sigma^{\pm 1}_{1,2} &= {\cal A}_2\hskip 1pt  Z_2^{\pm 1}(-k\eta)^{1/2}H_{i\mu_2}\, ,\\
\sigma^{\pm 1}_{2,2} &= \frac{i}{2} {\cal A}_2\hskip 1pt Z_2^{\pm 1}(-k\eta)^{-1/2}\Big[k\eta\big(H_{i\mu_2+1}-H_{i\mu_2-1}\big)-3H_{i\mu_2}\Big]\, ,\\
\sigma^{\pm 2}_{2,2} &= {\cal A}_2\hskip 1pt Z_2^{\pm 2} (-k\eta)^{-1/2}H_{i\mu_2}\, ,\label{sigma2}
\end{align}
for the higher-helicity modes. 

\vskip 4pt
To fix the normalization, we again impose orthonormality of the mode functions
\begin{align}
\left\langle \sigma^{(\lambda)}_{\mu \alpha}(\k,\eta)e^{i\k\cdot\x}, \sigma^{(\lambda')}_{\nu \beta}(\k',\eta)e^{i\k'\cdot\x}\right\rangle = \delta_{\lambda\lambda'}\delta(\k-\k')\, .\label{spin2ortho}
\end{align}
We have
\begin{align}
&\left\langle \sigma^{(0)}_{\mu \alpha}(\k,\eta)e^{i\k\cdot\x}, \sigma^{(0)}_{\nu \beta}(\k',\eta)e^{i\k'\cdot\x}\right\rangle = -\frac{i}{a^2}\eta^{\mu\nu}\eta^{\alpha\beta}\int \d^3 x\, \left[\sigma_{\mu\alpha}^{(0)} {\sigma_{\nu\beta}^{(0)*\prime}} - \sigma_{\mu\alpha}^{(0)\prime} \sigma_{\nu\beta}^{(0)*}\right]e^{i(\k-\k')\cdot\x}\nonumber\\
&\qquad\qquad= -\frac{i}{a^2}\left[\frac{4}{3}{\cal W}(\sigma^{0}_{0,2},\sigma^{0*}_{0,2})-2{\cal W}(\sigma^{0}_{1,2},\sigma^{0*}_{1,2})+\frac{3}{2}{\cal W}(\sigma^{0}_{2,2},\sigma^{0*}_{2,2})\right]\delta(\k-\k')\, ,\label{spin2inner}
\end{align}
where
\begin{align}
{\cal W}(\sigma^{0}_{0,2},\sigma^{0*}_{0,2}) &= \frac{4ik^3\eta^2}{\pi} \times N_2^2\, ,\\
{\cal W}(\sigma^{0}_{1,2},\sigma^{0*}_{1,2}) &=\frac{4ik(k^2\eta^2+1/4+\mu_2^2)}{\pi} \times N_2^2\, ,\\
{\cal W}(\sigma^{0}_{2,2},\sigma^{0*}_{2,2}) &=\frac{i[32k^4\eta^4+96k^2\eta^2(1/4+\mu_2^2)+72(1/4+\mu_2^2)(9/4+\mu_2^2)]}{18\pi k \eta^2} \times N_2^2\, .
\end{align}
The condition (\ref{spin2inner}) then sets the normalization constant to be
\begin{align}
N_2 = \sqrt{\frac{\pi}{3}}\frac{1}{\sqrt{2k}}\frac{k}{H}\frac{1}{\big[(1/4+\mu_2^2)(9/4+\mu_2^2)\big]^{1/2}}\ .
\end{align}
We see that this diverges at $m^2=0$ and $m^2=2H^2$. This is again to be expected.
For $m=0$, the action gains gauge invariance, in which case only the helicity-$\pm 2$ modes are physical. For $m^2 = 2 H^2$, the field becomes partially massless, and the number of propagating degrees of freedom becomes four. In both cases, the longitudinal mode becomes a pure gauge mode.
Finally, determining the normalizations of the transverse modes in an analogous way, we get \begin{align}
Z_2^{\pm 1} = \frac{\sqrt\pi}{3}\frac{1}{\sqrt{2k}}\frac{k}{H}\frac{1}{(9/4+\mu_2^2)^{1/2}}\ , \quad Z_2^{\pm 2}=\sqrt{\frac{\pi}{2}}\frac{1}{\sqrt{2k}}\frac{k}{H}\, .
\end{align}
In the massless limit, $Z_2^{\pm 1}$ diverges and only $Z_2^{\pm 2}$ remains finite.

\paragraph{Spin-${\boldsymbol s}$}
For spins higher than two, we need to solve the on-shell equations (\ref{eom}). In order to decouple these equations, we expand the field $\sigma_{\mu_1\cdots\mu_s}$ into its different helicity components
\begin{align}
\sigma_{\mu_1\cdots\mu_s} &= \sum_{\lambda=-s}^s \sigma^{(\lambda)}_{\mu_1\cdots\mu_s}\, .
\end{align}
A mode of helicity $\lambda$ and $n$ polarization directions can be written as
\begin{align}
\sigma^{(\lambda)}_{i_1\cdots i_n\eta\cdots \eta} = \sigma_{n,s}^{\lambda}\hskip 1pt\varepsilon_{i_1\cdots i_n}^{ \lambda}\, , \label{equ:SigmaS}
\end{align}
where $ \sigma_{n,s}^{\lambda}=0$ for $n<|\lambda|$. The helicity-$\lambda$ mode function with $n=|\lambda|$ number of polarization directions satisfies 
\begin{align}
{\sigma_{|\lambda|,s}^{\lambda}}'' -\frac{2(1-\lambda)}{\eta} \hskip 1pt{\sigma_{|\lambda|,s}^{\lambda}}' +\left(k^2+\frac{m^2/H^2-(s+\lambda-2)(s-\lambda+1)}{\eta^2}\right) \sigma_{|\lambda|,s}^{\lambda} &=0\, ,
\end{align}
whose solution is given by
\begin{align}
\sigma_{|\lambda|,s}^{\lambda} = {\cal A}_s\hskip 1pt Z_s^{\lambda} (-k\eta)^{3/2-\lambda}H_{i\mu_s}\, .\label{y0}
\end{align}
The other mode functions can then be obtained iteratively using the following recursion relation:
\begin{align}
\sigma_{n+1,s}^{\lambda} = -\frac{i}{k}\left({\sigma_{n,s}^{\lambda}}'-\frac{2}{\eta} \sigma_{n,s}^{\lambda}\right)-\sum_{m=|\lambda|}^n B_{m,n+1}\hskip 1pt \sigma_{m,s}^{\lambda}\, , \label{moderecur}
\end{align}
where 
\begin{align}
B_{m,n}\equiv \frac{2^n\hskip 1pt n!}{m!(n-m)!(2n-1)!!}\frac{\Gamma[\frac{1}{2}(1+m+n)]}{\Gamma[\frac{1}{2}(1+m-n)]}\, .
\end{align}
Having obtained the formula that enables us to compute the mode functions of arbitrary spin and helicity, let us now fix their normalization constants. In order to do so, we first define an inner product between two mode functions. Note that if $f_{\mu_1\cdots\mu_s}$ and $h_{\nu_1\cdots\mu_s}$ are two solutions to (\ref{eom}), then the current 
\begin{align}
J_\mu \equiv f^{\nu_1\cdots\nu_s}\nabla_\mu h_{\nu_1\cdots\nu_s}^* - h_{\nu_1\cdots\nu_s}^*\nabla_\mu f^{\nu_1\cdots\nu_s}\, ,\label{current}
\end{align}
is conserved, $\nabla^\mu J_\mu=0$. This means that we can define an inner product of two solutions 
\begin{align}
\left\langle f_{\mu_1\cdots\mu_s}, h_{\nu_1\cdots\nu_s} \right\rangle &\equiv -i g^{\mu_1\nu_1}\cdots g^{\mu_s\nu_s}\int \d\Sigma\, n^\lambda\sqrt{\hat g} \, \big[f_{\mu_1\cdots\mu_s} \nabla_\lambda h_{\nu_1\cdots\nu_s}^* - h_{\nu_1\cdots\nu_s}^* \nabla_\lambda f_{\mu_1\cdots\mu_s} \big]\, ,
\end{align}
where $\Sigma$ denotes a spacelike hypersurface, $\hat g$ is the determinant of the spatial metric, and $n^\mu$ is the timelike unit vector orthogonal to $\Sigma$. The conservation of the current (\ref{current}) implies that the inner product is time independent. For the FRW metric, the above inner product reduces to
\begin{align}
\left\langle f_{\mu_1\cdots\mu_s}, h_{\nu_1\cdots\nu_s} \right\rangle = -\frac{i}{a^{2(s-1)}}\eta^{\mu_1\nu_1}\cdots\eta^{\mu_s\nu_s}\int \d^3 x\, \big[f_{\mu_1\cdots\mu_s} {h_{\nu_1\cdots\nu_s}^{*\prime}} - f_{\mu_1\cdots\mu_s}^{\prime} h_{\nu_1\cdots\nu_s}^*  \big]\, .\label{inner}
\end{align}
The normalization in (\ref{y0}) is then determined by imposing orthonormality under the inner product~(\ref{inner}):
\begin{align}
\left\langle \sigma^{(\lambda)}_{\mu_1\cdots\mu_s}(\k,\eta)e^{i\k\cdot\x}, \sigma^{(\lambda')}_{\nu_1\cdots\nu_s}(\k',\eta)e^{i\k'\cdot\x}\right\rangle = \delta_{\lambda\lambda'}\delta(\k-\k')\, .\label{spinsortho}
\end{align}
Since the inner product is time independent, it does not matter which time slice we choose to evaluate the integral on. We will therefore evaluate the integral on the future boundary by taking the limit $\eta\to 0$. From (\ref{moderecur}), we note that $\sigma_{n_1,s}^{\lambda}$ is subleading compared to $\sigma_{n_2,s}^{\lambda}$ in the limit $\eta\to 0$ for all $n_1<n_2$, so we simply need to compute the Wronskian of the mode with the highest number of polarization directions, $\sigma_{s,s}^{\lambda}$. If we had kept all the Wronskians, then the subleading time-dependent terms would cancel. Note also that the trace terms in (\ref{moderecur}) become subleading in the limit $\eta\to 0$, so we will drop these terms. Since (\ref{inner}) is a constant, the leading term in the Wronskian must scale as $\eta^{2(1-s)}$ to cancel off the factor $a^{2(1-s)}$. In the insert below, we will show that the orthonormality condition fixes the normalization constant to be
\begin{align}
(Z_s^{\lambda})^2 &= \frac{1}{k}\left(\frac{k}{H}\right)^{2s-2}({\cal Z}_s^{\lambda})^2\, ,\label{Zs}\\
({\cal Z}_s^{\lambda})^2 &\equiv  \frac{\pi}{4}\frac{[(2\lambda-1)!!]^2s!(s-\lambda)!}{(2s-1)!!(s+\lambda)!}\frac{\Gamma(\frac{1}{2}+\lambda+i\mu_s)\Gamma(\frac{1}{2}+\lambda-i\mu_s)}{\Gamma(\frac{1}{2}+s+i\mu_s)\Gamma(\frac{1}{2}+s-i\mu_s)}\, .\label{Zs2}
\end{align}
Note that the normalization constant has poles at $\mu_s^2=\{-(n+\frac{1}{2})^2\}_{n=\lambda}^s$, at which the spinning field becomes (partially) massless and some of the helicity modes become unphysical. For convenience, we will denote the normalization of the longitudinal mode by $N_s\equiv Z_s^0$ (${\cal N}_s\equiv {\cal Z}_s^0$).

\begin{framed}
{\small
\noindent{\it Derivation of} (\ref{Zs}).---First, note that the $n$-th mode function can be cast in the form
\begin{align}
\sigma_{n,s}^{\lambda} = {\cal A}_sZ_s^{\lambda} (-k\eta)^{3/2-n}\Big[(x_n+iy_n)H_{i\mu_s} + (w_n+iz_n)k\eta H_{i\mu_s+1}\Big] ,
\end{align}
by use of the recursion relation $H_{i\mu_s+1}(x)+H_{i\mu_s-1}(x)= (2i\mu_s/x)H_{i\mu_s}(x)$.
The coefficients $x_n$, $y_n$, $w_n$, and $z_n$ can in general depend on time, but are constant in the limit $\eta\to 0$.
The Wronskian is 
\begin{align}
{\cal W}\big[\sigma_{n,s}^{\lambda},{\sigma_{n,s}^{\lambda*}}\big] = \frac{4ik (Z_s^{\lambda})^2}{\pi(k\eta)^{2(n-1)}}\Big[X_n-2\mu_s Y_n(\coth\pi\mu_s-1)\Big] \, ,\label{spinswrons}
\end{align}
where
\begin{align}
X_n\equiv x_n^2+y_n^2 \, , \quad Y_n\equiv x_nz_n-y_nw_n+(w_n^2+z_n^2)\mu_s\, .
\end{align}
Let us show that in fact $Y_n=0$ for any $n$-th order mode function. We do this by induction. First, it is trivial to check that this is satisfied by the mode (\ref{y0}). Now, assume that $Y_n=0$ is satisfied at some $n$-th order. Using the recursion relation (\ref{moderecur}), and taking the limit $\eta\to 0$, we get
\begin{align}
\sigma_{n+1,s}^{\lambda} = {\cal A}(\mu_s)Z_s^{\lambda} (-k\eta)^{1/2-n}\Big[(x_{n+1}+iy_{n+1})H_{i\mu_s} + (w_{n+1}+iz_{n+1})k\eta H_{i\mu_s+1}\Big] ,
\end{align}
where
\begin{align}
2x_{n+1}&=- 2\mu_s x_n+(2n+1)y_n\, , \qquad\qquad\,\hskip 1pt 2y_{n+1} = -(2n+1)x_n- 2\mu_s y_n\, ,\nonumber\\
2w_{n+1}&=-2y_n+ 2\mu_s w_n+ (2n+1)z_n\, , \quad 2z_{n+1}= 2x_n - (2n+1)w_n + 2\mu_s z_n\, .\label{coeff}
\end{align}
These coefficients then give
\begin{align}
X_{n+1}=
\left[(n+\tfrac{1}{2})^2+\mu_s^2\right]X_n\, , \quad Y_{n+1}=
\left[(n+\tfrac{1}{2})^2+\mu_s^2\right]Y_n\, .\label{coeff2}
\end{align}
Hence, $Y_{n+1}=0$. Since $n$ was arbitrary, we conclude that $Y_n=0$ for all $n$. Next, we show that the Wronskian of the $n$-th longitudinal mode function has the form
\begin{align}
{\cal W}\big[\sigma_{n,s}^{\lambda},\sigma_{n,s}^{\lambda*}\big] = \frac{4ik (Z_s^{\lambda})^2}{\pi(k\eta)^{2(n-1)}}\frac{\Gamma(\frac{1}{2}+n+i\mu_s)\Gamma(\frac{1}{2}+n-i\mu_s)}{\Gamma(\frac{1}{2}+\lambda+i\mu_s)\Gamma(\frac{1}{2}+\lambda-i\mu_s)}\, .\label{spinswrons0}
\end{align}
The Wronskian of the mode function (\ref{y0}) is
\begin{align}
{\cal W}\big[\sigma_{\lambda,s}^{\lambda},\sigma_{\lambda,s}^{\lambda*}\big] = \frac{4ik (Z_s^{\lambda})^2}{\pi(k\eta)^{2(\lambda-1)}} \, ,
\end{align}
and hence satisfies (\ref{spinswrons0}). Assuming that (\ref{spinswrons0}) is true at $n$-th order and using (\ref{coeff2}), we get
\begin{align}
{\cal W}\big[\sigma_{n+1,s}^{\lambda},\sigma_{n+1,s}^{\lambda*}\big] &= \frac{4ik(Z_s^{\lambda})^2}{\pi(k\eta)^{2n}}X_{n+1}=\frac{\left[(n+\tfrac{1}{2})^2+\mu_s^2\right]}{(k\eta)^2}\frac{4ik(Z_s^{\lambda})^2}{\pi(k\eta)^{2(n-1)}}X_{n}\nonumber\\
& =\frac{\left[(n+\tfrac{1}{2})^2+\mu_s^2\right]}{(k\eta)^2}{\cal W}\big[\sigma_{n,s}^{\lambda},\sigma_{n,s}^{\lambda*}\big] \nonumber\\
&=\frac{4ik (Z_s^{\lambda})^2}{\pi(k\eta)^{2n}}\frac{\Gamma(\frac{3}{2}+n+i\mu_s)\Gamma(\frac{3}{2}+n-i\mu_s)}{\Gamma(\frac{1}{2}+\lambda+i\mu_s)\Gamma(\frac{1}{2}+\lambda-i\mu_s)} \, ,
\end{align}
where in the last line we have use the fact that
\begin{align}
\frac{\Gamma(\frac{3}{2}+n+i\mu_s)\Gamma(\frac{3}{2}+n-i\mu_s)}{\Gamma(\frac{1}{2}+n+i\mu_s)\Gamma(\frac{1}{2}+n-i\mu_s)}=(n+\tfrac{1}{2})^2+\mu_s^2\, .
\end{align}
Thus, we have proven (\ref{spinswrons0}). Finally, the inner product (\ref{inner}) is given by
\begin{align}
&\left\langle \sigma^{(\lambda)}_{\mu_1\cdots\mu_s}(\k,\eta)e^{i\k\cdot\x}, \sigma^{(\lambda)}_{\nu_1\cdots\nu_s}(\k',\eta)e^{i\k'\cdot\x}\right\rangle  \nonumber\\[6pt]
&\qquad \qquad= -\frac{i}{a^{2(s-1)}}\eta^{\mu_1\nu_1}\cdots\eta^{\mu_s\nu_s}\int \d^3 x\, \left[\sigma_{\mu_1\cdots\mu_s}^{(\lambda)} {\sigma_{\nu_1\cdots\nu_s}^{(\lambda)*\prime}} - {\sigma_{\mu_1\cdots\mu_s}^{(\lambda)\prime}} \sigma_{\nu_1\cdots\nu_s}^{(\lambda)*} \right]e^{i(\k-\k')\cdot\x}\, \nonumber\\[2pt]
&\qquad \qquad=-i(-H\eta)^{2(s-1)}{\cal W}\big[\sigma_{s,s}^{\lambda},\sigma_{s,s}^{\lambda*}\big]\varepsilon^\lambda_{i_1\cdots i_s}\varepsilon^{\lambda *}_{i_1\cdots i_s}\delta(\k-\k')\nonumber\\[2pt]
&\qquad \qquad=\frac{4k(Z_s^{\lambda})^2}{\pi}\left(\frac{H}{k}\right)^{2(s-1)}\frac{\Gamma(\frac{1}{2}+s+i\mu_s)\Gamma(\frac{1}{2}+s-i\mu_s)}{\Gamma(\frac{1}{2}+\lambda+i\mu_s)\Gamma(\frac{1}{2}+\lambda-i\mu_s)} \varepsilon^\lambda_{i_1\cdots i_s}\varepsilon^{\lambda *}_{i_1\cdots i_s}\delta(\k-\k')\, .\label{spinsinner2}
\end{align}
Note that our final normalization depends on the normalization of the polarization tensors. This does not affect correlation functions, however, as we show in the next section. Plugging (\ref{selfcontraction}) into (\ref{spinsinner2}) and imposing (\ref{spinsortho}), we obtain (\ref{Zs}).
}
\end{framed}

\subsection{Two-Point Function}
\label{app:2pt}

In this section, we will compute the two-point functions of spinning fields. For this purpose, it will be convenient to contract free indices of the spinning fields with auxiliary vectors. In other words, we will compute
\begin{align}
\left\langle (n\cdot\sigma)^2\right\rangle_s' \equiv \left\langle \left(n_{i_1}\cdots n_{i_s}\sigma_{i_1\cdots i_s}(\eta)\right) \left(\tilde n_{j_1}\cdots \tilde n_{j_s}\sigma_{j_1\cdots j_s}(\eta')\right)\right\rangle'\, ,\label{2pt}
\end{align}
where the prime on the expectation value indicates the removal of the momentum-conserving delta function, and $\n \equiv (\cos\alpha,\sin\alpha,i)$ and $\tilde\n \equiv (\cos\beta,\sin\beta,-i)$ are null vectors. 
For generic $\eta$ and $\eta'$, the two-point function is
\begin{align}
\left\langle (n\cdot\sigma)^2\right\rangle_s' = \sum_{\lambda=-s}^s e^{is\chi}\left[\frac{(2s-1)!!}{(2\lambda-1)!!(s-\lambda)!}\right]^2\sigma_{s,s}^{\lambda}(-k\eta)\sigma_{s,s}^{\lambda*}(-k\eta')\, ,
\end{align}
where $\chi\equiv\alpha-\beta$. 
In the late-time limit (or the long-wavelength limit), the two-point function simplifies considerably. We get
\begin{align}
\lim_{\eta,\hskip 1pt \eta'\to 0}\left\langle (n\cdot\sigma)^2\right\rangle_s' = \frac{(H^2\eta\eta')^{3/2-s}}{4\pi H}\sum_{\lambda=-s}^s e^{i\lambda\chi}\left[\, {\cal C}(\mu_s,\lambda,s)\, \Gamma(-i\mu_s)^2\left(\frac{k^2\eta\eta'}{4}\right)^{i\mu_s}+c.c.\right] ,\label{2pt2}
\end{align}
where
\begin{align}
{\cal C}(\mu_s,\lambda,s) \equiv \frac{(2s-1)!!\hskip 1pt s!}{(s-\lambda)!(s+\lambda)!}\frac{\Gamma(\frac{1}{2}+s-i\mu_s)\Gamma(\frac{1}{2}+\lambda+i\mu_s)}{\Gamma(\frac{1}{2}+s+i\mu_s)\Gamma(\frac{1}{2}+\lambda-i\mu_s)}\, .
\end{align}
This late-time expectation value matches the two-point function of a spin-$s$ field of a conformal field theory living on the future boundary, which have been computed in \cite{Arkani-Hamed:2015bza}.

\begin{framed}
{\small
\noindent{\it Derivation of} (\ref{2pt2}).---The two-point function (\ref{2pt}) can be written as
\begin{align}
\langle (n\cdot\sigma)^2 \rangle_s' &= \sum_{\lambda=-s}^s(n_{i_1}\cdots n_{i_s}\varepsilon^{\lambda}_{i_1\cdots i_s})(\tilde n_{j_1}\cdots\tilde n_{j_s}\varepsilon^{\lambda *}_{j_1\cdots j_s})\,\sigma^{\lambda}_{s,s} \sigma^{\lambda*}_{s,s}\, .\label{appeq:2pt}
\end{align}
Let us compute $\sigma^{\lambda}_{s,s} \sigma^{\lambda*}_{s,s}$ in the late-time limit. First, recall that we can cast the mode function in the form
\begin{align}
\sigma_{n,s}^{\lambda} = {\cal A}_s\hskip 1pt Z_s^{\lambda} (-k\eta)^{3/2-n}\Big[(x_n+iy_n)H_{i\mu_s} + (w_n+iz_n)k\eta H_{i\mu_s+1}\Big]\, .\label{spinsmode}
\end{align}
Taking the asymptotic limits of the Hankel functions, we get
\begin{align}
\sigma_{n,s}^{\lambda}\sigma_{n,s}^{\lambda*}\Big|_{\eta,\eta'\to 0} = (Z_s^{\lambda})^2\frac{(k^2\eta\eta')^{3/2-n}}{\pi^2}\left[W_n\, \Gamma(-i\mu_s)^2\left(\frac{k^2\eta\eta'}{4}\right)^{i\mu_s}+c.c.\right] + \text{local terms}\, ,\label{appeq:2pt2}
\end{align}
where
\begin{align}
W_n\equiv x_n^2+y_n^2+2\mu_s(x_n+i y_n)(iw_n+z_n)\, .
\end{align}
Using (\ref{coeff}), we obtain the recursion relation
\begin{align}
W_{n+1} = \left(n+\tfrac{1}{2}-i\mu_s\right)^2W_n\, .
\end{align}
Following similar arguments as in the previous section, it can then be shown that
\begin{align}
W_n = \frac{\Gamma(\frac{1}{2}+s-i\mu_s)^2}{\Gamma(\frac{1}{2}+\lambda-i\mu_s)^2}\, .\label{gn}
\end{align}
Substituting (\ref{Zs}), (\ref{appeq:2pt2}), and (\ref{gn}) into (\ref{appeq:2pt}), we obtain
\begin{align}
\langle (n\cdot\sigma)^2 \rangle_s' &= \frac{(H^2\eta\eta')^{3/2-s}}{4\pi H}
\sum_{\lambda=-s}^s I^\lambda_s(\n,\tilde\n)\left[{\cal D}(s,\lambda,\mu_s) \Gamma(-i\mu_s)^2\left(\frac{k^2\eta\eta'}{4}\right)^{i\mu_s}+c.c.\right] ,\label{2pt3}
\end{align}
where we have dropped the local terms and defined
\begin{align}
I^\lambda_s(\n,\tilde\n) &\equiv \frac{(n_{i_1}\cdots n_{i_s}\varepsilon_{i_1\cdots i_s}^{\lambda})(\tilde n_{i_1}\cdots \tilde n_{i_s}\varepsilon_{i_1\cdots i_s}^{\lambda *})}{\varepsilon_{i_1\cdots i_s}^{\lambda}\varepsilon_{i_1\cdots i_s}^{\lambda *}} \, , \\[4pt]
{\cal D}(s,\lambda,\mu_s) &\equiv \frac{\Gamma(\frac{1}{2}+s-i\mu_s)\Gamma(\frac{1}{2}+\lambda+i\mu_s)}{\Gamma(\frac{1}{2}+s+i\mu_s)\Gamma(\frac{1}{2}+\lambda-i\mu_s)}\, .\label{Dfunc}
\end{align}
To obtain an expression for $I^\lambda_s$, let us first recall that the structure of the polarization tensors are given by the (associated) Legendre polynomials. Contracting with null vectors, only the term with the leading power in $k$ survives (with no Kronecker delta's), whose coefficient is $(2s-1)!!/[(2\lambda-1)!!(s-\lambda)!]$. This means that
\begin{align}
(n_{i_1}\cdots n_{i_s}\varepsilon_{i_1\cdots i_s}^{\lambda})(\tilde n_{i_1}\cdots \tilde n_{i_s}\varepsilon_{i_1\cdots i_s}^{\lambda *}) = \left[\frac{(2s-1)!!}{(2\lambda-1)!!(s-\lambda)!}\right]^2\, e^{is\chi}\, , \label{nullcontraction}
\end{align}
where we used the fact that we get one factor of $e^{i\alpha}$ for each contraction with a null vector, i.e.~$n_{i_1}\cdots n_{i_s}\varepsilon^s_{i_1\cdots i_s} = e^{is\alpha}$.
Combining (\ref{nullcontraction}) and (\ref{selfcontraction}), we get
\begin{align}
I^\lambda_s(\n,\tilde\n) = \frac{(2s-1)!!\hskip 1pt s!}{(s-\lambda)!(s+\lambda)!}e^{i\lambda\chi}\, .\label{I}
\end{align}
Substituting this into (\ref{2pt3}), we obtain (\ref{2pt2}).}
\end{framed}

\newpage
\section{In-In Results}
\label{app:inin}

In this appendix, we present details of the in-in computations of Section~\ref{sec:correlators}.
In particular, we will give explicit expressions for the shape functions introduced in (\ref{zeta3spinssingle}), (\ref{zeta3spin1double}), (\ref{zeta3spin1triple}) and (\ref{tss3ptspinssingle}). 

\paragraph{Preliminaries}
The expectation value of an operator $ {\cal Q}$ is computed by 
\begin{align}
\langle {\cal Q}(\eta)\rangle = \langle 0| \left[\bar{\rm T} e^{i\int_{-\infty}^\eta \d \eta' \hat H_I(\eta')}\right]{\cal Q}(\eta)\left[{\rm T} e^{-i\int_{-\infty}^\eta \d \eta' \hat H_I(\eta')}\right]|0\rangle\ ,\label{inin}
\end{align}
where $|0\rangle$ is the vacuum state of the free theory, $\rm T$ and $\bar{\rm T}$ denote time-ordering and anti-time-ordering, respectively, and $\hat H_I$ is the interaction Hamiltonian. To compute the quantum expectation values, we promote the fields $\pi$, $\gamma$, $\sigma$ to operators and expand in Fourier space
\begin{align}
\pi(\k,\eta) = \pi_k(\eta)  a(\k) + h.c.\, ,\quad \gamma_{ij}(\k,\eta) &= \sum_{\lambda=\pm 2} \varepsilon_{ij}^\lambda(\k)\gamma_k^\lambda(\eta)  b(\k,\lambda) + h.c. \ ,\\
\hat\sigma_{i_1\cdots i_s}(\k,\eta) &=  \sum_{\lambda=-s}^s \varepsilon_{i_1\cdots i_s}^\lambda(\k) \sigma_{s,s}^{\lambda}(k,\eta)  b_s(\k,\lambda) + h.c.\, ,
\end{align}
where the creation and annihilation operators obey the usual canonical commutation relations
\begin{align}
[ a(\k), a^\dagger(\k')] &= (2\pi)^3\delta(\k-\k') \ , \\ 
[ b(\k,\lambda), b^\dagger(\k',\lambda')] = [ b_s(\k,\lambda), b_s^\dagger(\k',\lambda')]&=(2\pi)^3\delta(\k-\k')\delta_{\lambda\lambda'} \ .
\end{align}
The mode functions for the Goldstone and the graviton are
\begin{align}
\pi_k(\eta)&=\frac{H}{f_\pi^2}\frac{i}{\sqrt{2k^3}}(1+i\cs k\eta)e^{-i\cs k\eta}\, ,\quad \gamma_k^\lambda(\eta) = \frac{H}{\Mp}\frac{i}{\sqrt{2k^3}}(1+ik\eta)e^{-ik\eta} \, .
\end{align}
The mode functions $\sigma_{s,s}^\lambda(k,\eta)$ were derived in Appendix~\ref{app:spindS}.  It will be convenient to write the longitudinal and helicity-$\pm 2$ mode functions as
\begin{align}
\sigma^0_{s,s}(-k\eta) &= N_s(-k\eta)^{3/2-s}\hskip 1pt G_{i\mu_s}^{(s)}(-k\eta)\, , \quad \sigma^{\pm 2}_{s,s}(-k\eta) = Z_s^{\pm 2}(-k\eta)^{3/2-s}\hskip 1pt \tilde G_{i\mu_s}^{(s)}(-k\eta)\, ,\label{Gmode} 
\end{align}
where the functions $G_{i\mu_s}^{(s)} \equiv G_{i\mu_s}^{(s,\lambda=0,n=s)}$ and $\tilde G_{i\mu_s}^{(s)} \equiv G_{i\mu_s}^{(s,\lambda=\pm 2,n=s)}$ can be obtained recursively using (\ref{moderecur}), or
\begin{align}
G_{i\mu_s}^{(s,\lambda,n+1)}(x) =\frac{i}{2}\left[2x \partial_x G_{i\mu_s}^{(s,\lambda,n)}(x)+(1-2n)G_{i\mu_s}^{(s,\lambda,n)}(x)\right] -\sum_{m=\lambda}^s B_{m,n+1}G_{i\mu_s}^{(s,\lambda,m)}(x)\, ,
\end{align}
given $G_{i\mu_s}^{(s,\lambda,\lambda)}(x)={\cal A}_s\hskip 1ptH_{i\mu_s}(x)$. For $s=1$ and 2, we get 
\begin{align}
G^{(1)}_{i\mu_1}(x) &\equiv \frac{i}{2}{\cal A}_1\hskip 1pt \Big[x\big(H_{i\mu_1-1}(x)-H_{i\mu_1+1}(x)\big)-H_{i\mu_1}(x)\Big]\, ,\\
G_{i\mu_2}^{(2)}(x)& \equiv \frac{1}{12}{\cal A}_2\hskip 1pt \Big[6x\big[(2-i\mu_2)H_{i\mu_2-1}(x)-(2+i\mu_2)H_{i\mu_2+1}(x)\big]-(9-8x^2)H_{i\mu_2}(x)\Big]\, .
\end{align}

\paragraph{Results}
In Section~\ref{sec:correlators}, the results for the bispectra were defined in terms of a number of momentum-dependent functions. In the following, we give explicit integral expressions for these functions: 

\vskip 4pt
$\bullet$\ \  For $s\ge 2$, the functions ${\cal I}^{(s)}$ in (\ref{zeta3spinssingle}) are given by
\begin{align}
{\cal I}^{(s)}&\equiv \sum_{j=1}^3\frac{2\pi^3 {\cal N}_s^2}{k_1^{3/2} k_2^{7/2} k_3}{\rm Re}[{\cal I}_{j}^{(s)}]\, ,\\[4pt]
{\cal I}^{(s)}_1&\equiv -\int_0^\infty\d x\, \tilde{\cal T}_{i\mu_s}^{(s)*}(\cs,k_1,k_2,k_3,x)\int_0^\infty\d y\hskip 1pt \, \tilde{\cal F}^{(s)}_{i\mu_s}(\cs,y)\, ,\\[2pt]
{\cal I}^{(s)}_2&\equiv \int_0^\infty\d x\, {\cal T}_{i\mu_s}^{(s)}(\cs,k_1,k_2,k_3,x)\int_{\kappa_{12}x/\cs}^\infty\d y\hskip 1pt \, \tilde{\cal F}^{(s)}_{i\mu_s}(\cs,y)\, ,\label{i2}\\[2pt]
{\cal I}^{(s)}_3&\equiv \int_0^\infty\d x\hskip 1pt \, {\cal F}_{i\mu_s}^{(s)}(\cs,x)\int_{\cs\kappa_{21}x}^\infty\d y\, \tilde{\cal T}_{i\mu_s}^{(s)}(\cs,k_1,k_2,k_3,y)\, , 
\end{align}
where $\kappa_{ij}\equiv k_i/k_j$ and ${\cal N}_s\equiv {\cal Z}_s^0$ is the normalization constant defined in (\ref{Zs2}). The integrands are represented by the functions
\begin{align}
{\cal F}_{i\mu_s}^{(s)}(\cs,x)&\equiv x^{s-5/2}(1+i\cs x)G_{i\mu_s}^{(s)}(y)e^{-i\cs x}\, ,\\
\tilde{\cal F}_{i\mu_s}^{(s)}(\cs,x)&\equiv x^{s-5/2}(1+i\cs x)G_{i\mu_s}^{(s)*}(y)e^{-i\cs x}\, ,\\
{\cal T}_{i\mu_s}^{(s)}(\cs,k_1,k_2,k_3,x)&\equiv x^{s-1/2}\hskip 1pt (1+i x)G_{i\mu_s}^{(s)}(x k_1/\cs k_2)e^{-ix (1+k_3/k_2)}\, ,\\
\tilde {\cal T}_{i\mu_s}^{(s)}(\cs,k_1,k_2,k_3,x)&\equiv x^{s-1/2}\hskip 1pt (1+i x)G_{i\mu_s}^{(s)*}(x k_1/\cs k_2)e^{-ix (1+k_3/k_2)}\, ,
\end{align}
where $G_{i\mu_s}^{(s)}$ was defined in (\ref{Gmode}). The integral $\int_0^\infty \d x\, {\cal F}^{(1)}_{i\mu_1}$ is in fact IR divergent. To avoid this issue, we integrate by parts and work with ${\cal F}^{(1)}_{i\mu_1} \to x^{1/2}H_{i\mu_1}(x)e^{ix}$ and $\tilde{\cal F}^{(1)}_{i\mu_1} \to x^{1/2}H_{i\mu_1}(x)e^{-ix}$.
\vskip 4pt
\noindent
$\bullet$\ \  The functions ${\cal J}^{(s)}$ in 
(\ref{zeta3spin1double}) are given by 
\begin{align}
{\cal J}^{(s)}&=\sum_{j=1}^6\frac{2\pi^3 {\cal N}_s^4}{k_1^3 k_2^{3/2} k_3^{3/2}}{\rm Im}[{\cal J}_{j}^{(s)}]\, ,\\[4pt]
{\cal J}_{1}^{(s)} &\equiv-\int_0^\infty\d x\, \hat{\cal G}^{(s)*}_{i\mu_s}(\cs,k_1,k_2,k_3,x)\int_0^\infty\d y\,\tilde{\cal F}^{(s)}_{i\mu_s}(\cs,y)\int_{\kappa_{32}y}^\infty\d z\, \tilde{\cal F}^{(s)}_{i\mu_s}(\cs,z)\, ,\\
{\cal J}_2^{(s)} &\equiv-\int_0^\infty\d x\, \tilde{\cal F}^{(s)*}_{i\mu_s}(\cs,x)\int_0^\infty\d y\,  \tilde{\cal G}^{(s)}_{i\mu_s}(\cs,k_1,k_2,k_3,y)\int_{\kappa_{31}y/\cs}^\infty\d z\, \tilde{\cal F}^{(s)}_{i\mu_s}(\cs,z)\, ,\\
{\cal J}_3^{(s)} &\equiv-
\int_0^\infty\d x\, \tilde{\cal F}^{(s)*}_{i\mu_s}(\cs,x)\int_0^\infty\d y\, \tilde{\cal F}^{(s)}_{i\mu_s}(\cs,y)\int_{\cs\kappa_{12}y}^\infty\d z\,\hat{\cal G}^{(s)}_{i\mu_s}(\cs,k_1,k_2,k_3,z)\, ,\\
{\cal J}_4^{(s)} &\equiv\int_0^\infty\d x\, {\cal G}^{(s)}_{i\mu_s}(\cs,k_1,k_2,k_3,x)\int_{\kappa_{12}x/\cs}^\infty\d y\,  \tilde{\cal F}^{(s)}_{i\mu_s}(\cs,y)\int_{\kappa_{31}y}^\infty\d z\,  \tilde{\cal F}^{(s)}_{i\mu_s}(\cs,z)\, ,\\
{\cal J}_5^{(s)} &\equiv\int_0^\infty\d x\, {\cal F}^{(s)}_{i\mu_s}(\cs,x)\int_{\cs\kappa_{32}x}^\infty\d y\,  \tilde{\cal G}^{(s)}_{i\mu_s}(\cs,k_1,k_2,k_3,y)\int_{\kappa_{12}y/\cs}^\infty\d z\, \tilde{\cal F}^{(s)}_{i\mu_s}(\cs,z)\, ,\\
{\cal J}_6^{(s)} &\equiv
\int_0^\infty\d x\, {\cal F}^{(s)}_{i\mu_s}(\cs,x)\int_{\kappa_{21}x}^\infty\d y\, {\cal F}^{(s)}_{i\mu_s}(\cs,y)\int_{\cs\kappa_{32}y}^\infty\d z\,  \hat{\cal G}^{(s)}_{i\mu_s}(\cs,k_1,k_2,k_3,z)\, ,
\end{align}
where
\begin{align}
{\cal G}^{(s)}_{i\mu_s}(\cs,k_1,k_2,k_3,x) &\equiv x\, G_{i\mu_s}^{(s)}(x k_2 / \cs k_1)G_{i\mu_s}^{(s)}(x k_3 / \cs k_1)e^{-i x}\, ,\\
\tilde{\cal G}^{(s)}_{i\mu_s}(\cs,k_1,k_2,k_3,x) &\equiv x\, G_{i\mu_s}^{(s)*}(x k_2 / \cs k_1)G_{i\mu_s}^{(s)}(x k_3 / \cs k_1)e^{-i x}\, ,\\
\hat{\cal G}^{(s)}_{i\mu_s}(\cs,k_1,k_2,k_3,x) &\equiv x\, G_{i\mu_s}^{(s)*}(x k_2 / \cs k_1)G_{i\mu_s}^{(s)*}(x k_3 / \cs k_1)e^{-i x}\, .
\end{align}

\vskip 4pt
\noindent
$\bullet$\ \  The functions ${\cal K}^{(s)}$ in 
(\ref{zeta3spin1triple}) are given by
\begin{align}
{\cal K}^{(s)}&= \sum_{j=1}^{10} \frac{2\pi^3 {\cal N}_s^6}{ k_1^3 k_2^{3/2} k_3^{3/2}}{\rm Re}[{\cal K}_{j}^{(s)}]\, ,\\[4pt]
{\cal K}_1^{(s)} &\equiv -\int_0^\infty \d w\, {\cal H}_{i\mu_s}^{(s)}(k_1,k_2,k_3,w)\int_{w}^\infty \d x\, {\cal F}_{i\mu_s}^{(s)*}\int_0^\infty \d y\, {\cal F}_{i\mu_s}^{(s)*}\int_{\kappa_{32}y}^\infty \d z\, {\cal F}_{i\mu_s}^{(s)*}\, ,\\
{\cal K}_2^{(s)} &\equiv -\int_0^\infty \d w\,\tilde{\cal F}_{i\mu_s}^{(s)*}\int_{w}^\infty \d x\, \tilde{\cal H}_{i\mu_s}^{(s)}(k_1,k_2,k_3,x)\int_0^\infty \d y\, {\cal F}_{i\mu_s}^{(s)*}\int_{\kappa_{32}y}^\infty \d z\, {\cal F}_{i\mu_s}^{(s)*}\, ,\\
{\cal K}_3^{(s)} &\equiv \int_0^\infty \d w\, {\cal H}_{i\mu_s}^{(s)}(k_1,k_2,k_3,w)\int_0^\infty \d x\, \tilde{\cal F}_{i\mu_s}^{(s)}\int_{\kappa_{21}x}^\infty \d y\, \tilde{\cal F}_{i\mu_s}^{(s)}\int_{\kappa_{32}y}^\infty \d z\, \tilde{\cal F}_{i\mu_s}^{(s)}\, ,\\
{\cal K}_4^{(s)} &\equiv \int_0^\infty \d w\,  \tilde{\cal F}_{i\mu_s}^{(s)*} \int_0^\infty \d x\, \tilde{\cal H}_{i\mu_s}^{(s)}(k_1,k_2,k_3,x)\int_{\kappa_{21}x}^\infty \d y\, \tilde{\cal F}_{i\mu_s}^{(s)}\int_{\kappa_{32}y}^\infty \d z\, \tilde{\cal F}_{i\mu_s}^{(s)}\, ,\\
{\cal K}_5^{(s)} &\equiv \int_0^\infty \d w\, \tilde{\cal F}_{i\mu_s}^{(s)*}\int_0^\infty \d x\, {\cal F}_{i\mu_s}^{(s)}\int_{\kappa_{12}x}^\infty \d y\, \hat{\cal H}_{i\mu_s}^{(s)}(k_1,k_2,k_3,y)\int_{\kappa_{31}y}^\infty \d z\, \tilde{\cal F}_{i\mu_s}^{(s)}\, ,\\
{\cal K}_6^{(s)} &\equiv \int_0^\infty \d w\, \tilde{\cal F}_{i\mu_s}^{(s)*}\int_0^\infty \d x\, {\cal F}_{i\mu_s}^{(s)}\int_{\kappa_{32}x}^\infty \d y\,{\cal F}_{i\mu_s}^{(s)}\int_{\kappa_{13}y}^\infty \d z\, \bar{\cal H}_{i\mu_s}^{(s)}(k_1,k_2,k_3,w)\, ,\\
{\cal K}_7^{(s)} &\equiv -\int_0^\infty \d w\,{\cal H}_{i\mu_s}^{(s)}(k_1,k_2,k_3,w)\int_{w}^\infty \d x\, \tilde{\cal F}_{i\mu_s}^{(s)}\int_{\kappa_{21}x}^\infty \d y\, \tilde{\cal F}_{i\mu_s}^{(s)}\int_{\kappa_{32}y}^\infty \d z\, \tilde{\cal F}_{i\mu_s}^{(s)}\, ,\\
{\cal K}_8^{(s)} &\equiv -\int_0^\infty \d w\, {\cal F}_{i\mu_s}^{(s)} \int_{w}^\infty \d x\, \tilde{\cal H}_{i\mu_s}^{(s)}(k_1,k_2,k_3,x)\int_{\kappa_{21}x}^\infty \d y\,\tilde{\cal F}_{i\mu_s}^{(s)}\int_{\kappa_{32}y}^\infty \d z\,\tilde{\cal F}_{i\mu_s}^{(s)}\, ,\\
{\cal K}_9^{(s)} &\equiv -\int_0^\infty \d w\, {\cal F}_{i\mu_s}^{(s)}\int_{\kappa_{21}w}^\infty \d x\, {\cal F}_{i\mu_s}^{(s)}\int_{\kappa_{12}x}^\infty \d y\, \hat{\cal H}_{i\mu_s}^{(s)}(k_1,k_2,k_3,y)\int_{\kappa_{31}y}^\infty \d z\, \tilde{\cal F}_{i\mu_s}^{(s)}\, ,\\
{\cal K}_{10}^{(s)} &\equiv - \int_0^\infty \d w\, {\cal F}_{i\mu_s}^{(s)}\int_{\kappa_{21}w}^\infty \d x\, {\cal F}_{i\mu_s}^{(s)}\int_{\kappa_{32}x}^\infty \d y\, {\cal F}_{i\mu_s}^{(s)}\int_{\kappa_{13}y}^\infty \d z\, \bar{\cal H}_{i\mu_s}^{(s)}(k_1,k_2,k_3,w)\, ,
\end{align}
where we have suppressed some arguments and defined
\begin{align}
{\cal H}_{i\mu_s}^{(s)}(k_1,k_2,k_3,x) &\equiv x^{1/2}\hskip 1pt G_{i\mu_s}^{(s)}(x)\hskip 1pt G_{i\mu_s}^{(s)}(k_2x/k_1)\hskip 1pt G_{i\mu_s}^{(s)}(k_3x/k_1)\, ,\\
\tilde{\cal H}_{i\mu_s}^{(s)}(k_1,k_2,k_3,x) &\equiv x^{1/2}\hskip 1pt G_{i\mu_s}^{(s)*}(x)\hskip 1pt G_{i\mu_s}^{(s)}(k_2x/k_1)\hskip 1pt G_{i\mu_s}^{(s)}(k_3x/k_1)\, ,\\
\hat{\cal H}_{i\mu_s}^{(s)}(k_1,k_2,k_3,x) &\equiv x^{1/2}\hskip 1pt G_{i\mu_s}^{(s)*}(x)\hskip 1pt G_{i\mu_s}^{(s)*}(k_2x/k_1)\hskip 1pt G_{i\mu_s}^{(s)}(k_3x/k_1)\, ,\\
\bar{\cal H}_{i\mu_s}^{(s)}(k_1,k_2,k_3,x) &\equiv x^{1/2}\hskip 1pt G_{i\mu_s}^{(s)*}(x)\hskip 1pt G_{i\mu_s}^{(s)*}(k_2x/k_1)\hskip 1pt G_{i\mu_s}^{(s)*}(k_3x/k_1)\, .
\end{align}

\vskip 4pt
\noindent
$\bullet$\ \  The functions ${\cal B}^{(s)}$ in (\ref{tss3ptspinssingle}) are given by
\begin{align}
{\cal B}^{(s)} &\equiv \sum_{i=1}^3\frac{\pi^3 \tilde{\cal N}_s^2}{4k_1^{3/2}k_2^{7/2}k_3}{\rm Re}[{\cal B}^{(s)}_i] \, ,\\
{\cal B}^{(s)}_1&=-\int_0^\infty\d x\, \tilde{\cal R}_{i\mu_s}^{(s)*}(\cs,k_1,k_2,k_3,x)\int_0^\infty\d y\, y^{s-5/2} \tilde G^{(s)*}_{i\mu_2}(y)e^{-iy}\, , \\
{\cal B}^{(s)}_2&=\int_0^\infty\d x\, {\cal R}_{i\mu_s}^{(s)}(\cs,k_1,k_2,k_3,x)\int_{\kappa_{13} x/\cs}^\infty\d y\,y^{s-5/2}\tilde G^{(s)*}_{i\mu_2}(y)e^{-iy}\, ,\\
{\cal B}^{(s)}_3&=\int_0^\infty\d x\,x^{s-5/2} \tilde G^{(s)}_{i\mu_2}(x)e^{-ix}\int_{\cs\kappa_{31}x}^\infty\d y\, \tilde{\cal R}_{i\mu_s}^{(s)}(\cs,k_1,k_2,k_3,x)\, .
\end{align}
where $\tilde {\cal N}_s \equiv {\cal Z}_s^{\pm 2}$ is the normalization constant defined in (\ref{Zs2}) and 
\begin{align}
	{\cal R}_{i\mu_s}^{(s)}(\cs,k_1,k_2,k_3,x)&\equiv x^{s-1/2}\hskip 1pt (1+i x)\tilde G_{i\mu_s}^{(s)}(x k_1/\cs k_2)e^{-ix (1+k_3/k_2)}\, ,\\
\tilde {\cal R}_{i\mu_s}^{(s)}(\cs,k_1,k_2,k_3,x)&\equiv x^{s-1/2}\hskip 1pt (1+i x)\tilde G_{i\mu_s}^{(s)*}(x k_1/\cs k_2)e^{-ix (1+k_3/k_2)}\, .
\end{align}

\newpage
\section{Soft Limits}
\label{app:squeezed}

In this appendix, we will derive analytic formulas for the soft limits of the non-analytic parts of all correlation functions that we considered in this work.

\subsection[${\langle \zeta\zeta\zeta\rangle}$]{$\boldsymbol{\langle \zeta\zeta\zeta\rangle}$}
We will focus on the squeezed limit of the scalar three-point function for the single-exchange diagram (cf.~Fig.~\ref{fig:3pt_a}), and consider even spins first. 
This leads to a non-analytic behavior if the quadratic mixing leg is taken to be soft. In the squeezed limit, $k_1\ll k_2\approx k_3$, this contribution is given by 
\begin{align}
\lim_{k_1\ll k_3} \frac{\langle\zeta_{\k_1}\zeta_{\k_2}\zeta_{\k_3}\rangle'}{\Delta_\zeta^4}&\, = \, \alpha_s\Delta_\zeta^{-1}\times P_s(\hat\k_1\cdot\hat\k_3)\times {\cal I}^{(s)}(\mu_s,\cs,k_1,k_3,k_3) + (\k_2\leftrightarrow \k_3)\, ,\label{eq:singleSq}
\end{align}
where the functions ${\cal I}^{(s)}$ are given by
\begin{align}
{\cal I}^{(s)} &\equiv - \frac{(2\pi)^3\cs^{s-3/2}H^{5-2s}}{8}\sum_{\pm\pm} (\pm ik_1^{s-3})(\pm i\cs^2 k_3^{s-4})\, {\cal I}^{(s)}_{\pm\pm}\, ,\label{Is}\\
{\cal I}^{(s)}_{\pm\pm} &\equiv \int_{-\infty}^0\frac{\d\eta}{a^{2s-3}}\, \eta\hskip 1pt(1\mp i\cs k_3\eta)e^{\pm 2i\cs k_3\eta}\int_{-\infty}^0\frac{\d\tilde\eta}{a^{2s-4}}\, (1\mp i \cs k_1\tilde\eta)e^{\pm i \cs q\tilde\eta} \, G_{\pm\pm}(k_1,\eta,\tilde\eta)\, . \label{C3}
\end{align}
In (\ref{C3}) we introduced the time-ordered Green's functions on the Schwinger-Keldysh contours
\begin{align}
G_{++}(k,\eta,\tilde\eta) &= G^>(k,\eta,\tilde\eta)\hskip 1pt\Theta(\eta-\tilde\eta)+G^<(k,\eta,\tilde\eta)\hskip 1pt\Theta(\tilde\eta-\eta)\, ,\\
G_{+-}(k,\eta,\tilde\eta) &= G^<(k,\eta,\tilde\eta)\, ,\\
G_{-+}(k,\eta,\tilde\eta) &= G^>(k,\eta,\tilde\eta)\, ,\\
G_{--}(k,\eta,\tilde\eta) &= G^<(k,\eta,\tilde\eta)\hskip 1pt\Theta(\eta-\tilde\eta)+G^>(k,\eta,\tilde\eta)\hskip 1pt\Theta(\tilde\eta-\eta)\, ,
\end{align}
where
\begin{align}
G^>(k,\eta,\tilde\eta) = \sigma^0_{s,s}(-k\eta)\hskip 1pt\sigma^{0*}_{s,s}(-k\tilde\eta)\, ,\quad G^<(k,\eta,\tilde\eta) = \sigma^{0*}_{s,s}(-k\eta)\hskip 1pt\sigma^0_{s,s}(-k\tilde\eta)\, ,
\end{align}
denote the Wightman functions of the longitudinal mode of a spin-$s$ field, and $\pm$ indicates the (anti-)time-ordering along the integration contour. The non-local part of the Green's function is independent of the sign of the time difference, in which case the time-ordered Green's can be replaced with the non-time-ordered ones, $G_{+\pm}=G_{-\pm}$. The integrals thus factorize, and substituting for the $\sigma$ mode functions, the integral (\ref{Is}) becomes
\begin{align}
{\cal I}^{(s)} =N^{(s)}\sum_\pm (\pm \hskip 1pt k_1^{s-3}\hskip 1ptk_3^{s-4})\,{\cal P}^{(s)}_\pm (k_1,\cs k_3) {\cal Q}^{(s)*} (\cs ,k_1)+ c.c.\, ,
\end{align}
where we used the fact that ${\cal I}^{(s)}_{+\pm}={\cal I}^{(s)*}_{-\mp}$ and defined
\begin{align}
N^{(s)}&\equiv -\frac{s! \pi^5 \cs^{s+1/2}}{4(2s-1)!!} \frac{{\rm sech}\hskip 1pt\pi\mu_s}{\Gamma(\frac{1}{2}+s-i\mu_s)\Gamma(\frac{1}{2}+s+i\mu_s)}\, ,\\
{\cal P}^{(s)}_\pm (k_1,\cs k_3) &\equiv  e^{-\pi\mu_s/2}\int_0^{\infty}\d x\, x^{s-1/2}(1\mp i\cs k_3x)G_{i\mu_s}^{(s)}(k_1x)e^{\pm 2i\cs k_3x}\, ,\label{app:3ptM1} \\
{\cal Q}^{(s)*}(\cs ,k_1) &\equiv e^{-\pi\mu_s/2}\int_0^{\infty}\d x\, x^{s-5/2}(1+ i\cs k_1x)G_{i\mu_s}^{(s)*}(k_1x)e^{-i\cs k_1x}\, ,\label{app:3ptM}
\end{align}
with $G^{(s)}_{i\mu_s}$ introduced in (\ref{Gmode}). The integrals in (\ref{app:3ptM1}) and (\ref{app:3ptM}) can be computed analytically for arbitrary $\cs$. To derive the results below, we will use the formula
 \begin{align}
e^{-\pi\mu_s/2}\int_0^\infty \d x\, x^n H_{i\mu_s}(b x)e^{iax} & = \frac{(i/2)^n}{\sqrt\pi\hskip 1pt b^{n+1}}F_{21}(n+3/2,\mu_s,(b-a)/2b)\, ,
\end{align}
where
\begin{align}
F_{21}(a,\mu_s,z)\equiv \frac{\Gamma(a-\frac{1}{2}-i\mu_s)\Gamma(a-\frac{1}{2}+i\mu_s)}{\Gamma(a)}  \,{}_2F_1\Big(a-\frac{1}{2}-i\mu_s,a-\frac{1}{2}+i\mu_s,a+s,z\Big)\, .
\end{align}
In the squeezed limit, $k_1\ll \cs k_3$, the result for (\ref{app:3ptM1}) is
\begin{align}
\lim_{k_1\ll \cs k_3}{\cal P}^{(s)}_\pm(k_1,\cs k_3) &=  \frac{(-i)^{1/2}e^{(1\mp 1)\pi\mu_s/2}\Gamma(\frac{1}{2}+s+i\mu_s)\Gamma(\frac{1}{2}+s-i\mu_s)}{4\pi(\pm 2 \cs k_3)^{1/2+s}}\nonumber\\
&\ \ \ \times \left(\frac{k_1}{4\cs k_3}\right)^{i\mu_s}(5+2s+2i\mu_s)\frac{\Gamma(-i\mu_s)}{\Gamma(\frac{1}{2}-i\mu_s)}\mp i e^{-(1\mp 1)\pi\mu_s} \times c.c.\, .
\end{align}
Since we cannot take a soft limit of the integral (\ref{app:3ptM}), its general expression is rather complicated. For simplicity, let us display the results for the two limiting cases, $\cs=1$ and $\cs\ll 1$, for which (\ref{app:3ptM}) reduces to
\begin{align}
{\cal Q}^{(s)*}(\cs=1,k_1)&= f^{(s)}(1)\times\frac{i(2ik_1)^{3/2-s}}{\sqrt\pi\hskip 1pt\Gamma(s)}\frac{\Gamma(\frac{1}{2}+s-i\mu_s)\Gamma(\frac{1}{2}+s+i\mu_s)}{(s-\frac{3}{2})^2+\mu_s^2}\, ,\label{app:3ptMcs1}\\
{\cal Q}^{(s)*}(\cs\ll 1,k_1) &= f^{(s)}(0)\times \frac{2i(ik_1/2)^{3/2-s}}{\pi}\frac{\Gamma[\frac{1}{2}(\frac{1}{2}+s+i\mu_s)]\Gamma[\frac{1}{2}(\frac{1}{2}+s-i\mu_s)]}{(s-\frac{3}{2})^2+\mu_s^2}\, .\label{app:3ptMcs0}
\end{align}
Notice that the mixing integral becomes independent of $\cs$ in the small $\cs$ limit. The function $f^{(s)}(\cs)$ is precisely the difference between evaluating the integral (\ref{app:3ptM}) with the mode function $G^{(s)}_{i\mu_s}$ and a simple Hankel function $H_{i\mu_s}$. Since the mode function is a linear combination of Hankel functions, $f^{(s)}$ is a simple polynomial. The result for $s=2$ is
\begin{align}
	f^{(2)}(1) &= -\frac{985-664\mu_2^2+16\mu_2^4}{576}\, ,\\
	f^{(2)}(0) &= -\frac{23-4\mu_2^2}{12}\, .
\end{align}
Summing (\ref{app:3ptMcs1}) and (\ref{app:3ptM1}) and focusing on terms which are non-analytic in momentum, we find 
\begin{align}
\lim_{k_1\ll \cs k_3}{\cal I}^{(s)}(\mu_s,\cs,k_1,k_3,k_3) & =  \frac{A_s}{k_1^3k_3^3}\left(\frac{k_1}{k_3}\right)^{3/2}\cos\left[\mu_s\ln \left(\frac{k_1}{k_3}\right)+\phi_s\right] ,
\label{I1cs1}
\end{align}
where the amplitude and the phase are given by 
\begin{align}
A_s &= |\tilde A_s|\times \begin{cases} \displaystyle f^{(s)}(1)\times\frac{\sqrt{\pi}}{2^{2s-2}\hskip 1pt\Gamma(s)}\frac{\Gamma(\frac{1}{2}+s-i\mu_s)\Gamma(\frac{1}{2}+s+i\mu_s)}{(s-\frac{3}{2})^2+\mu_s^2}\, \propto\, e^{-\pi\mu_s} & \cs=1 \\[1em] \displaystyle f^{(s)}(0)\times\frac{\Gamma[\frac{1}{2}(\frac{1}{2}+s+i\mu_s)]\Gamma[\frac{1}{2}(\frac{1}{2}+s-i\mu_s)]}{(s-\frac{3}{2})^2+\mu_s^2} \, \propto\, e^{-\pi\mu_s/2} & \cs\ll 1 \end{cases}\, ,\label{As}\\[5pt]
\phi_s &\equiv \arg \tilde A_s-\mu_s\ln 4\hskip 1pt \cs
\, ,
\end{align}
with
\begin{align}
\tilde A_s\equiv \frac{i^s\pi^3\hskip 1pt s!}{8(2s-1)!!}\frac{(5+2s+2i\mu_s)(1+i\hskip 1pt{\rm sinh}\hskip 1pt \pi\mu_s)}{\cosh\pi\mu_s}\frac{\Gamma(-i\mu_s)}{\Gamma(\frac{1}{2}-i\mu_s)}\, ,
\end{align}
for even spins, whereas the result for odd spins is given by replacing $1+i\hskip 1pt{\rm sinh}\hskip 1pt \pi\mu_s\to i\cosh\pi\mu_s$. The final answer is then obtained by summing the permutations $(\k_2\leftrightarrow \k_3)$ in (\ref{eq:singleSq}). Momentum conservation implies
\begin{align}
	\hat\k_1\cdot \hat\k_2 = -\hat\k_1\cdot \hat\k_3 -\frac{k_1}{k_3}\big[1-(\hat\k_1\cdot \hat\k_3)^2\big] + {\cal O}(k_1^2/k_3^2)\, . 
\end{align}
Writing the spin as $s=2\ell+1$ for odd spins, with $\ell$ an integer, we get
\begin{align}
P_{2\ell+1}(\hat\k_1\cdot\hat\k_3) +(\k_2\leftrightarrow\k_3) &= \nonumber \\
&\hspace{-2cm}=-(2\ell+1)\frac{k_1}{k_3}\big[P_{2\ell}(\hat\k_1\cdot\hat\k_3)-(\hat\k_1\cdot\hat\k_3)P_{2\ell+1}(\hat\k_1\cdot\hat\k_3)\big] + {\cal O}(k_1^2/k_3^2)\, .\label{LegendreOdd}
\end{align}
For odd spins, the leading terms cancel in the sum over the two permutations, and the squeezed limit scales as $(k_1/k_3)^{5/2}$. Note that the right-hand side of (\ref{LegendreOdd}) is an even-degree polynomial of the angle. For even spin $s=2\ell$, we have instead (see also \cite{Assassi:2015jqa})
\begin{align}
P_{2\ell}(\hat\k_1\cdot\hat\k_3) +(\k_2\leftrightarrow\k_3) &= \nonumber \\
&\hspace{-2cm}=2P_{2\ell}(\hat\k_1\cdot\hat\k_3) -2\ell\frac{k_1}{k_3}\big[P_{2\ell-1}(\hat\k_1\cdot\hat\k_3)-(\hat\k_1\cdot\hat\k_3)P_{2\ell}(\hat\k_1\cdot\hat\k_3)\big] + {\cal O}(k_1^2/k_3^2)\, ,\label{LegendreEven}
\end{align}
where the leading terms add up and thus scale as $(k_1/k_3)^{3/2}$. The next-to-leading term at order $(k_1/k_3)^{5/2}$ is an odd-degree polynomial of the angle.

\subsection[${\langle \gamma\zeta\zeta\rangle}$]{$\boldsymbol{\langle \gamma\zeta\zeta\rangle}$}

We also studied the tensor-scalar-scalar bispectrum $\langle \gamma\zeta\zeta\rangle$.
Its squeezed limit can be written as
\begin{align}
\lim_{k_1 \ll k_3} \frac{\langle\gamma^\lambda_{\k_1}\zeta_{\k_2}^{\phantom\lambda}\zeta_{\k_3}^{\phantom\lambda}\rangle'}{\Delta_\gamma\Delta_\zeta^{3}}&\,=\,  \alpha_s\sqrt{r}\Delta_\zeta^{-1} \, {\cal E}_2^\lambda(\hat\k_1\cdot\hat\k_3)\hat P_s^\lambda(\hat\k_1\cdot\hat\k_3)\, \nonumber \\[-6pt]
&\hspace{2.3cm}\times {\cal B}^{(s)}(\mu_s,\cs,k_1,k_3,k_3)+(\k_2\leftrightarrow \k_3)\, ,
\end{align}
where
\begin{align}
{\cal B}^{(s)} &\equiv - \frac{\pi^3\cs^{s-3/2}}{8}\sum_{\pm\pm} (\pm ik_1^{s-3})(\pm i\cs^2k_3^{s-4})\, {\cal B}_{\pm\pm}\, ,\label{eq:B}\\
{\cal B}^{(s)}_{\pm\pm} &\equiv \int_{-\infty}^0\frac{\d\eta}{\eta^{2-2s}} (1\mp i\cs k_3\eta)e^{\pm 2i\cs k_3\eta}\int_{-\infty}^0\d\tilde\eta\, e^{\pm i k_1\tilde\eta} \, \tilde G_{\pm\pm}(k_1,\eta,\tilde\eta)\, ,
\end{align}
with $\tilde G_{\pm\pm}$ the Green's functions for the helicity-$\pm 2$ mode, $\sigma^{\pm 2}_{s,s}$. Following the same steps as in the scalar case, we obtain the following factorized form of the integrals
\begin{align}
{\cal B}^{(s)} &= \tilde N^{(s)}\sum_\pm (\pm k_1^{s-3}k_3^{s-4})\tilde{\cal P}_\pm (k_1,\cs k_3) \tilde{\cal Q}^* (k_1)+ c.c.\, ,
\end{align}
where
\begin{align}
\tilde N^{(s)} &\equiv -\frac{9\pi^4\cs^{s+1/2}}{32}\frac{(s-2)!s!}{(s+2)!(2s-1)!!}\frac{\Gamma(\frac{5}{2}-i\mu_s)\Gamma(\frac{5}{2}+i\mu_s)}{\Gamma(\frac{1}{2}+s-i\mu_s)\Gamma(\frac{1}{2}+s+i\mu_s)} \, ,\\
\tilde{\cal P}_\pm (k_1,\cs k_3) &\equiv  e^{-\pi\mu_s/2}\int_0^{\infty}\d x\, x^{s-1/2}(1\mp i\cs k_3x) \tilde G^{(s)}_{i\mu_s}(k_1x)e^{\pm 2i\cs k_3x}\, ,\label{app:3ptPtilde} \\
\tilde{\cal Q}^*(k_1) &\equiv e^{-\pi\mu_s/2}\int_0^{\infty}\d x\, x^{s-5/2} \tilde G^{(s)*}_{i\mu_s}(k_1x)e^{-ik_1x}\, ,\label{app:3ptQtilde}
\end{align}
with $\tilde G^{(s)}_{i\mu_s}$ defined in (\ref{Gmode}). In the squeezed limit, $k_1\ll \cs k_3$, the integral (\ref{app:3ptPtilde}) becomes
\begin{align}
\lim_{k_1\ll \cs k_3}\tilde {\cal P}^{(s)}_\pm(k_1,\cs k_3) &=  \frac{i^{3/2}e^{(1\mp 1)\pi\mu_s/2}\Gamma(\frac{1}{2}+s+i\mu_s)\Gamma(\frac{1}{2}+s-i\mu_s)}{4\pi(\pm 2 \cs k_3)^{1/2+s}}\nonumber\\
&\ \ \ \times \left(\frac{k_1}{4\cs k_3}\right)^{i\mu_s}(5+2s+2i\mu_s)\frac{\Gamma(-i\mu_s)}{\Gamma(\frac{5}{2}-i\mu_s)}\mp i e^{-(1\mp 1)\pi\mu_s} \times c.c.\, .
\end{align}
The integral (\ref{app:3ptQtilde}) is given by
\begin{align}
{\tilde Q}^{(s)*}(k_1) =\tilde f^{(s)}\times \frac{2i(2ik_1)^{3/2-s}}{\sqrt{\pi}\hskip 1pt\Gamma(s-1)}\frac{\Gamma(\frac{1}{2}+s-i\mu_2)\Gamma(\frac{1}{2}+s-i\mu_2)}{\big((s-\frac{3}{2})^2+\mu_s^2\big)\big((s-\frac{1}{2})^2+\mu_s^2\big)}\, ,
\end{align}
where $\tilde f^{(s)}$ a polynomial of $\mu_s$ that encodes the difference between evaluating the integral with $\tilde G^{(s)}_{i\mu_s}$ and $H_{i\mu_s}$. For spin-2, this is simply $\tilde f^{(2)}=1$. The bispectrum is then given by
\begin{align}
{\cal B}^{(s)}(\mu_s,\cs,k_1,k_3,k_3) & =  \frac{|B_s|}{k_1^3k_3^3}\left(\frac{k_1}{k_3}\right)^{3/2}\cos\left[\mu_s\ln \left(\frac{k_1}{k_3}\right)+\tilde\phi_s\right] ,
\label{Bcs1}
\end{align}
where
\begin{align}
B_s&\equiv  -\tilde f^{(s)}\times \frac{9i^s\pi^{7/2}}{2^{2s+4}}\frac{(5+2s+2i\mu_s)(1+i\sinh\pi\mu_s)}{(s+1)(s+2)(2s-1)!!\cosh\pi\mu_s}\frac{\Gamma(-\frac{3}{2}+s-i\mu_s)\Gamma(-\frac{3}{2}+s+i\mu_s)}{ \Gamma(-i\mu_s)^{-1}\Gamma(-\frac{3}{2}-i\mu_s)}\, ,\\[4pt] 
\tilde\phi_s &\equiv \arg B_s - \mu_s\ln 4\hskip 1pt \cs\, ,
\end{align}
for even spins. The result for odd spins requires the replacement $1+i\hskip 1pt{\rm sinh}\hskip 1pt \pi\mu_s\to \cosh\pi\mu_s$. The final bispectrum is then obtained by summing over the permutations $(\k_2\leftrightarrow\k_3)$.

\newpage
\bibliographystyle{utphys}
\bibliography{HigherSpin}

\end{document}